\documentclass[slac_one]{revtex4}
\usepackage{graphicx}
\usepackage{fancyhdr}
\usepackage{epsfig}
\def\oo{\'o}
\def\pp{p\bar p}
\def\q2{Q^2}
\def\kt{k_T}
\def\as{\alpha_s}
\def\xo{x_{\gamma}^{\rm obs}}
\def\etjet{E_T^{\rm jet}}
\def\etajet{\eta^{\rm jet}}
\def\xpo{x_p^{\rm obs}}
\def\g2{GeV$^2$}
\def\asz{\as(\mz)}
\def\mz{M_Z}
\def\oas{{\cal O}(\as)}
\def\oass{{\cal O}(\as^2)}
\def\oasss{{\cal O}(\as^3)}
\def\colab#1{#1 Coll.}
\def\Journal#1#2#3#4{{#1} {#2} (#3) #4}

\def\NPB{{\em Nucl. Phys.} {\bf B}}
\def\PLB{{\em Phys. Lett.}  {\bf B}}
\def\PRL{{\em Phys. Rev. Lett.}}
\def\PRD{{\em Phys. Rev.} {\bf D}}

\def\EPC{{\em Eur. Phys. Jour.} {\bf C}}

\def\etal{et al.}

\begin{document}

\title{Precision tests of QCD with jets and vector bosons at HERA
  and TevaTron}

\author{C. Glasman}
\affiliation{Universidad Aut\oo noma de Madrid, Spain}

\begin{abstract}
Recent results from HERA and TevaTron on precision tests of QCD with
jets, $W$ and $Z$ bosons and photons associated with jets and heavy
flavours are presented. The measurements were used to probe QCD at the
highest energies, to provide experimental constraints on SM processes
that constitute background to new physics, to extract values of the
coupling of the strong interaction and to constrain the proton parton
distribution functions. The implications of the results on LHC physics
are discussed.
\end{abstract}

\maketitle

\thispagestyle{fancy}

\section{Introduction}
Parton-parton scattering via QCD interactions is the main process
at hadron colliders. Besides their intrinsic interest, these processes
represent sometimes an irreducible background to other Standard Model
(SM) processes and to new physics.

The $ep$ collider HERA, which provides a simple hadronic enviroment to
test QCD, and the $\pp$ collider TevaTron, which provides the highest
available energies, have produced over the years complementary
precision measurements of the parameters of the theory and precise
tests of its underlying dynamics. All this knowledge will be crucial
for understanding any physics at LHC. For instance, the gluon density
at low $x$ is a necessary input ingredient for calculating the Higgs
production cross section, whereas the predictions for $W$ or $Z$
bosons production cross sections require knowledge of the quark
densities.

This report summarises the most recent results from jet cross
sections, forward jets, photons and $W$ and $Z$ bosons and the
underlying event presented by the CDF, D\O, H1 and ZEUS experiments at
ICHEP08.

\section{Studies of the underlying event}
The main experimental uncertainties for jet cross sections arise from
the jet energy scale and the underlying event (UE), which contributes 
with additional energy density on top of the hard interaction but is
not related to it; this is a non-perturbative effect which is not
included in the calculations. These effects can be simulated with
Monte Carlo simulations and are extremely model dependent, so a good
understanding of the UE at the available energies is crucial to model
its effects at LHC energies.

\begin{figure}[h]
\setlength{\unitlength}{1.0cm}
\begin{picture} (18.0,4.5)
\put (2.0,-0.5){\epsfig{figure=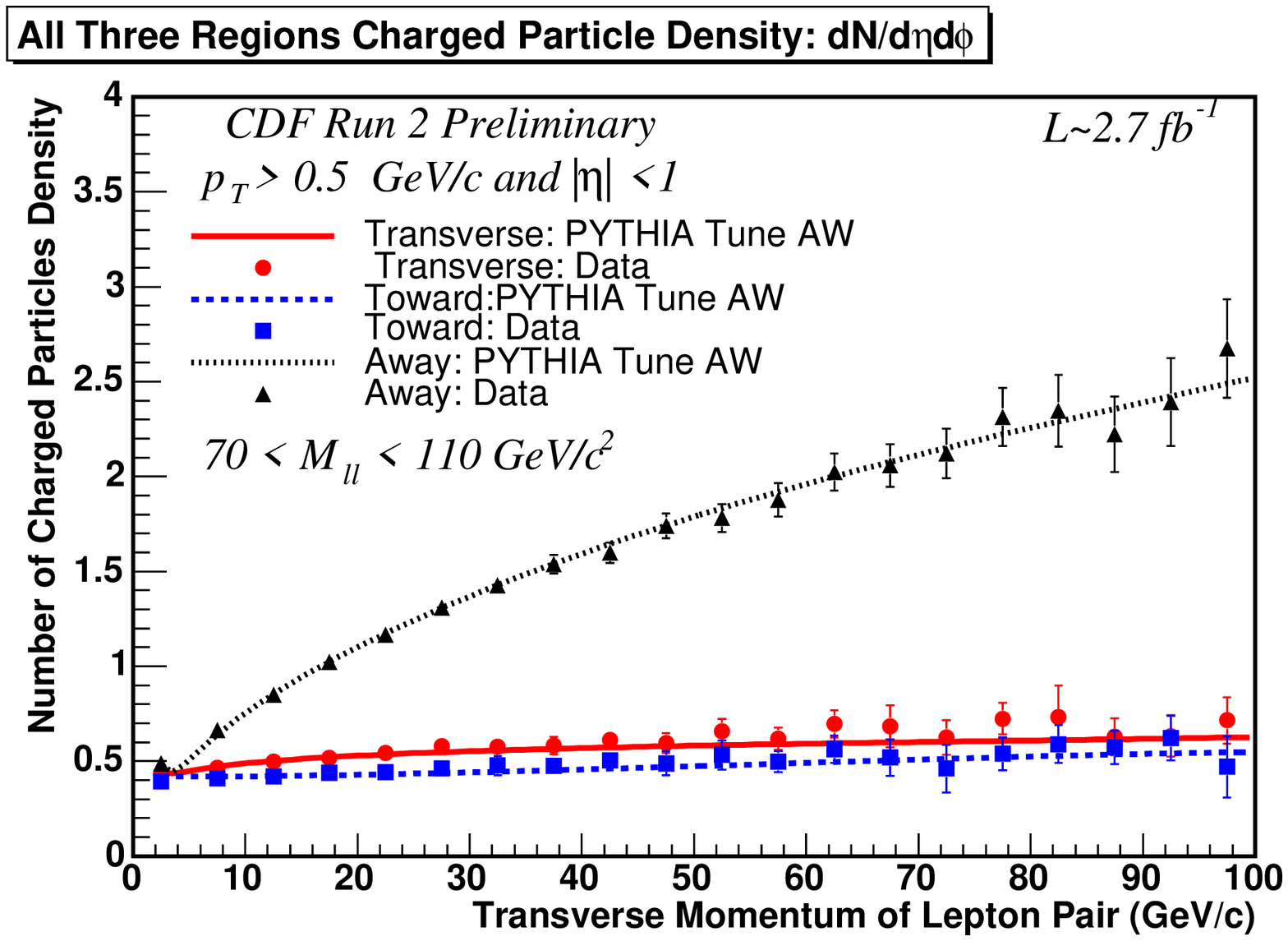,width=7cm}}
\put (8.5,-0.5){\epsfig{figure=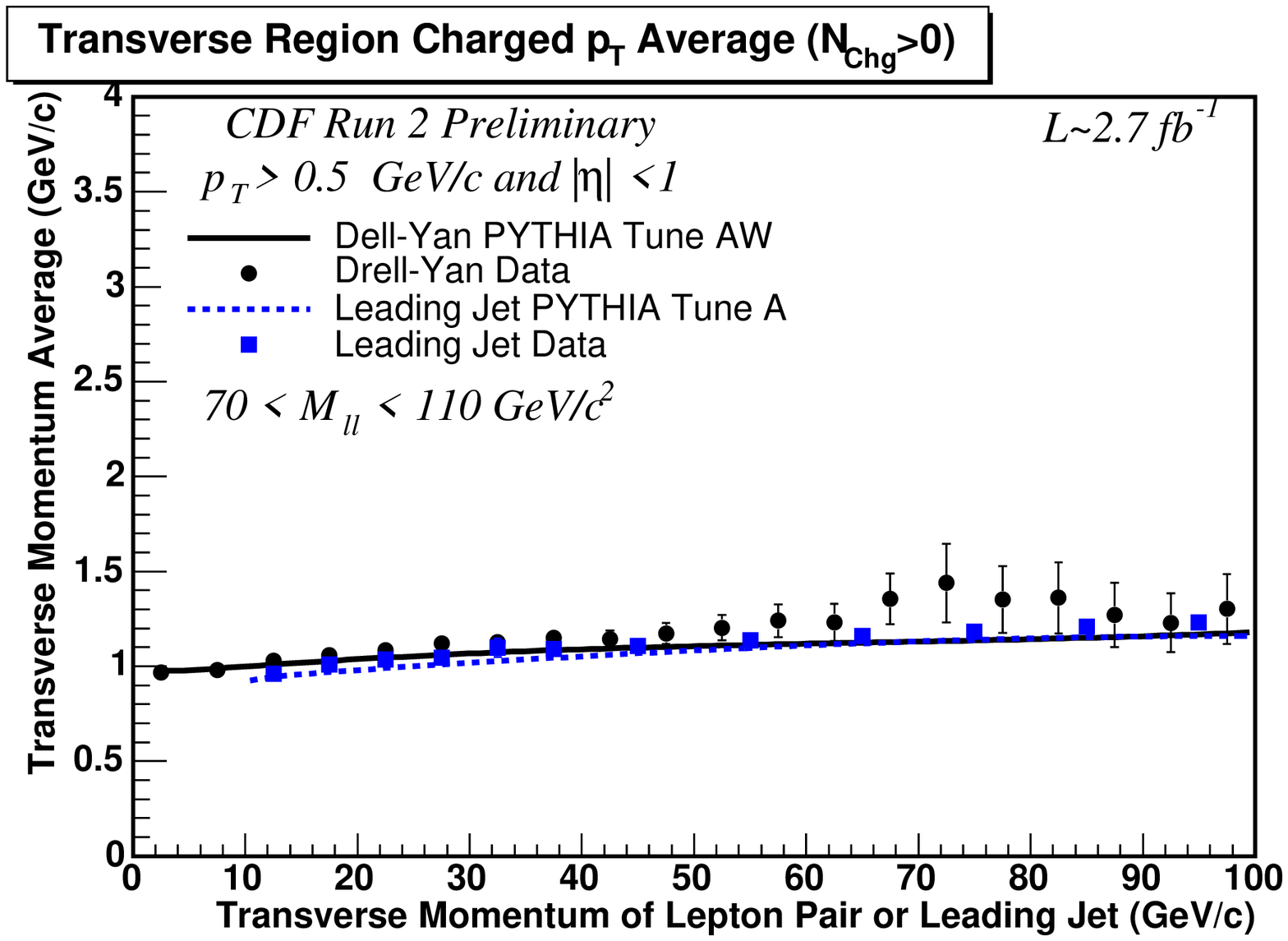,width=7cm}}
\put (7.0,3.0){\small (a)}
\put (13.5,3.0){\small (b)}
\end{picture}
\caption{\label{fig2}
{(a) Charged particle density vs. transverse momentum of lepton pair;
  (b) transverse momentum average vs. transverse momentum of lepton
  pair or leading jet.}}
\end{figure}

The CDF collaboration has performed a new study of the UE using
Drell-Yan processes~\cite{cdf419}. In this analysis, all particles in
the final state, except the lepton pair, can be considered as the
UE. Thus, the Drell-Yan events constitute a very clean probe of the
UE. The study of the UE observables was done for lepton-pair masses
around that of the Z boson ($70<M_{l\bar l}<110$ GeV). The transverse
plane was separated in four regions: the toward region, which
corresponds to the Z-boson direction, the opposite direction, which is
called the away region, and the two transverse regions. The transverse
regions are most sensitive to the UE. Several observables, such as the
charged particle density shown in Fig.~\ref{fig2}a, were studied as
functions of the transverse momentum of the lepton pair. The
measurements are approximately constant for the transverse and toward
regions and increase with lepton-pair $p_T$ in the away region. The
{\sc PYTHIA} tune AW MC predictions describe very well the
data. Figure~\ref{fig2}b shows the comparison of the UE observables in
Drell-Yan events with the same observable for the leading-jet analysis
in the transverse region. The results are in good agreement so the
Drell-Yan studies provide insight into the UE in a cleaner environment.

\begin{figure}[h]
\setlength{\unitlength}{1.0cm}
\begin{picture} (18.0,7.0)
\put (0.5,3.5){\epsfig{figure=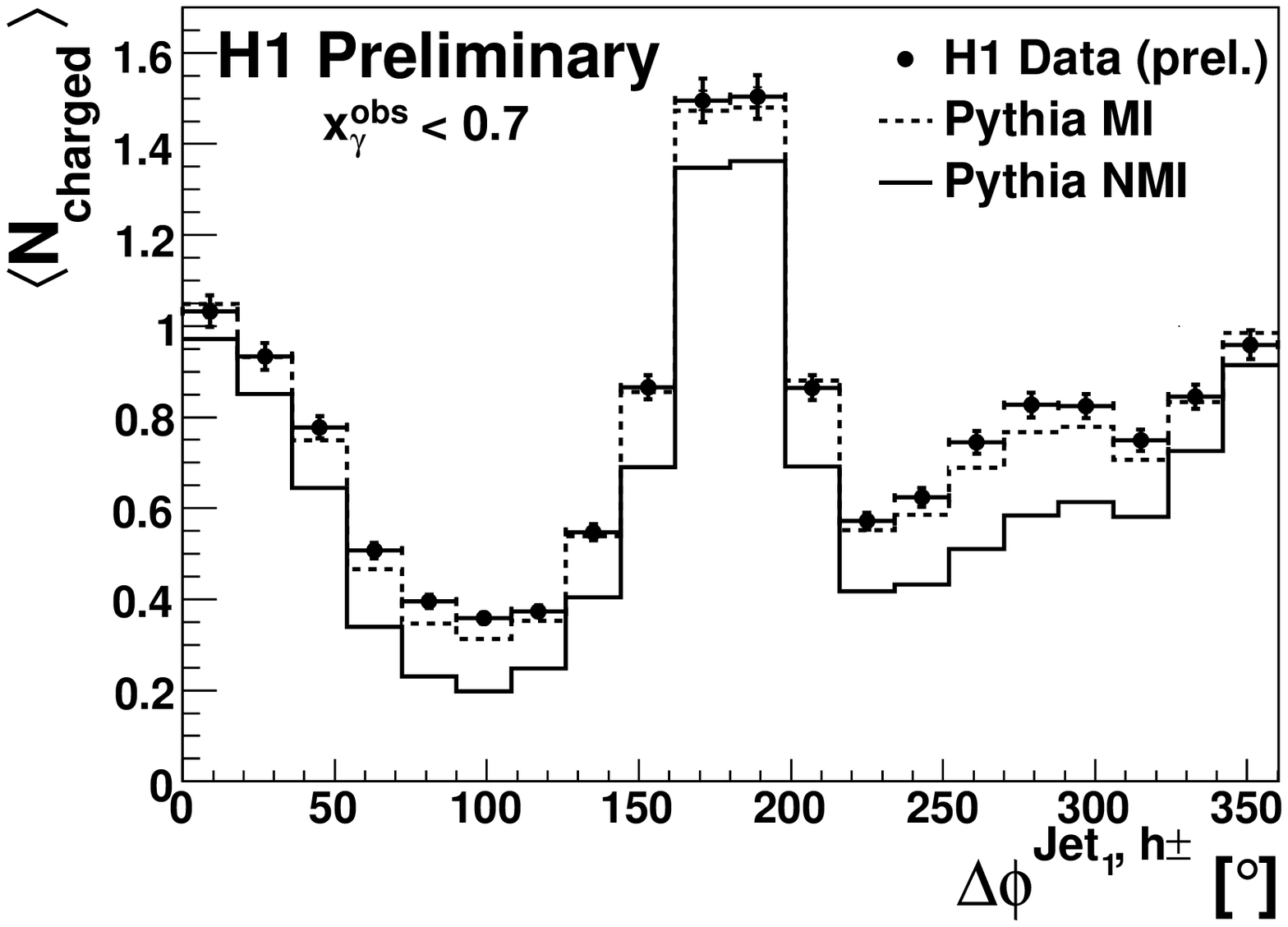,width=5cm}}
\put (6.0,3.5){\epsfig{figure=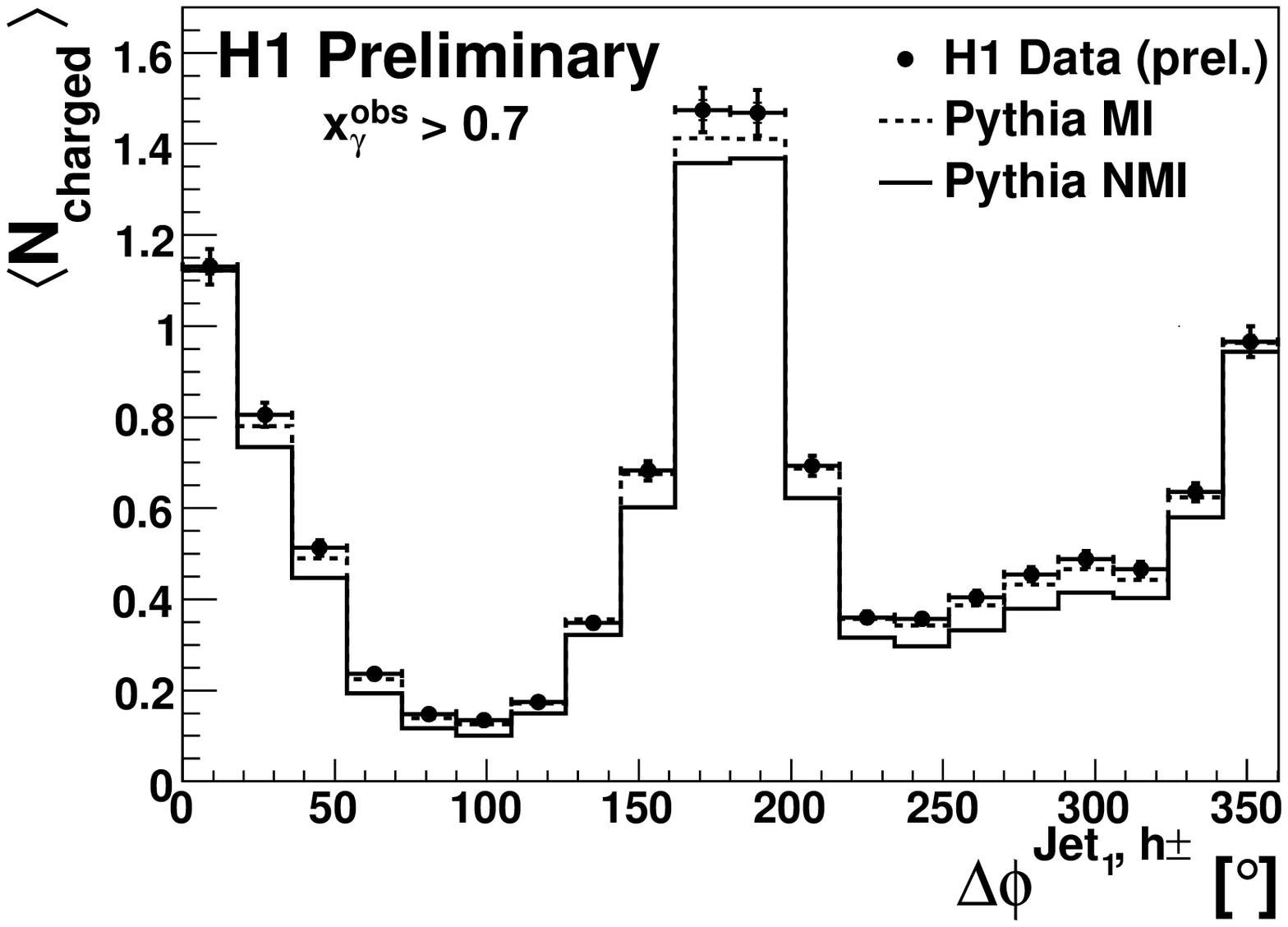,width=5cm}}
\put (11.5,3.5){\epsfig{figure=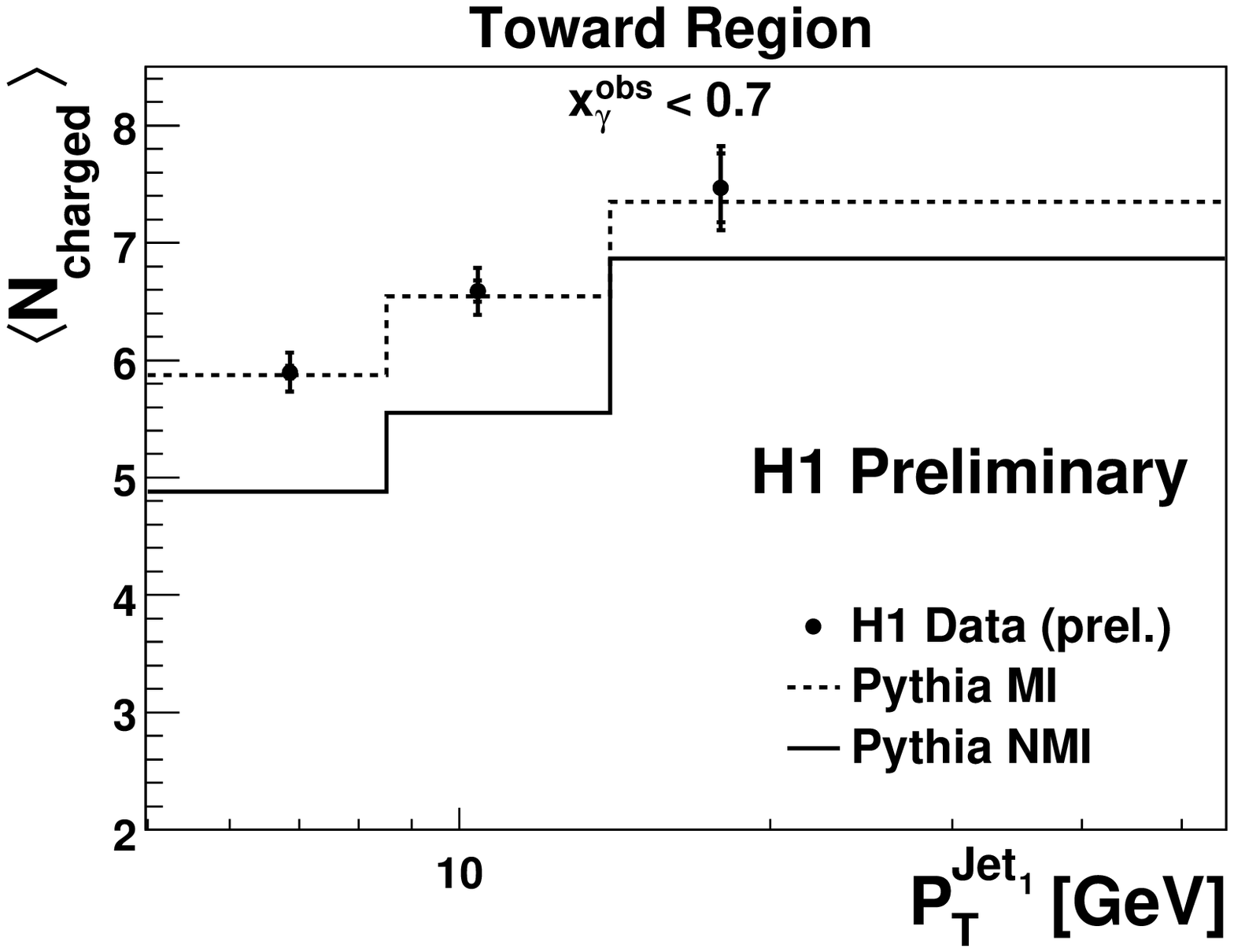,width=5cm}}
\put (0.5,-0.5){\epsfig{figure=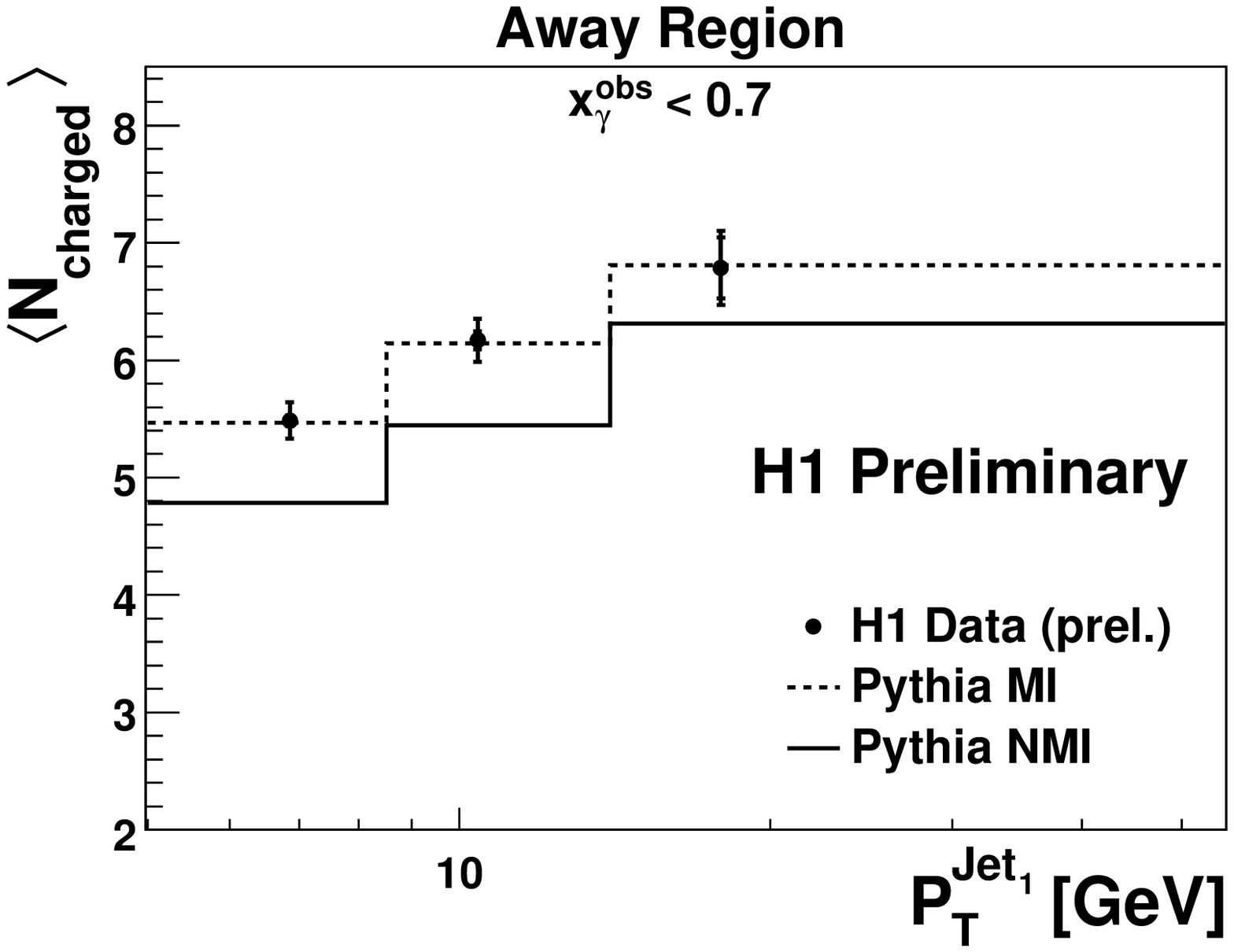,width=5cm}}
\put (6.0,-0.5){\epsfig{figure=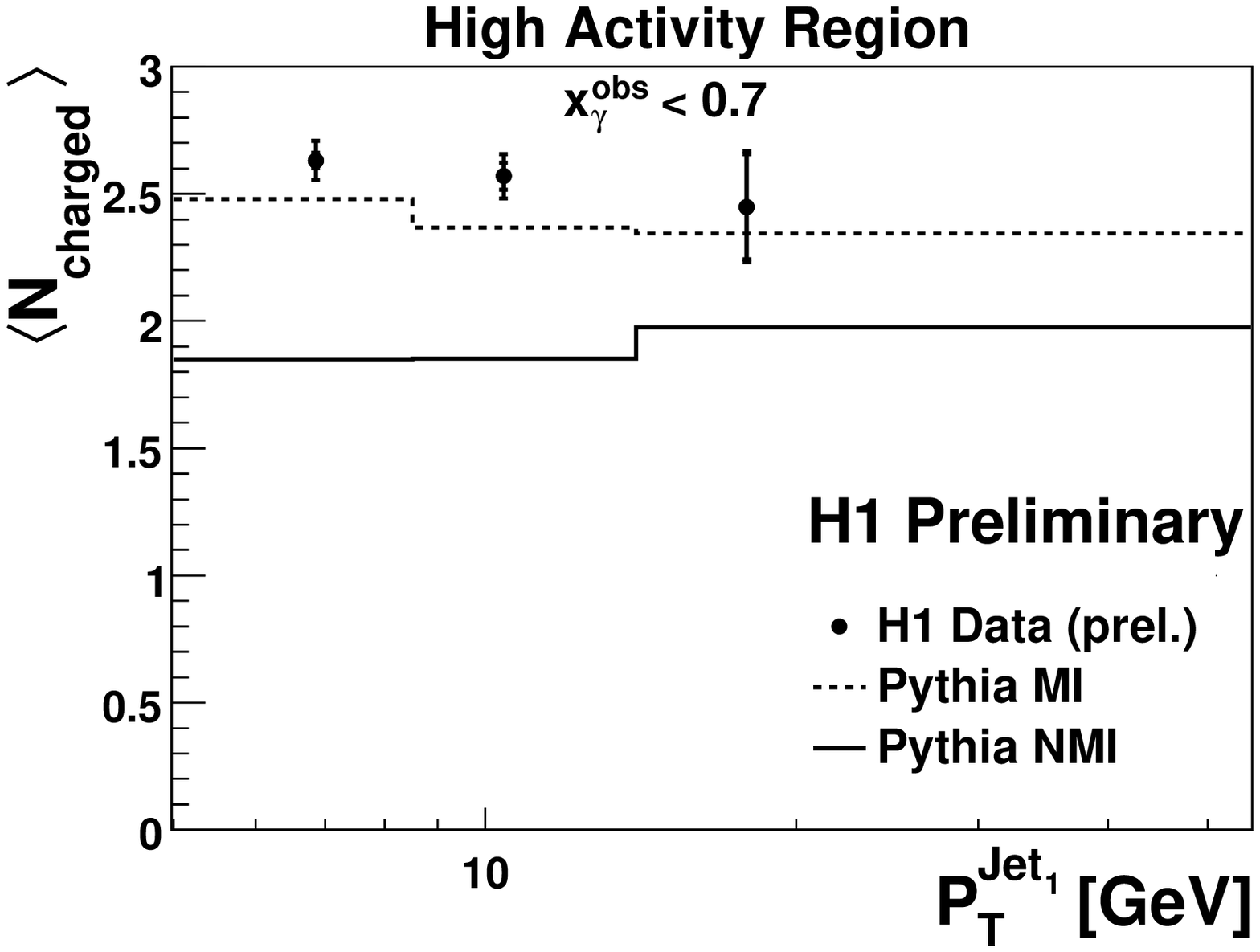,width=5cm}}
\put (11.5,-0.5){\epsfig{figure=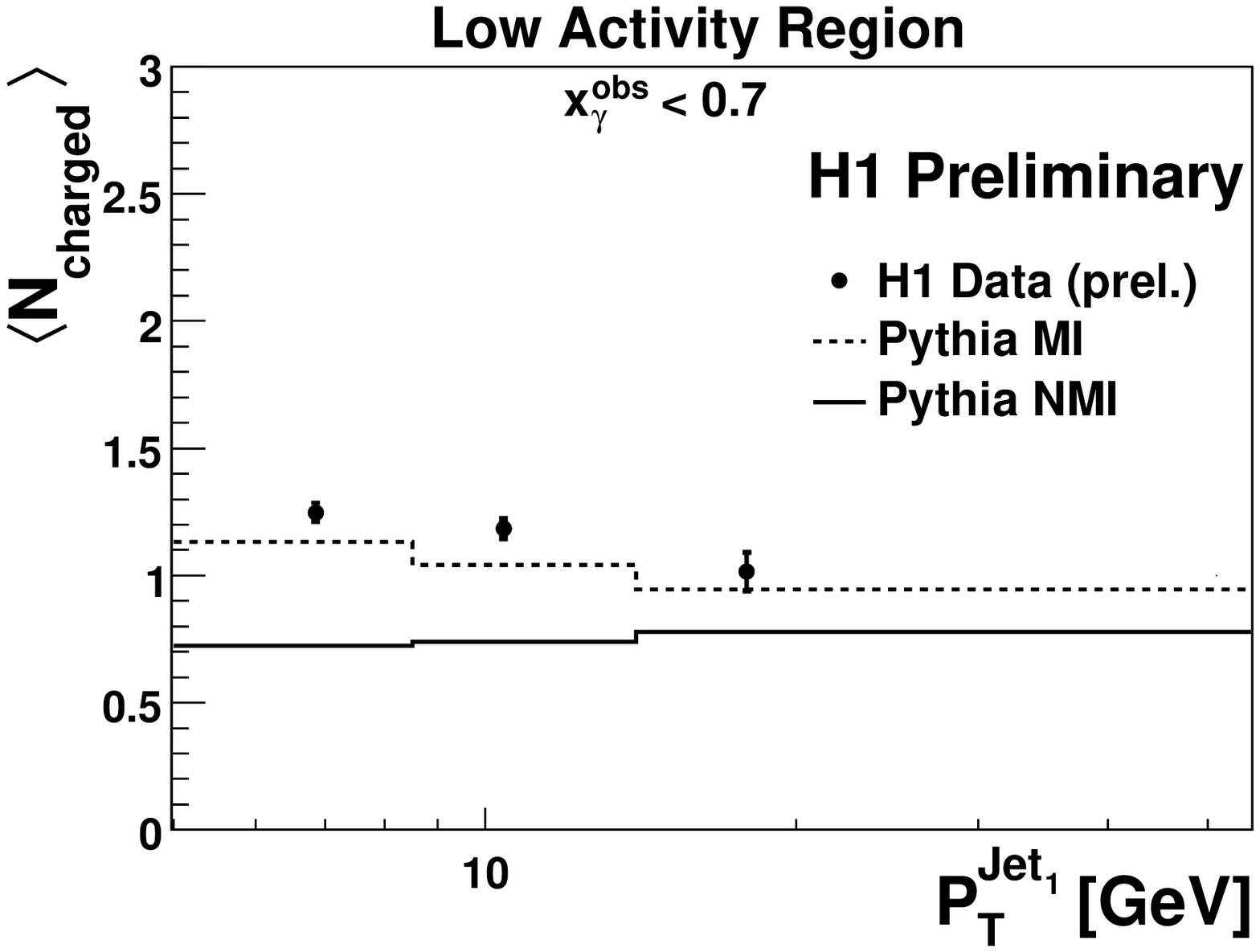,width=5cm}}
\put (4.0,6.0){\small (a)}
\put (9.5,6.0){\small (b)}
\put (15.0,6.8){\small (c)}
\put (4.0,2.8){\small (d)}
\put (9.5,2.8){\small (e)}
\put (15.0,2.8){\small (f)}
\end{picture}
\caption{\label{fig3}
{Mean charged multiplicity as a function of $\Delta\phi^{\rm jet}$ for
  (a) $\xo<0.7$ and (b) $\xo>0.7$; mean charged multiplicity as a
  function of $p_T^{\rm jet}$ in the (c) toward, (d) away, (e)
  high-activity and (f) low-activity regions.}}
\end{figure}

\begin{figure}[h]
\setlength{\unitlength}{1.0cm}
\begin{picture} (18.0,5.0)
\put (2.5,-0.5){\epsfig{figure=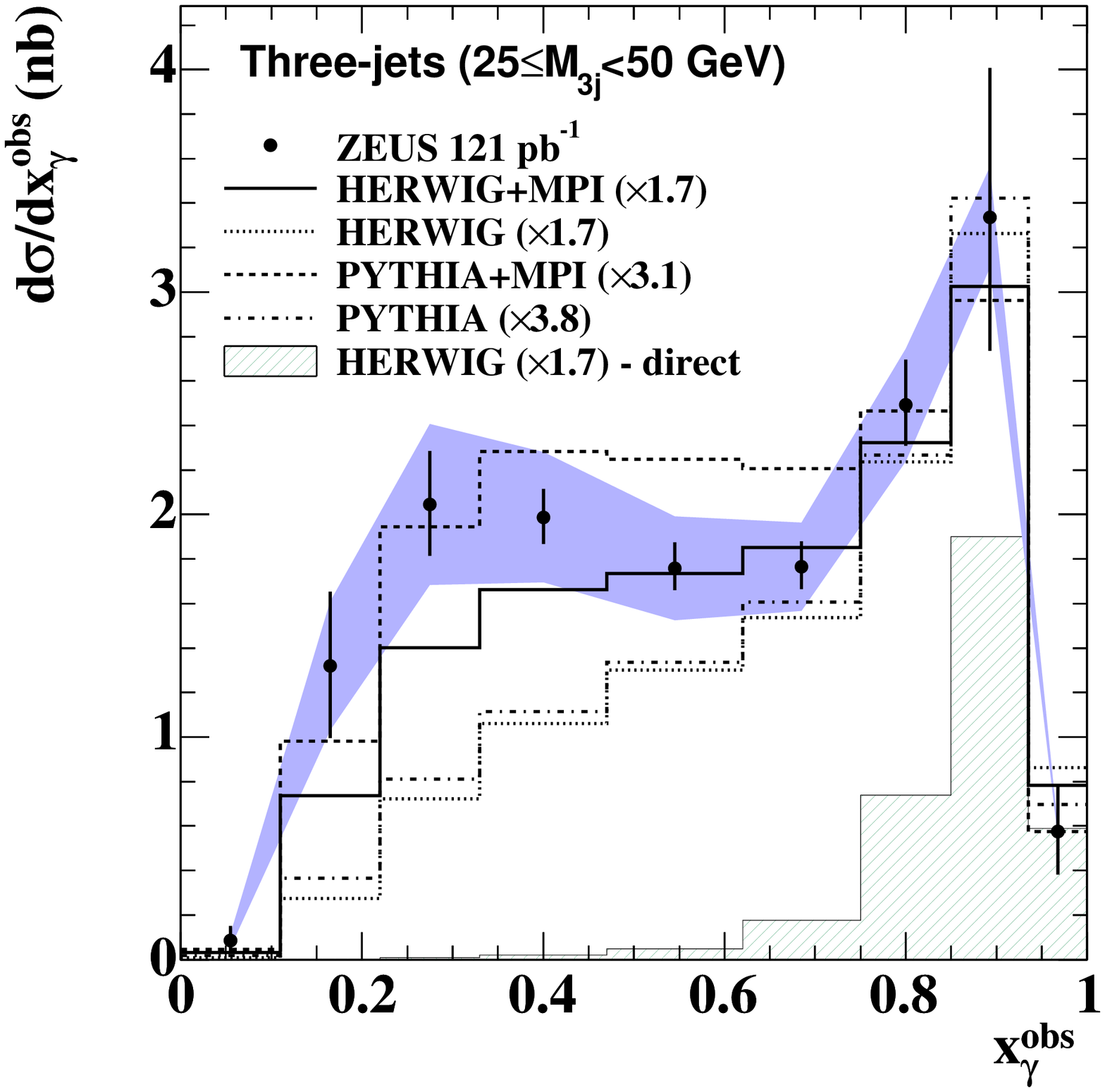,width=6cm}}
\put (8.5,-0.5){\epsfig{figure=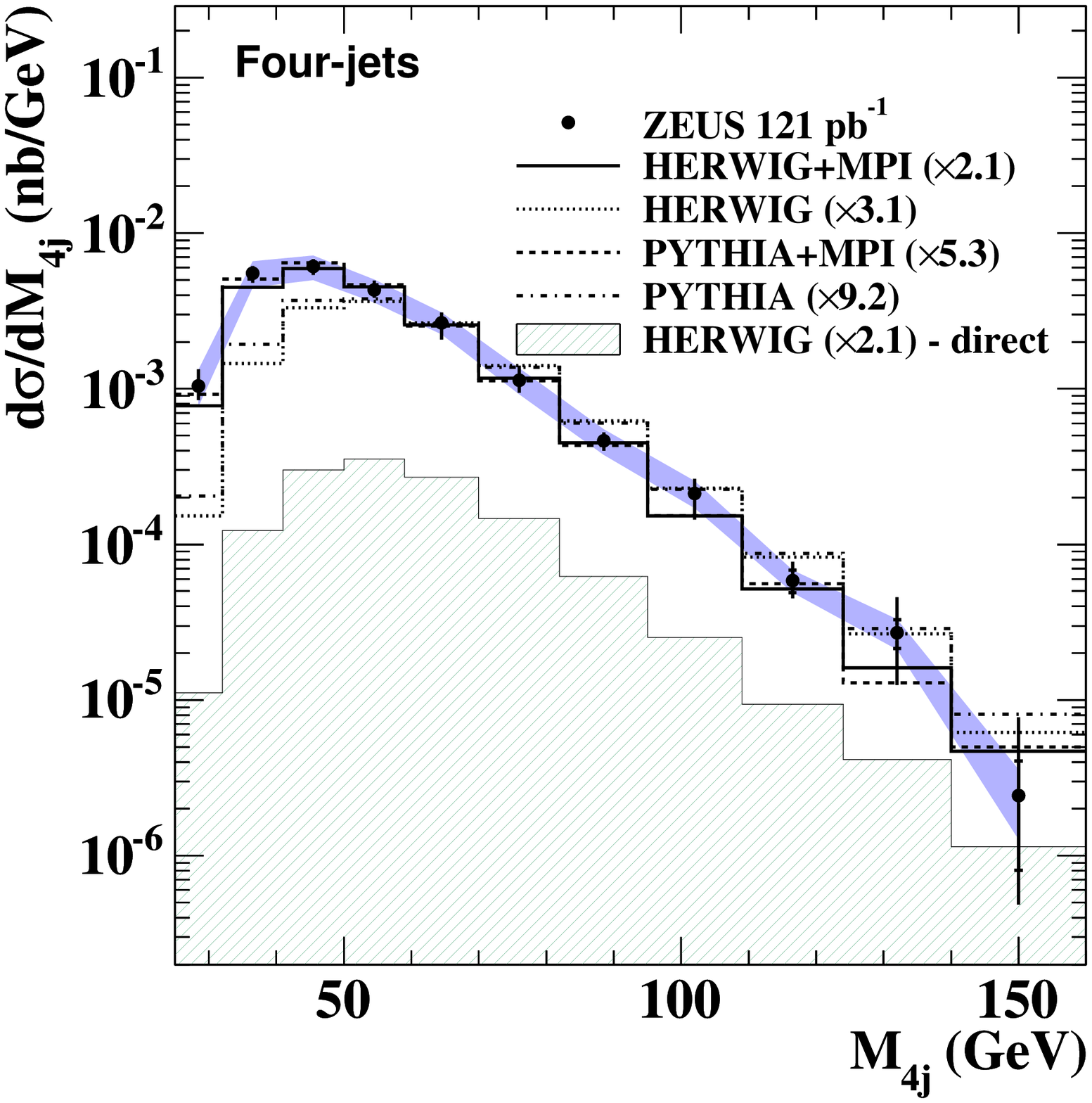,width=6cm}}
\put (6.5,4.0){\small (a)}
\put (12.5,3.0){\small (b)}
\end{picture}
\caption{\label{fig5}
{(a) Three-jet cross section as a function of $\xo$; (b) four-jet
  cross section as a function of the four-jet invariant mass.}}
\end{figure}

Photoproduction at HERA also allows the study of the UE. At leading
order (LO), there are two processes that contribute to jet
photoproduction: direct, in which the photon interacts as a point-like
particle with the partons from the proton, and resolved, in
which the photon interacts via its partonic structure, giving rise to
a hadronic-like final state. The observable 
$\xo=(1/E_{\gamma})({\sum_{\rm jets}\etjet e^{-\etajet}})$, where
$\etjet$ is the jet transverse energy, $\etajet$ is the jet
pseudorapidity and $E_{\gamma}$ is the energy of the incoming photon,
measures the fraction of the photon energy invested in the hard
interaction and can be used to separate both contributions since
direct processes take high values of $\xo$, whereas resolved processes
take smaller values. Measurements of multijet production in
photoproduction are directly sensitive to high orders and can be
used to test the parton-shower models. Resolved processes allow tests
of multiparton interactions and the UE.
The UE has been studied at HERA using dijet events in
photoproduction. Jet profiles have been measured~\cite{h1849} by the
H1 Collaboration in terms of the mean charged multiplicity as a
function of the azimuthal angle, separately for $\xo<0.7$, dominated
by resolved processes, and $\xo>0.7$, dominated by direct (see
Fig.~\ref{fig3}a and b). For $\xo<0.7$, a significant contribution
from the UE is expected. The mean charged multiplicity has also been
measured as a function of the leading-jet $\etjet$ for $\xo<0.7$ in
four regions of $\phi$ (see Fig.~\ref{fig3}c, d, e and f). The mean
charged multiplicity increases with increasing $\etjet$ in the toward
and away regions and decreases in the transverse regions. The
inclusion of multiparton interactions (MPIs) in the {\sc PYTHIA} MC
prediction improves the description of the data in all four
regions. The ZEUS Collaboration has measured~\cite{zeus130} the
three-jet cross section in photoproduction as a function of $\xo$ for
invariant masses below 50 GeV. The measurement is shown in
Fig.~\ref{fig5}a and displays a two-peak structure at low and high
$\xo$ values, which can be identified with the resolved and direct
processes, respectively. Figure~\ref{fig5}b shows the four-jet
invariant mass cross section. The data are compared with the
predictions of {\sc PYTHIA} and {\sc HERWIG} MC models with and
without MPIs; the models without MPIs describe the data at high
invariant mass for all $\xo$ values, but fail at low invariant mass
and low $\xo$. The models which include MPIs give an improved
description of the data at low $\xo$ and low mass.

\section{Probing QCD at the highest energies}
\begin{figure}[h]
\setlength{\unitlength}{1.0cm}
\begin{picture} (18.0,5.5)
\put (2.0,-0.5){\epsfig{figure=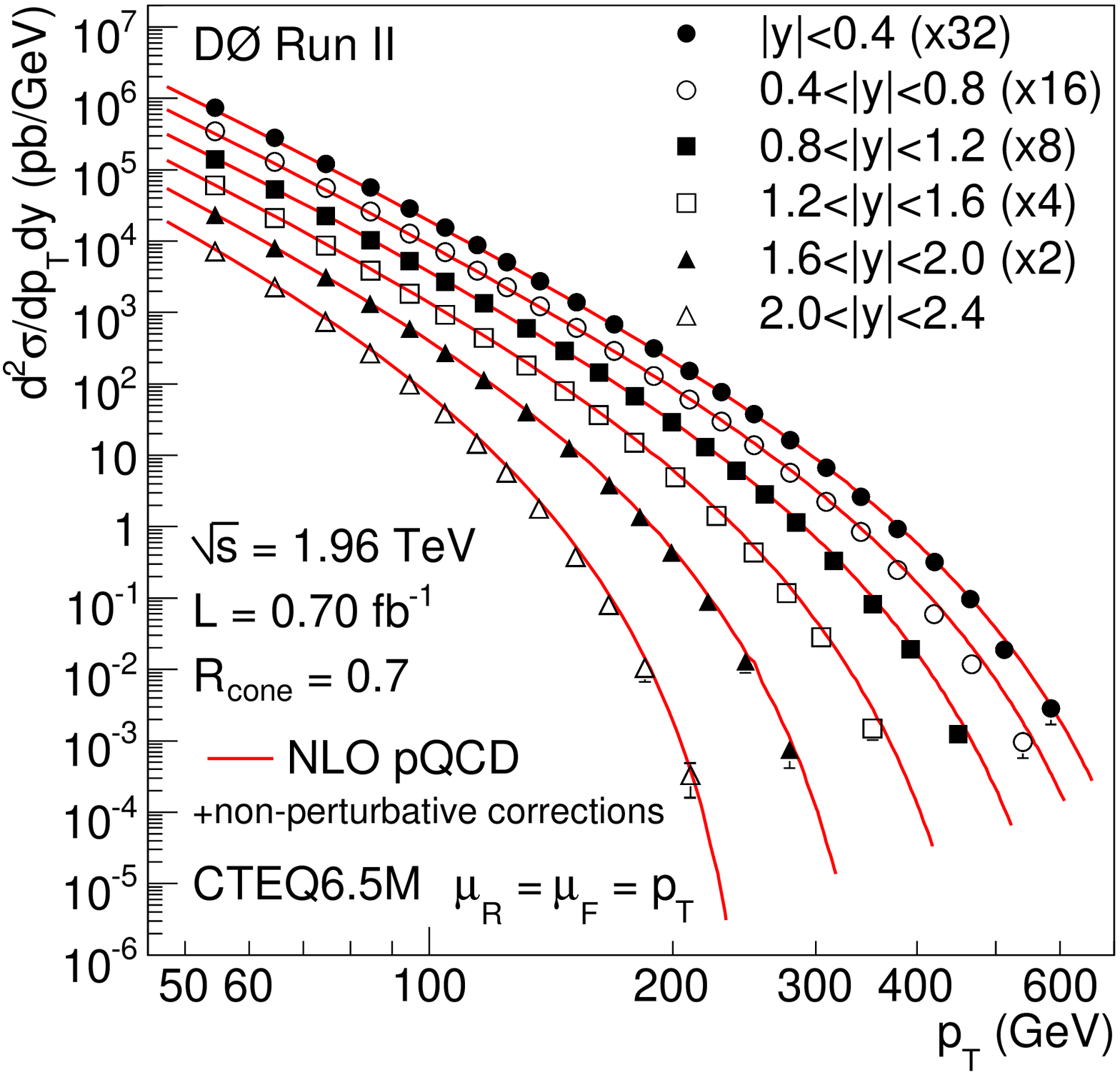,width=6cm}}
\put (8.0,-0.8){\epsfig{figure=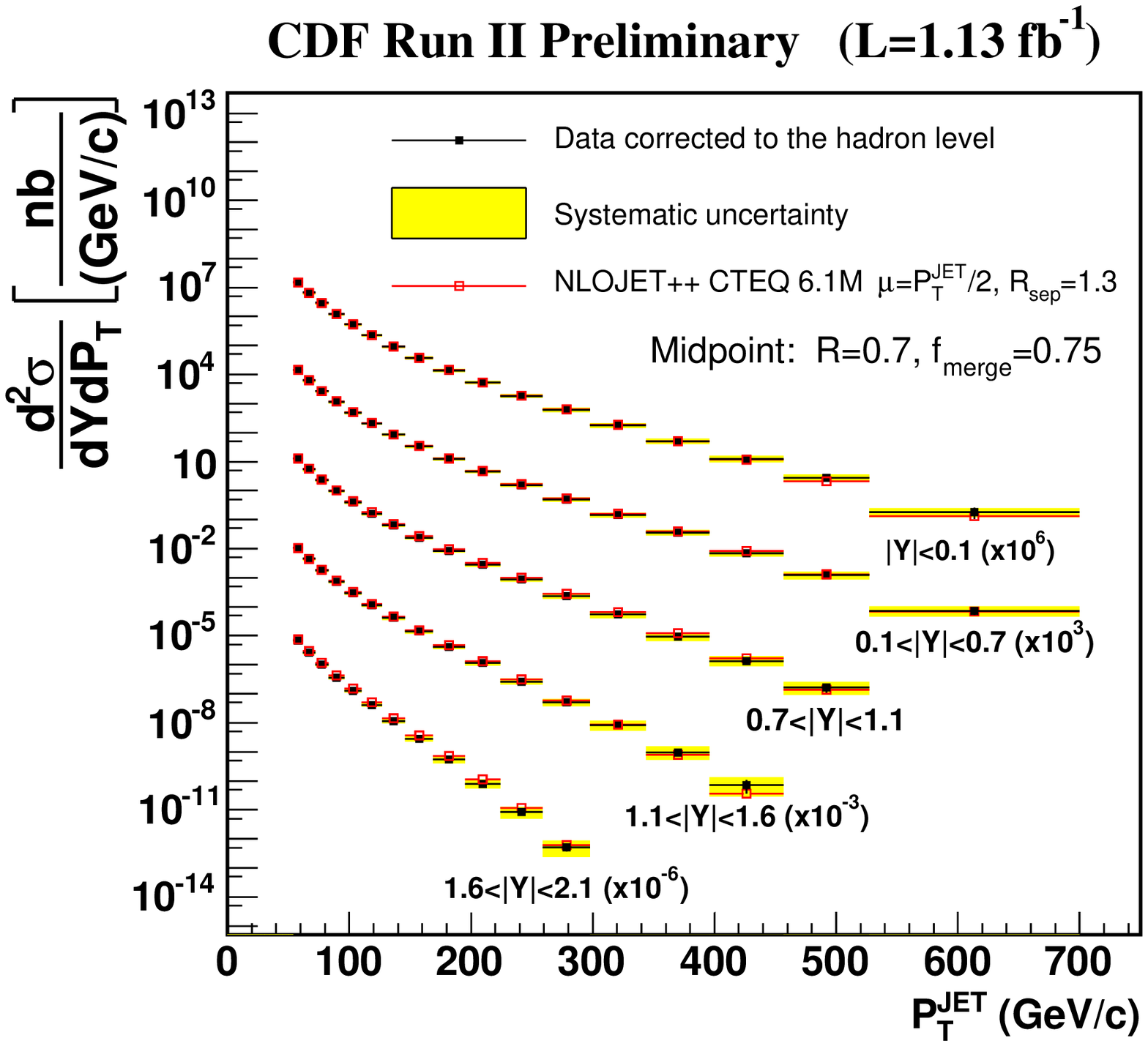,width=8cm}}
\put (7.0,3.0){\small (a)}
\put (14.5,3.0){\small (b)}
\end{picture}
\caption{\label{fig6}
{Inclusive-jet cross sections as functions of $\etjet$ in different
  rapidity regions measured by (a) D\O\ and (b) CDF.}}
\end{figure}

The CDF~\cite{cdf452} and the D\O~\cite{d0506} Collaborations have
measured the inclusive-jet cross section in $\pp$ collisions at 
$\sqrt s=1.96$ TeV as a function of $\etjet$ for different regions of
rapidity using the mid-point jet algorithm (see Fig.~\ref{fig6}). 
These are high precision measurements, especially at high $\etjet$,
where new physics might show up. These cross sections are also
sensitive to the gluon density at high $x$. The next-to-leading-order
(NLO) QCD calculations give a good description of the data. These
measurements constitute the most stringent test of pQCD at the highest
available energies so far.

\begin{figure}[h]
\setlength{\unitlength}{1.0cm}
\begin{picture} (18.0,10.0)
\put (0.7,4.5){\epsfig{figure=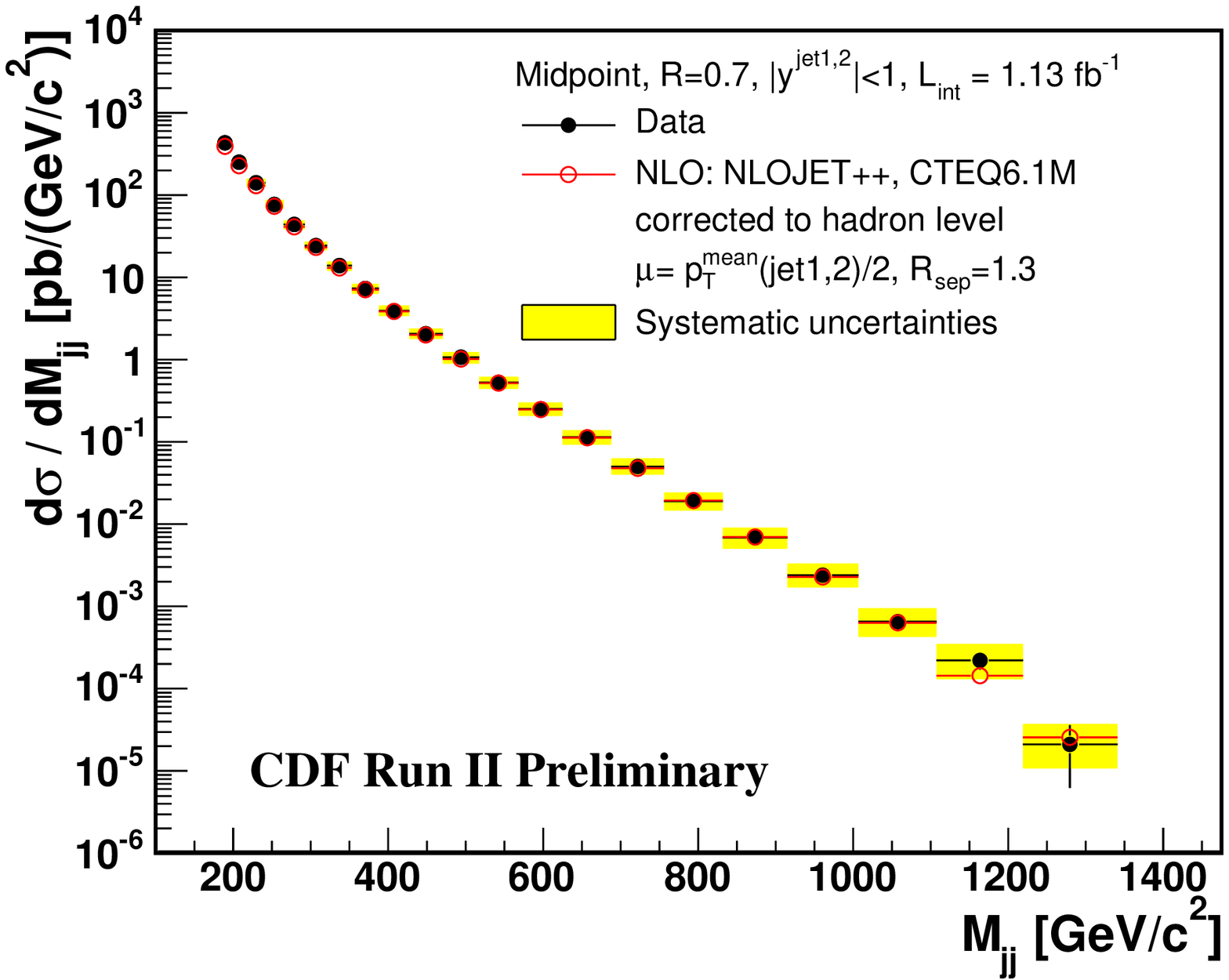,width=7cm}}
\put (8.7,4.5){\epsfig{figure=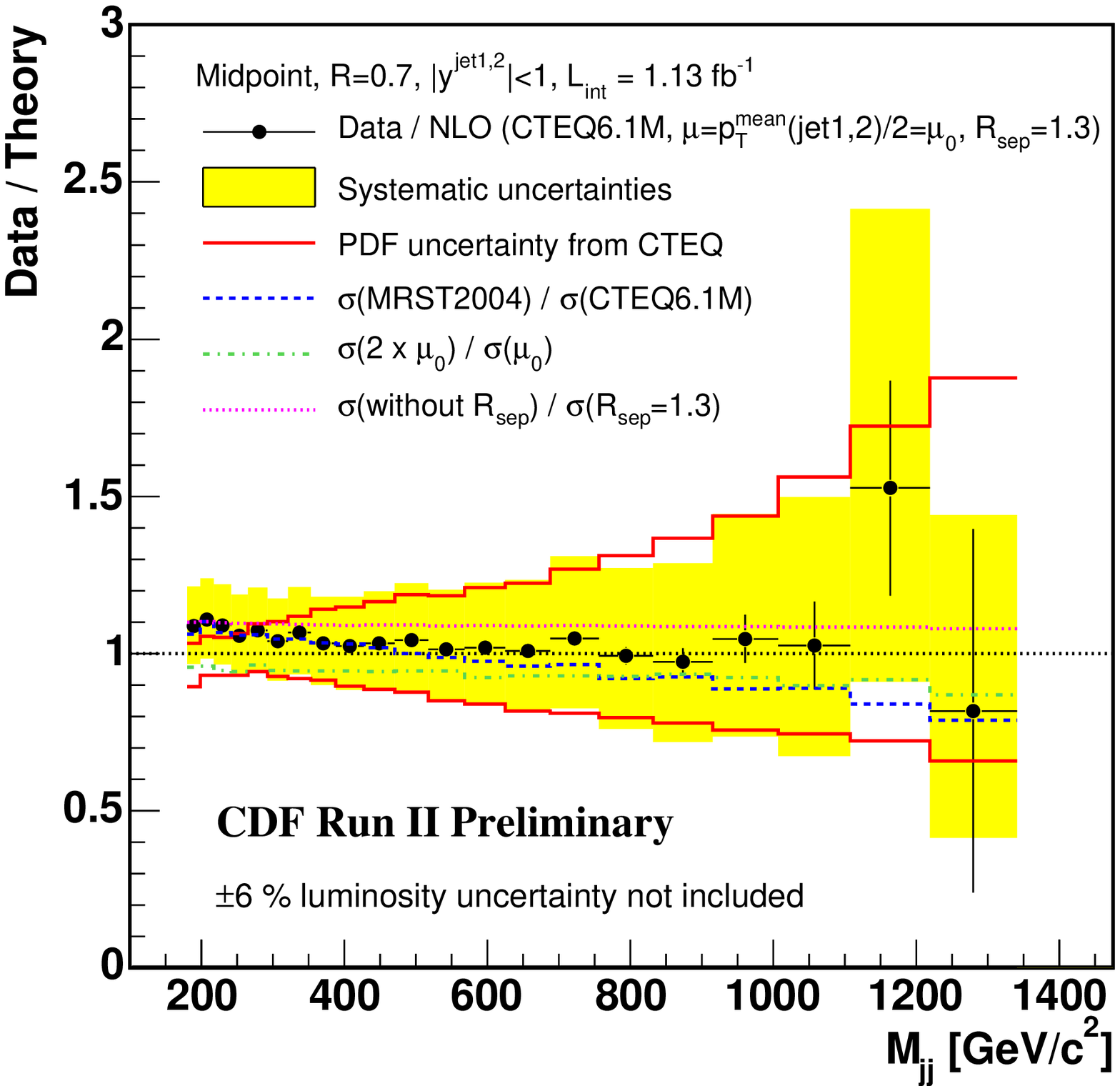,width=7cm,height=5.5cm}}
\put (1.0,-0.5){\epsfig{figure=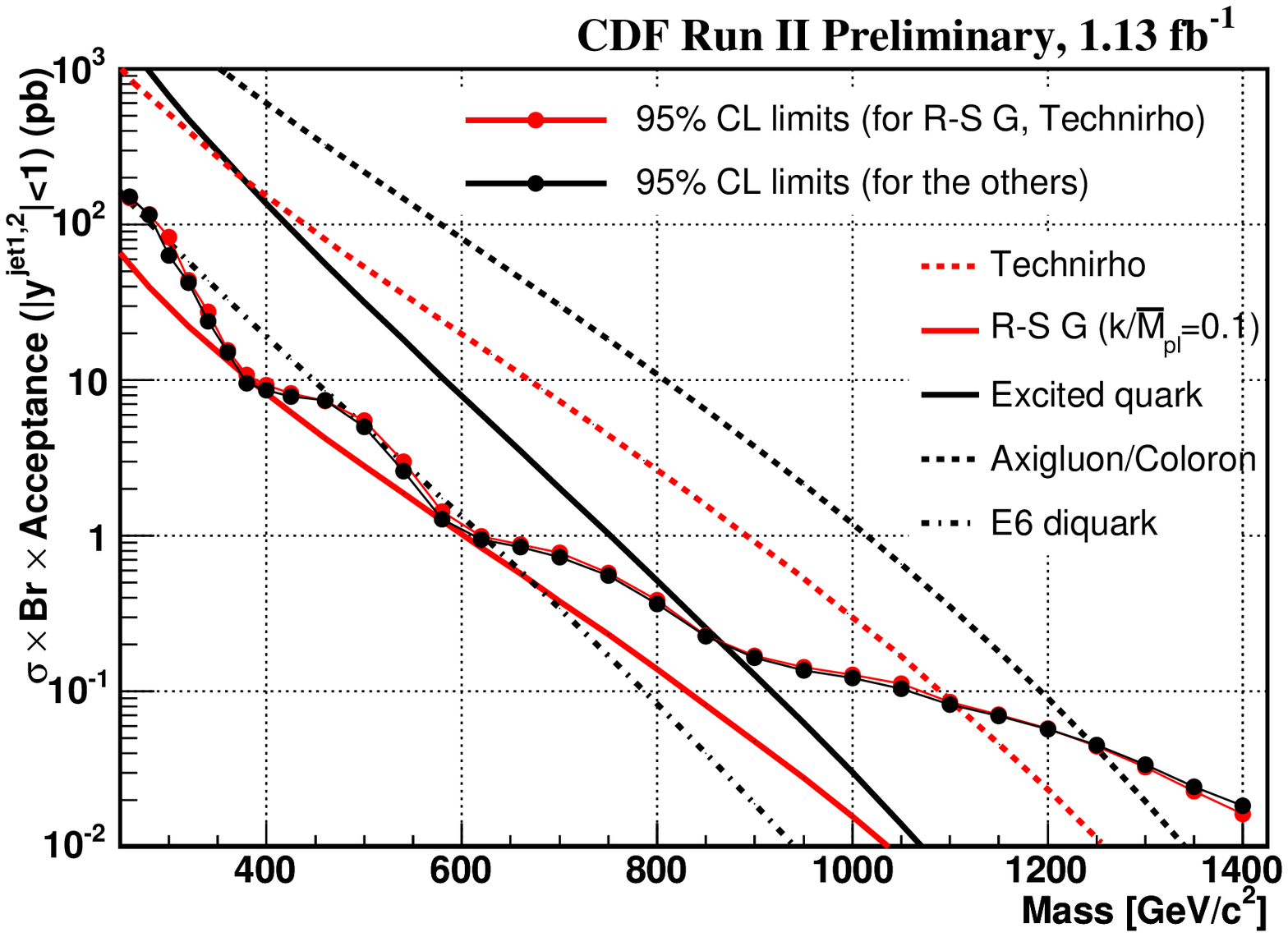,width=7cm}}
\put (9.0,-0.5){\epsfig{figure=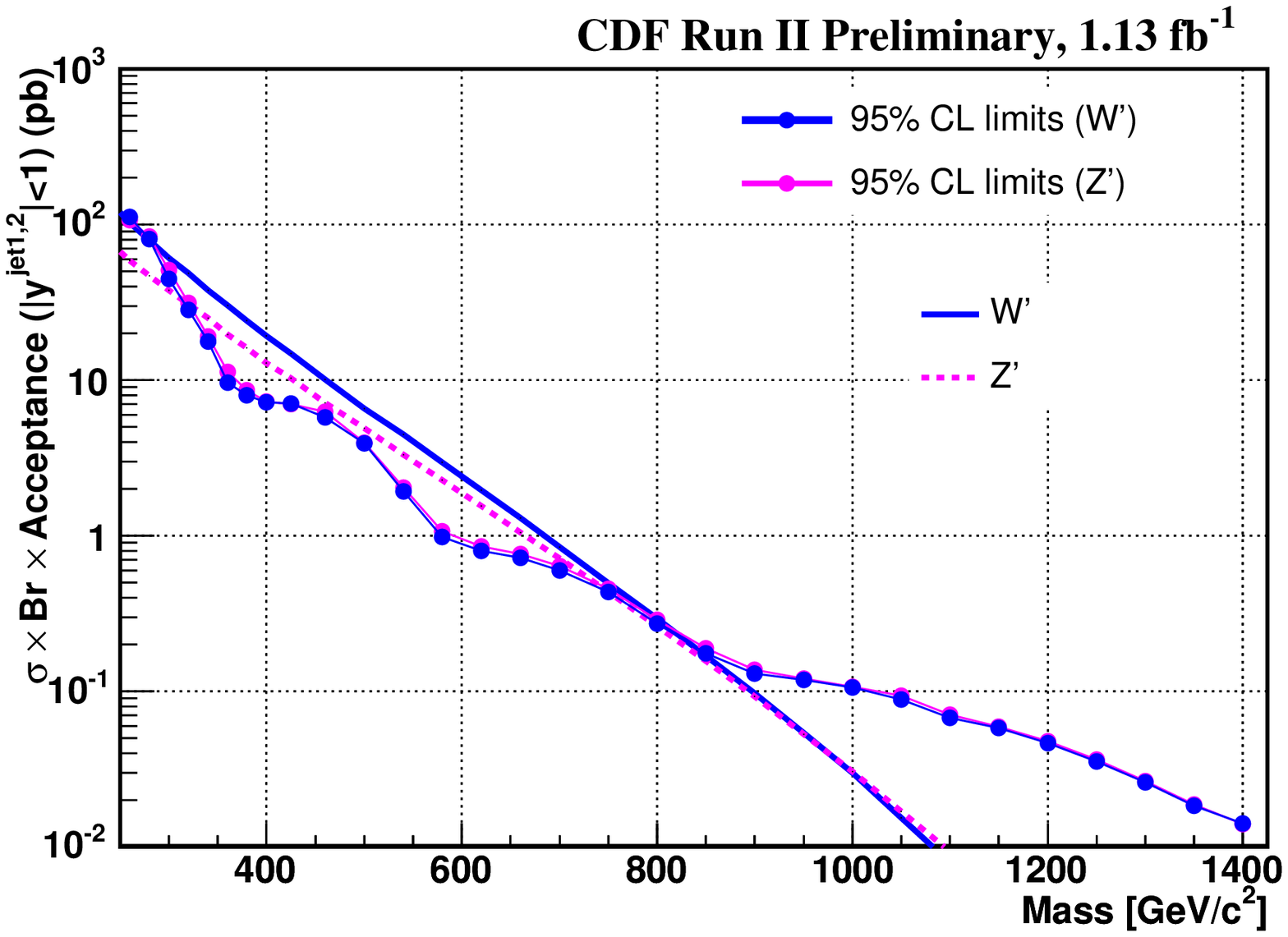,width=7cm}}
\put (6.5,8.0){\small (a)}
\put (14.5,8.0){\small (b)}
\put (4.5,2.5){\small (c)}
\put (12.5,2.5){\small (d)}
\end{picture}
\caption{\label{fig7}
{(a,b) Dijet invariant mass cross section; (c,d) $95\%$ CL limits as
functions of the particle mass.}}
\end{figure}

Jet production provides the highest energy reach with highest
statistics. In particular, dijet production is ideal to test the
SM and search for new physics, which might show up as narrow
resonances in the dijet invariant mass spectrum. Figures~\ref{fig7}a
and \ref{fig7}b show the measurement of the dijet invariant mass from
CDF~\cite{cdf452}; the data reach values of up to 1.4 TeV, which
constitute the highest energies measured so far. The NLO calculations
give a good description of the data in the whole measured range. Since
no evidence for new physics is observed, $95\%$ CL limits were set for
various models, which include excited quarks, new heavy vector bosons
and gravitons. Figure~\ref{fig7}c and \ref{fig7}d show the limits
obtained from the data as functions of the new particle mass.

\begin{figure}[h]
\setlength{\unitlength}{1.0cm}
\begin{picture} (18.0,6.0)
\put (0.0,-0.5){\epsfig{figure=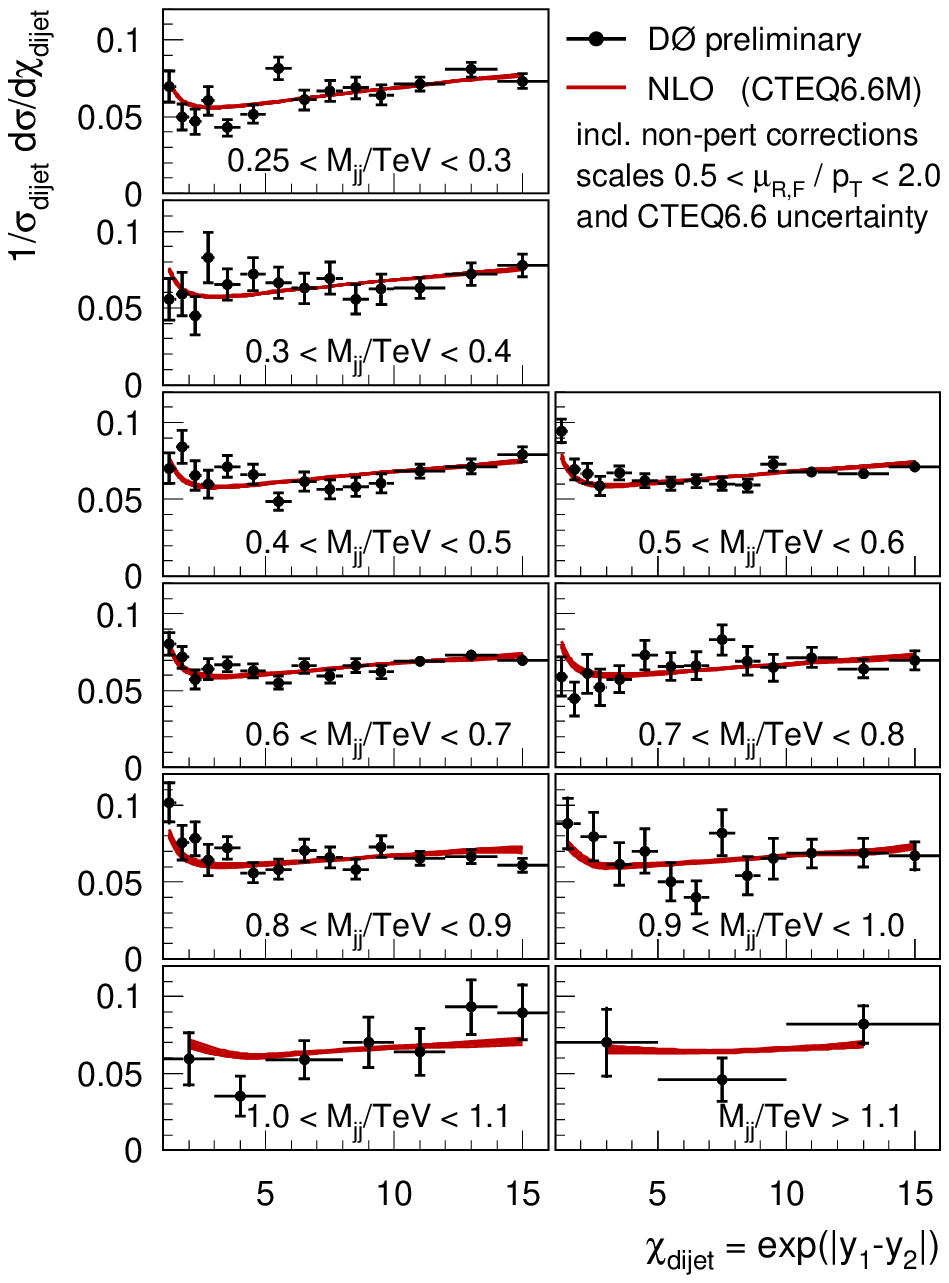,width=5cm}}
\put (12.7,-0.5){\epsfig{figure=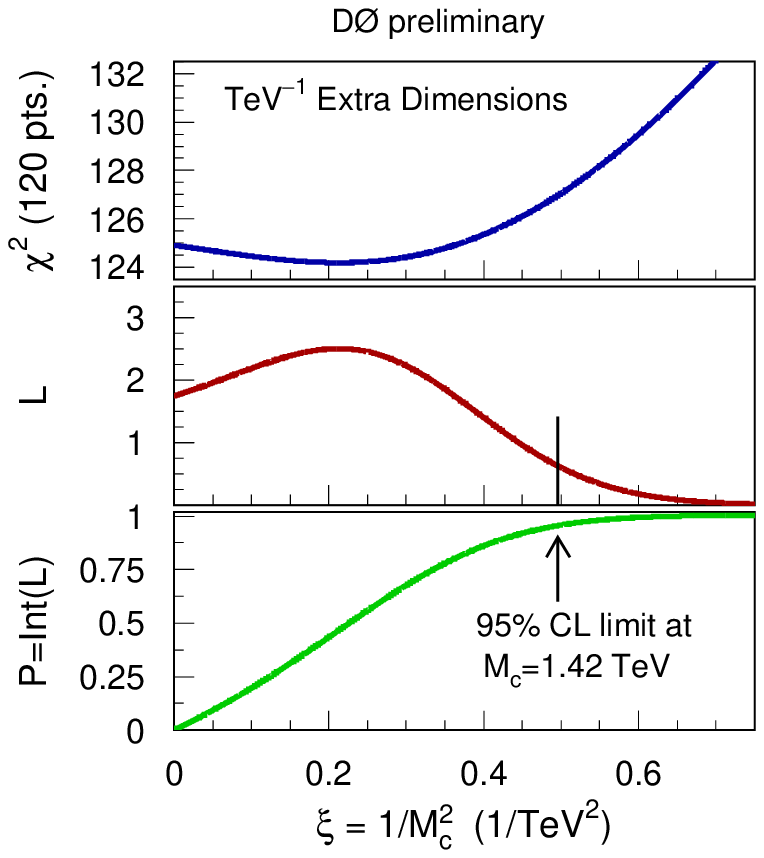,width=5cm}}
\put (12.7,0.0){\epsfig{figure=white.eps,width=1.0cm,height=5cm}}
\put (9.0,-0.5){\epsfig{figure=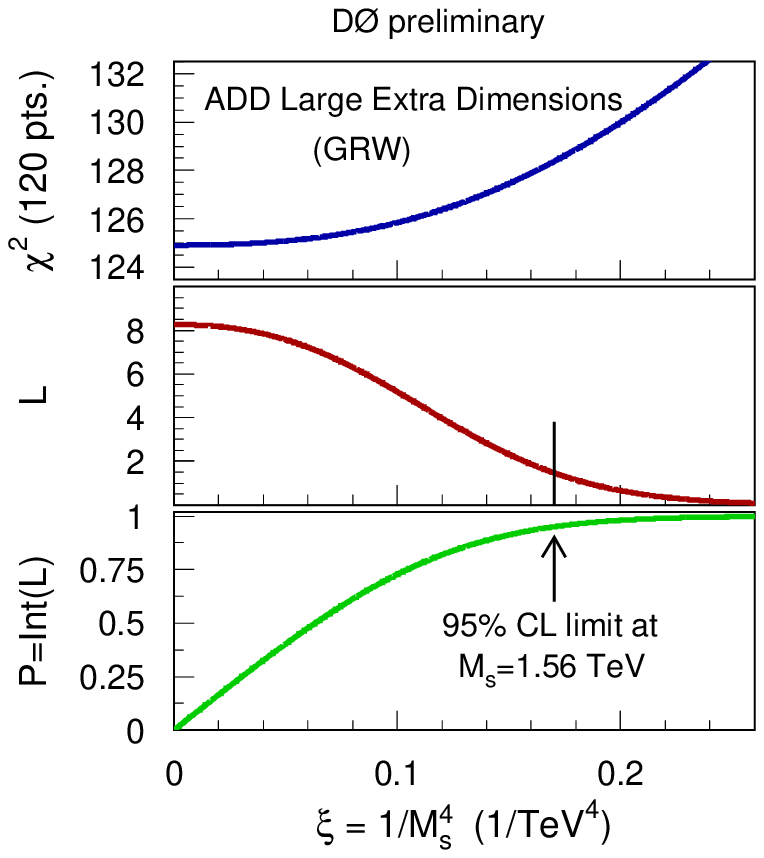,width=5cm}}
\put (9.0,0.0){\epsfig{figure=white.eps,width=1.0cm,height=5cm}}
\put (5.3,-0.5){\epsfig{figure=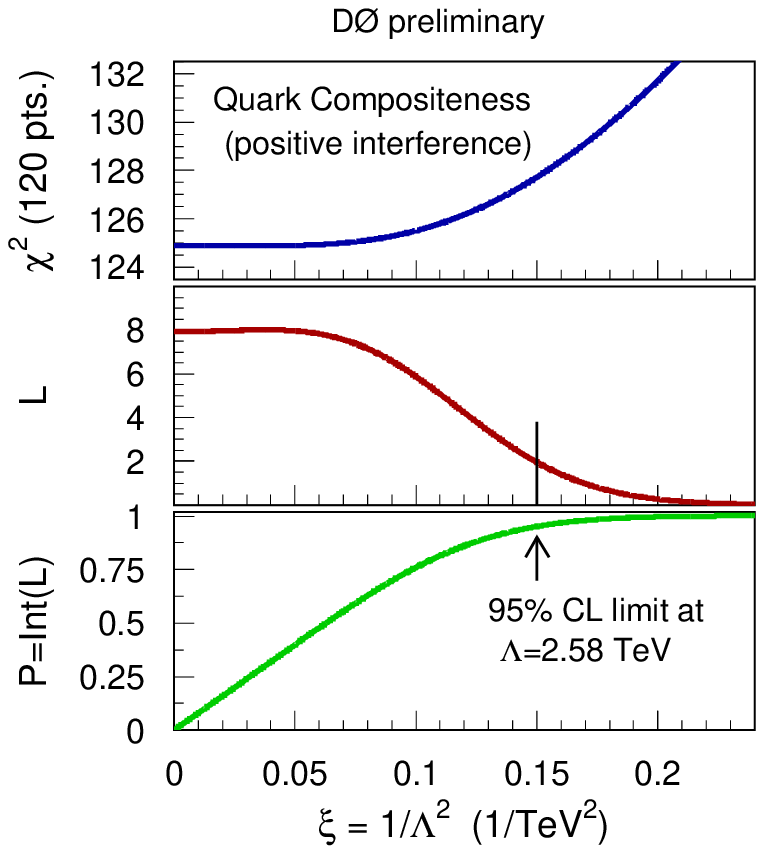,width=5cm}}
\put (4.0,4.5){\small (a)}
\put (9.0,3.5){\small (b)}
\put (12.7,3.5){\small (c)}
\put (16.5,3.5){\small (d)}
\end{picture}
\caption{\label{fig8}
{(a) Dijet cross sections as functions of $\chi_{\rm dijets}$; (b,c,d)
$\chi^2$, likelihood and probability as functions of the
characteristic parameter of each model.}}
\end{figure}

Angular correlations in the dijet system are directly sensitive to
the underlying parton dynamics and can also be used to search for new
physics. The angular correlation, defined as 
$\chi_{\rm dijets}=\exp(|y_1-y_2|)$, is expected to be constant for
Rutherford scattering; QCD processes induce a small deviation from
this behaviour, but new physics is expected to have a very different
shape at low values of $\chi_{\rm dijets}$. The D\O\ Collaboration has
measured~\cite{d0507} $\chi_{\rm dijets}$ in different regions of dijet
invariant mass from $0.25$ to above $1.1$~TeV, as shown in
Fig.~\ref{fig8}a. The NLO predictions give a good description of the
data in the whole measured range. Since there is no indication for the
presence of new physics, $95\%$ CL limits were set for various models
which include quark compositeness and extra
dimensions. Figures~\ref{fig8}b, \ref{fig8}c and \ref{fig8}d show the
$\chi^2$, likelihood and probability as a function of the
characteristic parameter of each model. These constitute the most
stringent limits for these models from hadron colliders up to date.

\section{Precision tests of QCD}

\begin{figure}[h]
\setlength{\unitlength}{1.0cm}
\begin{picture} (18.0,5.0)
\put (3.0,-0.6){\epsfig{figure=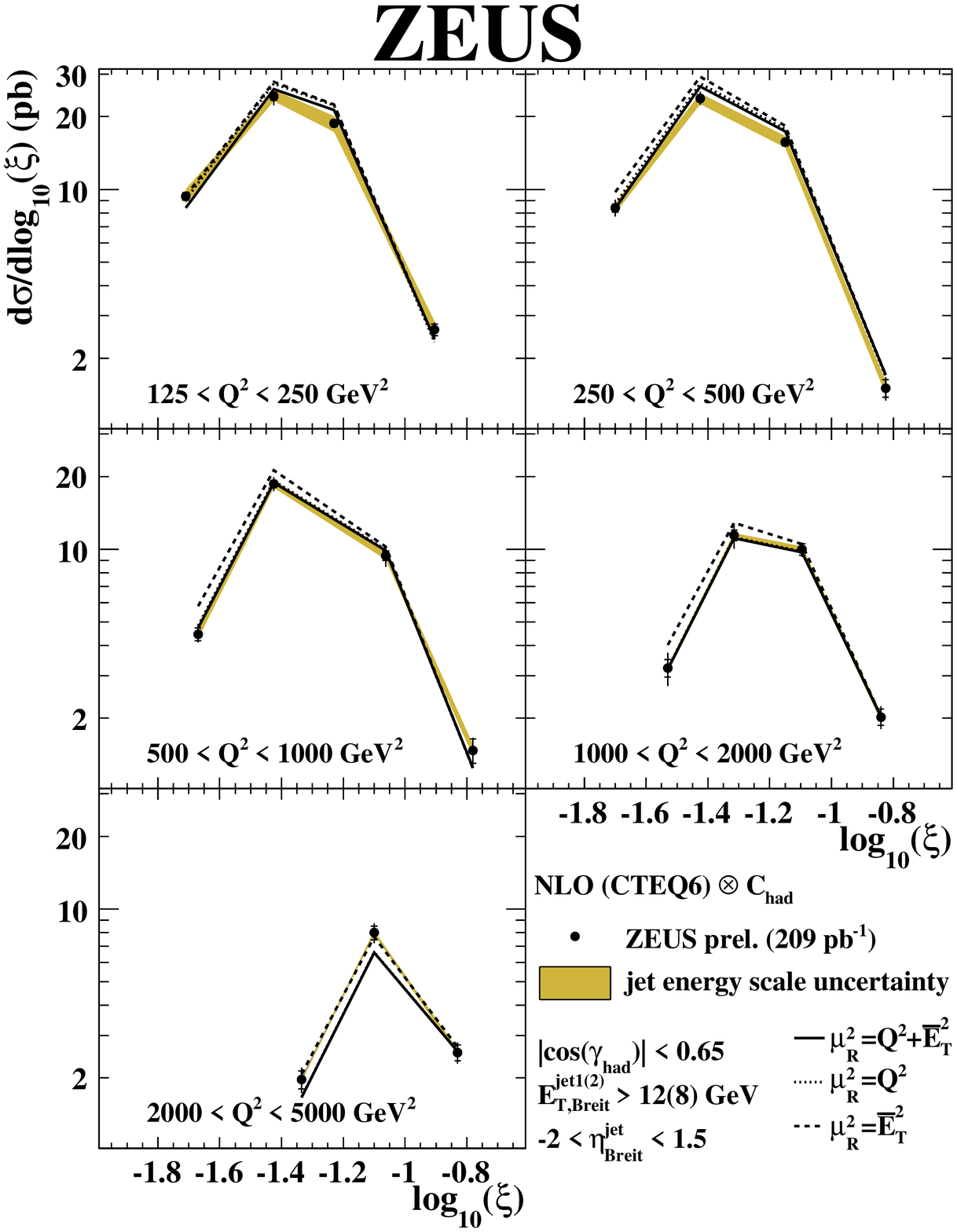,width=4.5cm}}
\put (9.0,-0.5){\epsfig{figure=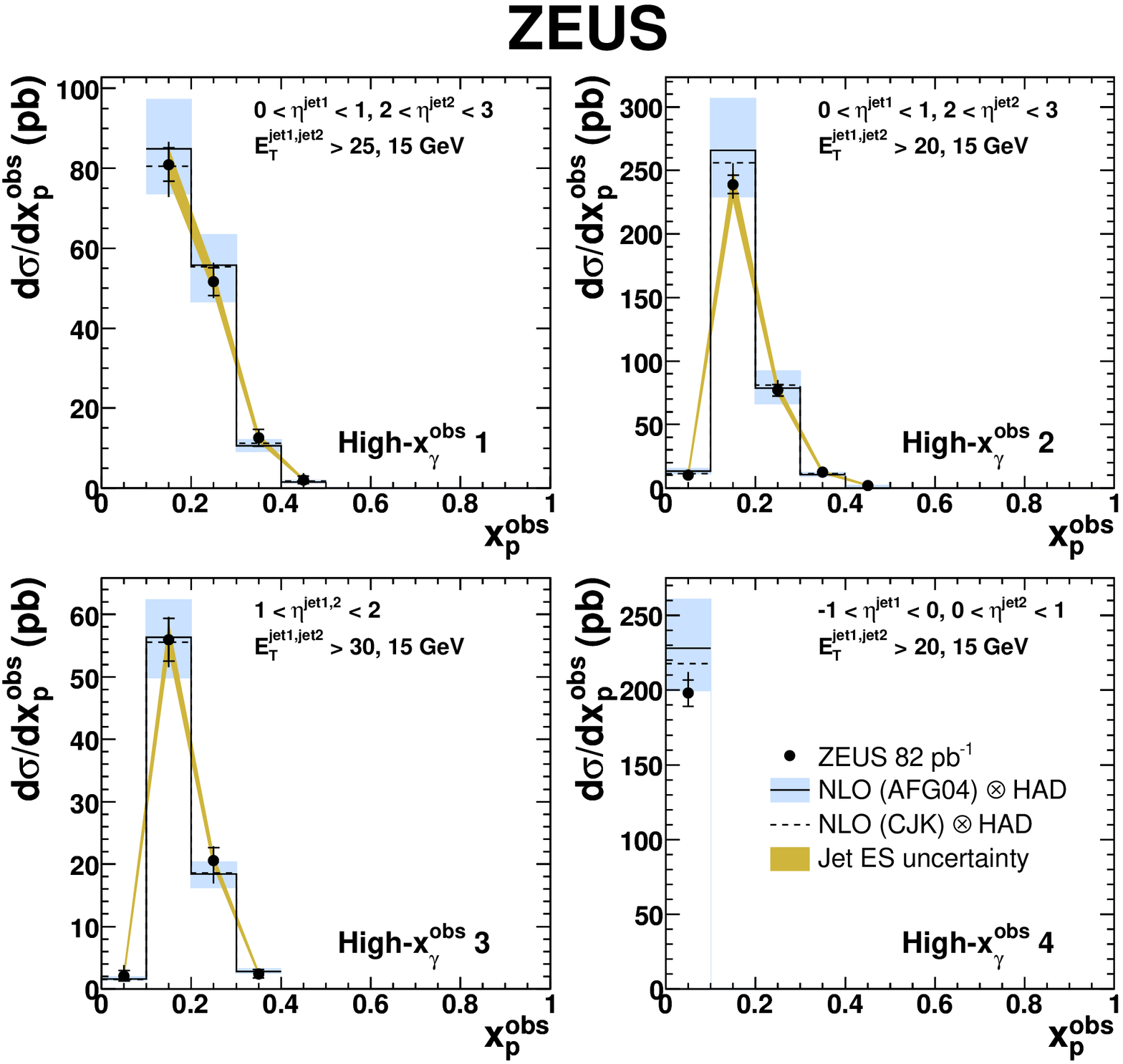,width=6cm}}
\end{picture}
\caption{\label{fig9}
{Dijet cross sections as functions of $\xi$ in different $\q2$ regions
in NC DIS (left); dijet cross sections as functions of $\xpo$ in
photoproduction (right).}}
\end{figure}

Recently, very precise measurements of jet cross sections in neutral
current (NC) DIS and photoproduction, which are directly sensitive to
the gluon content of the proton, have been incorporated in a QCD fit
to determine the proton PDFs. The result was an improved determination
of the gluon density for mid- to high-$x$ values, a region relevant
for new physics searches at LHC. In some regions of phase space the
uncertainty in the gluon density decreased by up to a factor of
two. Now the H1 and ZEUS Collaborations are making new and more
precise jet measurements with full HERA luminosity and extended phase
space to take full advantage of this technique. The ZEUS Collaboration
has measured double-differential dijet cross sections in NC
DIS~\cite{zeus143} and photoproduction~\cite{zeus125} as functions of
$\xi=x_{\rm Bj}(1+M_{\rm jj}^2/\q2)$ and 
$\xpo=(1/2E_p)({\sum_{\rm jets}\etjet e^{\etajet}})$, respectively,
which are both estimators of the fractional momentum carried
by the struck parton. The measurements are shown in Fig.~\ref{fig9}
and are well described by the NLO calculations. These analyses provide
a stringent test of QCD and were optimised to obtain the best
sensitivity to the gluon density in the proton.

\begin{figure}[h]
\setlength{\unitlength}{1.0cm}
\begin{picture} (18.0,5.5)
\put (1.0,-0.5){\epsfig{figure=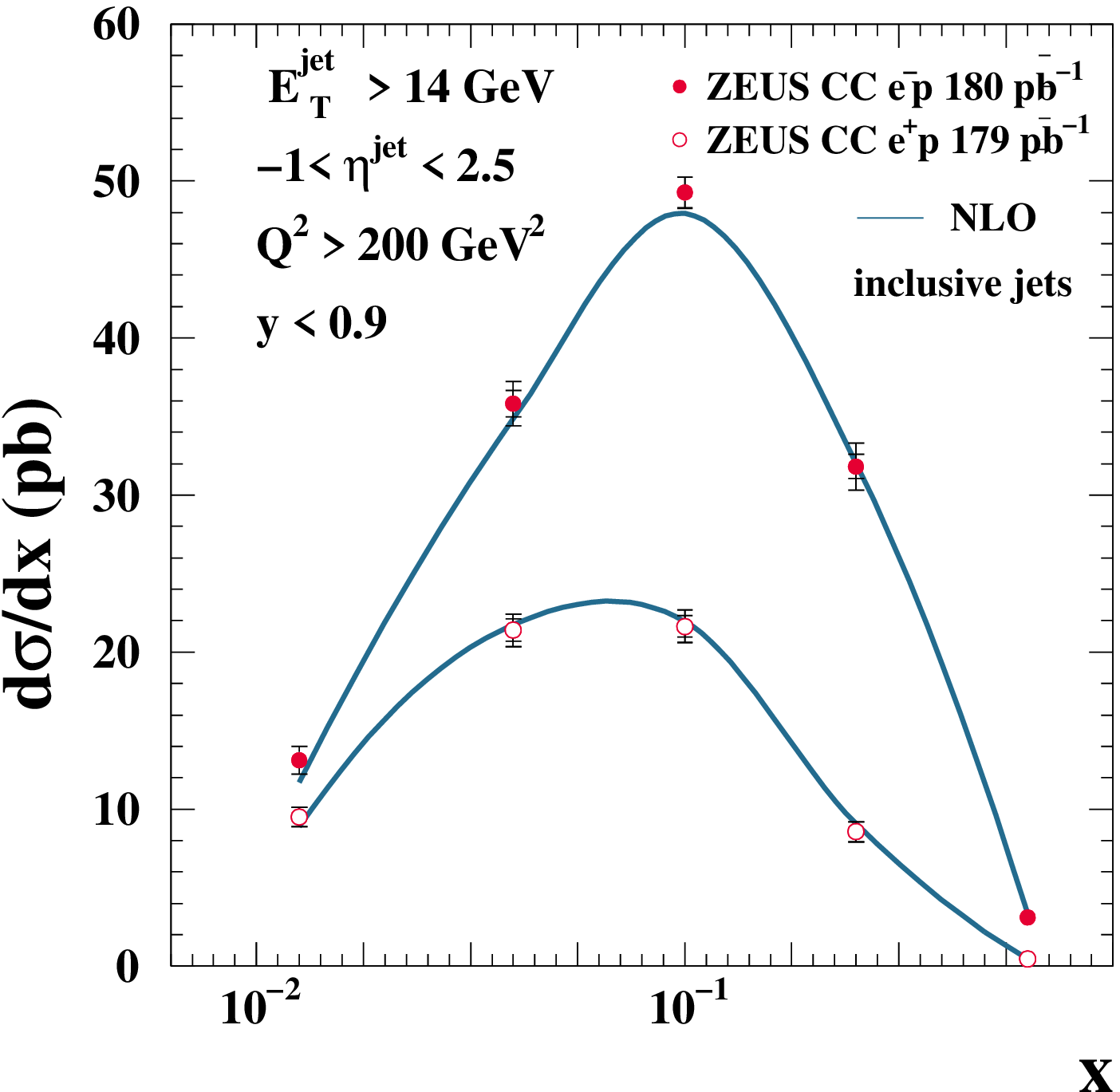,width=6cm}}
\put (8.0,-1.6){\epsfig{figure=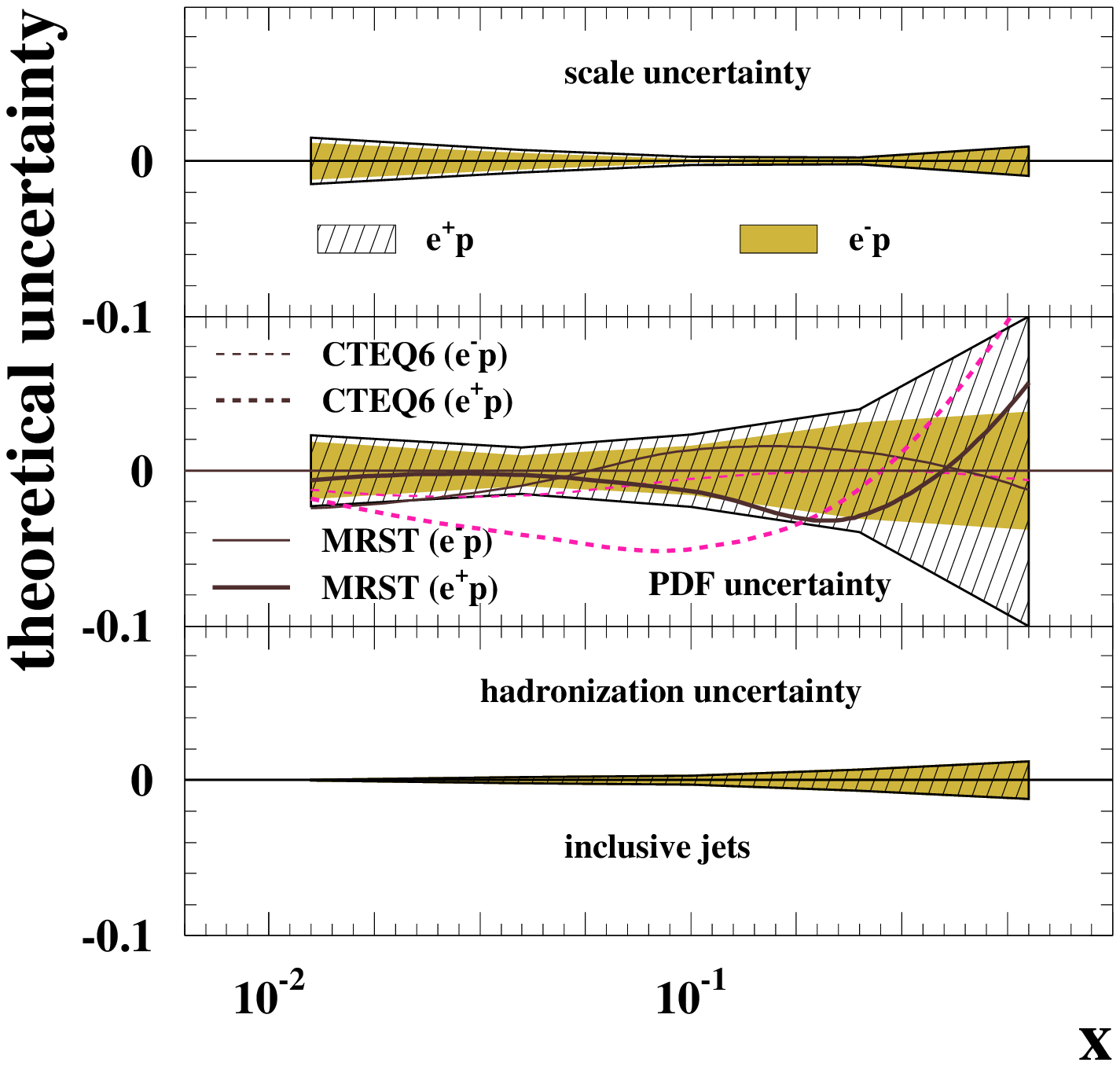,width=9cm}}
\put (6.0,3.0){\small (a)}
\put (14.0,5.0){\small (b)}
\end{picture}
\caption{\label{fig10}
{(a) Inclusive-jet cross section as a function of $x$ in CC DIS; (b)
  theoretical uncertainties of the inclusive-jet cross-section
  prediction as a function of $x$ in CC DIS.}}
\end{figure}

Inclusive-jet cross sections in charged current DIS have been
measured~\cite{zeus132} by the ZEUS Collaboration. Figure~\ref{fig10}a
shows the cross section as a function of $x$ for electron beams
(sensitive to the $u$-quark density) and positron beams (sensitive to
the $d$-quark density). The NLO calculations give a good description
of the data. Figure~\ref{fig10}b shows the theoretical uncertainties,
clearly dominated by the PDF uncertainty, which is largest for
positron beams at high $x$. Therefore, these measurements have the
potential to constrain further the valence-quark PDFs if included in
global fits.

\begin{figure}[h]
\setlength{\unitlength}{1.0cm}
\begin{picture} (18.0,5.5)
\put (0.0,6.5){\epsfig{figure=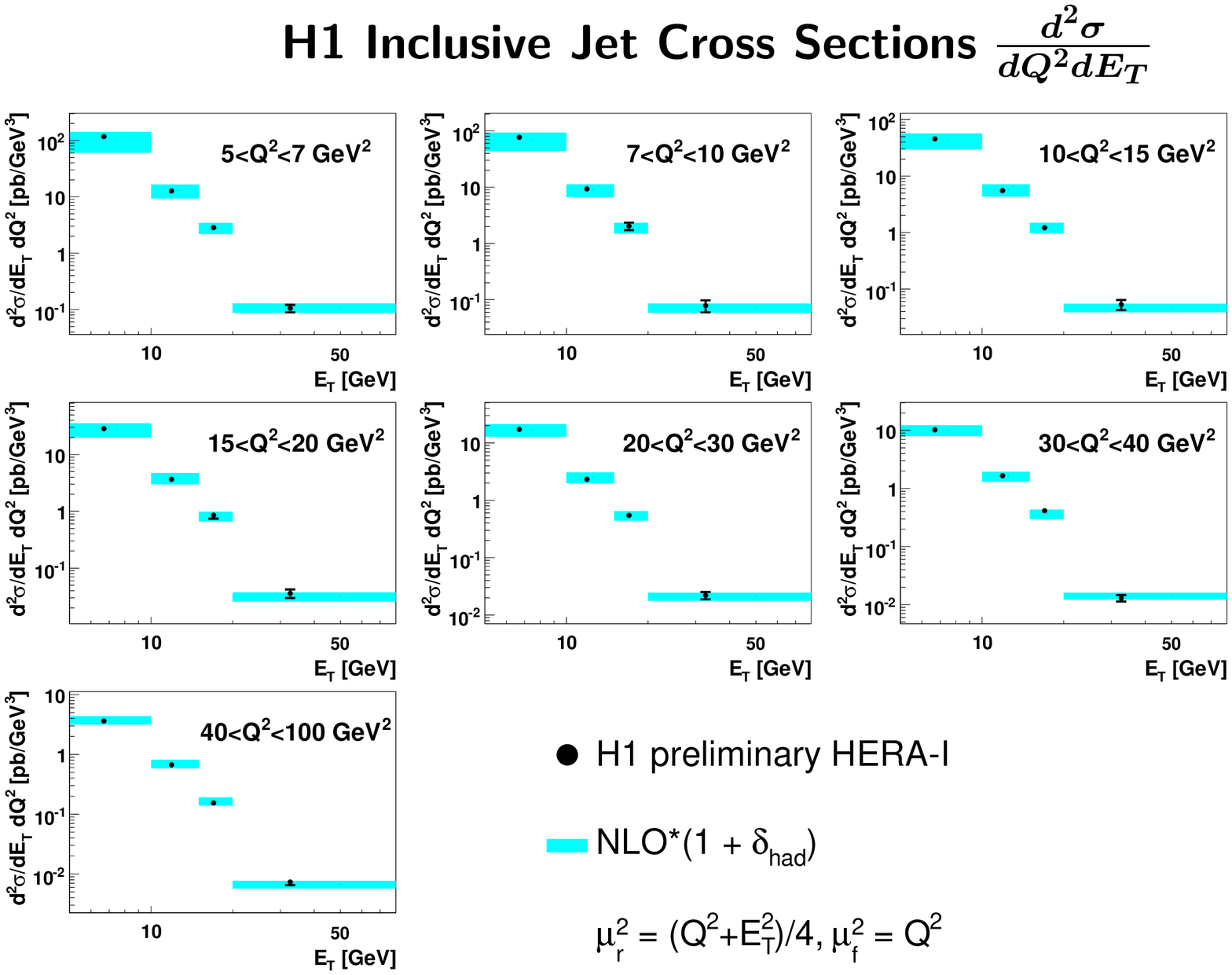,width=7cm,angle=270}}
\put (8.0,6.7){\epsfig{figure=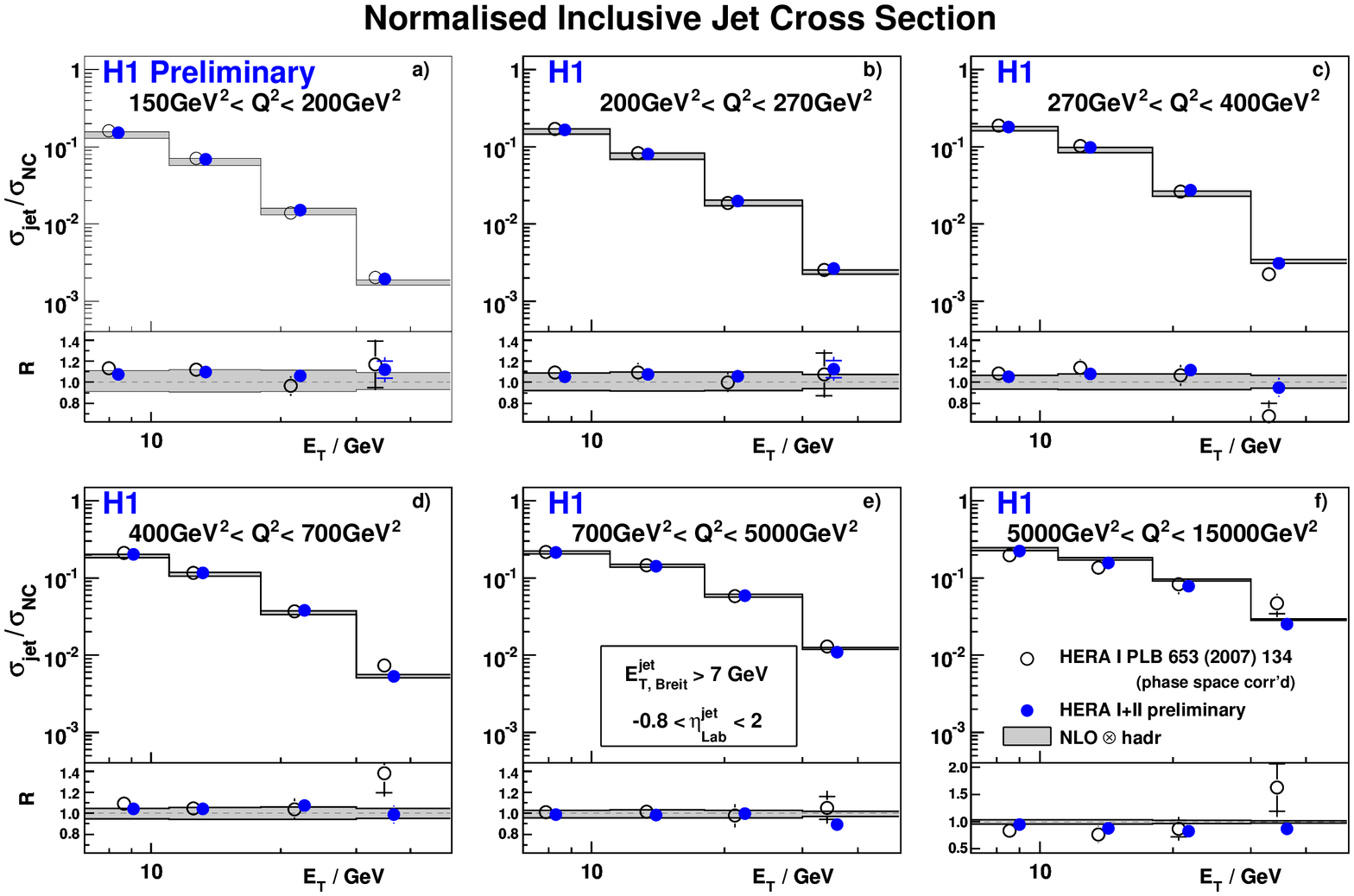,width=7cm,angle=270}}
\put (7.5,4.5){\small (a)}
\put (16.5,4.5){\small (b)}
\end{picture}
\caption{\label{fig11}
{Inclusive-jet cross sections as functions of $\etjet$ in different
  regions of $\q2$ in NC DIS.}}
\end{figure}

Inclusive-jet cross sections in NC DIS were measured~\cite{h1845} by
the H1 Collaboration as a function of $\etjet$ in different regions of
$\q2$ in the range $5<\q2<100$~\g2\ (see Fig.~\ref{fig11}a). The NLO
predictions give a good description of the data. However, the
theoretical uncertainty, which is dominated by terms beyond NLO, is
large: it reaches up to $30\%$ at the lowest $\q2$ values. This shows
the need for higher-order corrections. These measurements can help to
constrain the gluon PDF at low $\q2$ (low $x$) when higher-order
calculations become available for NC DIS.

\subsection{Precision measurements of $\as$}
\begin{figure}[h]
\setlength{\unitlength}{1.0cm}
\begin{picture} (18.0,6.0)
\put (2.0,-1.0){\epsfig{figure=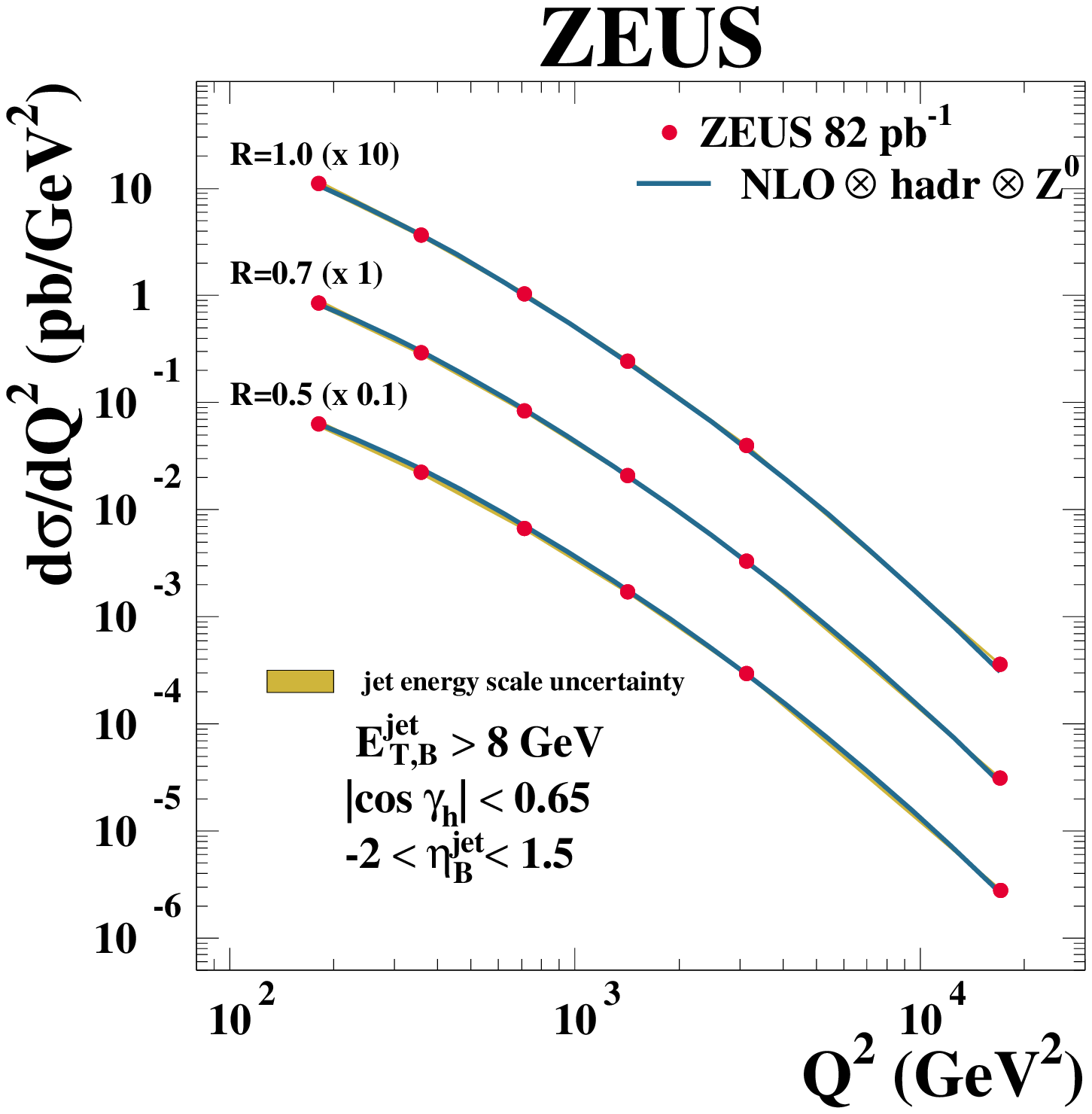,width=8cm}}
\put (9.0,-0.5){\epsfig{figure=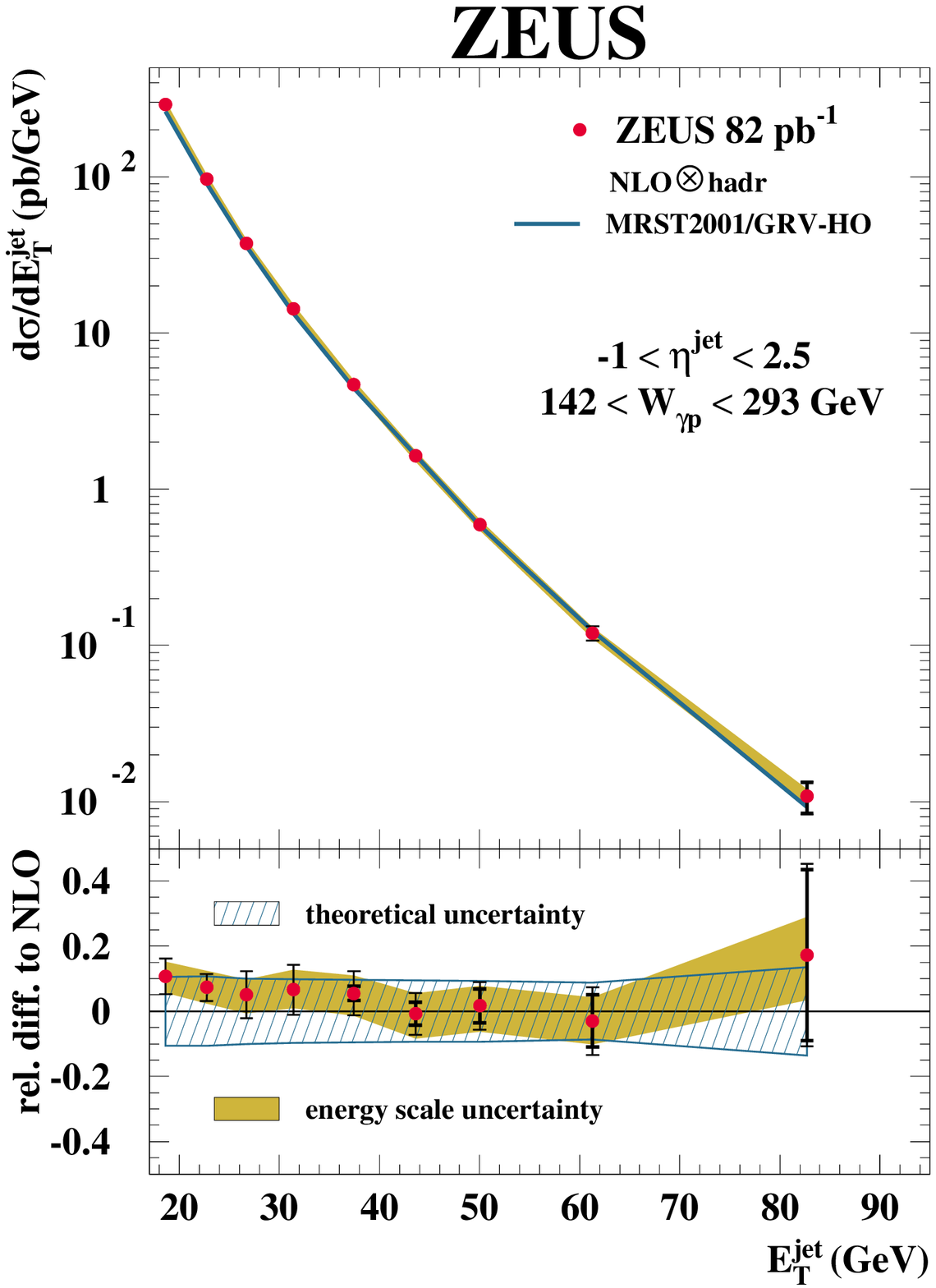,width=7cm}}
\put (7.0,4.0){\small (a)}
\put (13.5,3.5){\small (b)}
\end{picture}
\caption{\label{fig12}
{(a) Inclusive-jet cross sections as functions of $\q2$ for different
  values of the jet radius $R$ in NC DIS; (b) inclusive-jet cross
  section as a function of $\etjet$ in photoproduction.}}
\end{figure}

The success of pQCD lies on the precise and consistent determinations
of $\as$ from many diverse phenomena, such as structure functions,
$\tau$ decays, $Z$-line shape, lattice, jets, etc. At HERA, many
determinations of $\as$, mostly extracted from jet observables, give
values as precise as those from more inclusive measurements. The H1
and ZEUS Collaborations have made new determinations of $\as$,
focusing on decreasing the uncertainties further. 
The H1 Collaboration has determined~\cite{h1845,h1844} these values of
$\as$: $\asz=0.1186\pm 0.0014({\rm exp.})\pm 0.0134({\rm th.})$
and $\asz=0.1196\pm 0.0010({\rm exp.})\pm 0.0053({\rm th.})$
from the inclusive-jet cross sections at low $\q2$ and
the normalised inclusive-jet cross sections (see Fig.~\ref{fig11}b) in
the regions $5<\q2<100$~\g2\ and $150<\q2<15000$~\g2, respectively. In
this way, a region of phase space was selected in which experimental
uncertainties are well under control and also, the use of normalised
cross sections yields a cancellation of correlated uncertainties.
New determinations
of $\as$ were performed by ZEUS from the inclusive-jet cross sections
in NC DIS~\cite{zeus121} and photoproduction~\cite{zeus152} at high
$\q2$ and high $\etjet$, respectively, where the theoretical
uncertainties are minimised. Figure~\ref{fig12}a shows the cross
section in NC DIS as a function of $\q2$ for different values of the
jet-radius parameter $R$ in the $\kt$ algorithm. The NLO calculations
give a good description of the data for $R=0.5-1.0$ with similar
accuracy. Figure~\ref{fig12}b shows the cross section as a function of
$\etjet$ in photoproduction; the NLO calculation also gives a good
description of the data for these processes. Values of $\as$ were
extracted from these measurements: 
$\asz=0.1207\pm 0.0014({\rm stat.})_{-0.0033}^{+0.0035}({\rm syst.})_{-0.0023}^{+0.0022}({\rm th.})$
(NC DIS, $\q2>500$~\g2) and $\asz=0.1223\pm 0.0001({\rm stat.})_{-0.0021}^{+0.0023}({\rm syst.})\pm 0.0030({\rm th.})$
(photoproduction, $\etjet>17$~GeV).

To reduce the uncertainties even further and to take advantage of the
cross-calibration between experiments, a simultaneous fit to the
inclusive-jet cross sections in NC DIS from H1 and ZEUS has been
performed~\cite{h1zeus628} to give 

\vspace{0.1cm}
\centerline{$\asz=0.1198\pm 0.0019\ {\rm (exp.)}\pm 0.0026\ {\rm (th.)}$.}

The total uncertainty of the combined value, $\pm 2.7\%$, is very
competitive with the most recent result from LEP.

\section{Probing QCD with vector bosons}

\begin{figure}[h]
\setlength{\unitlength}{1.0cm}
\begin{picture} (18.0,5.5)
\put (-0.5,-0.5){\epsfig{figure=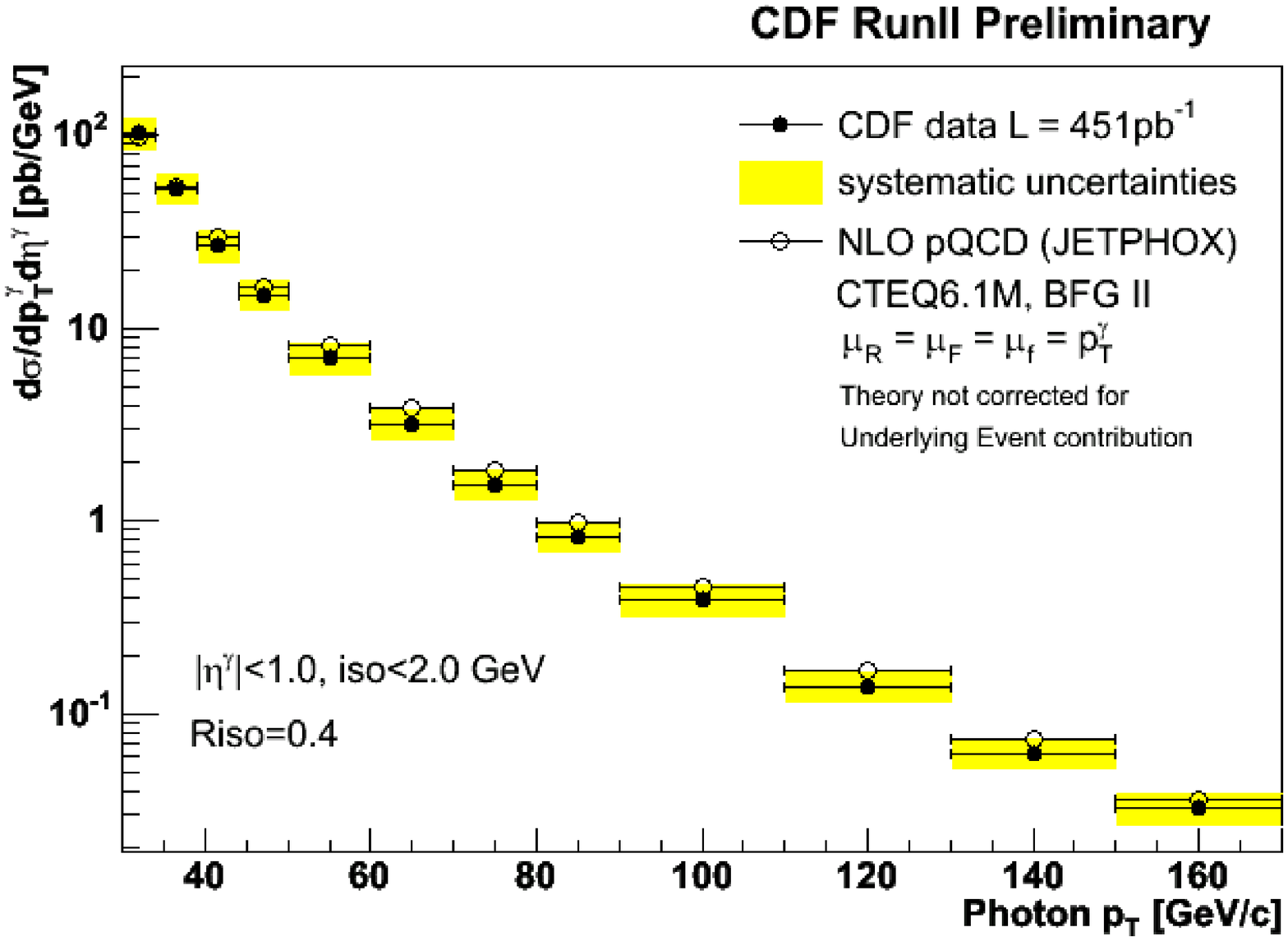,width=8.5cm}}
\put (7.0,-0.8){\epsfig{figure=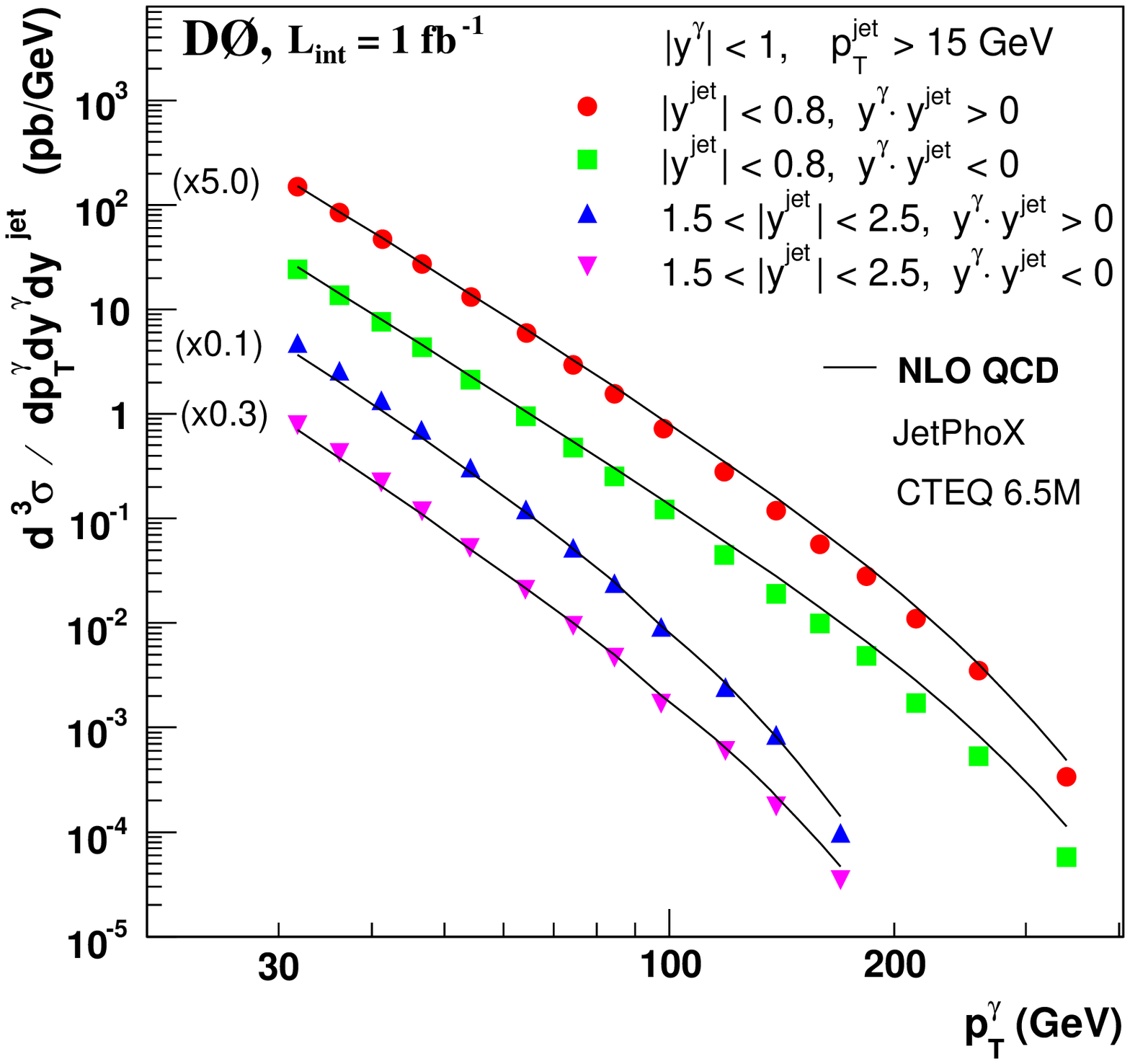,width=6.5cm}}
\put (13.0,-0.2){\epsfig{figure=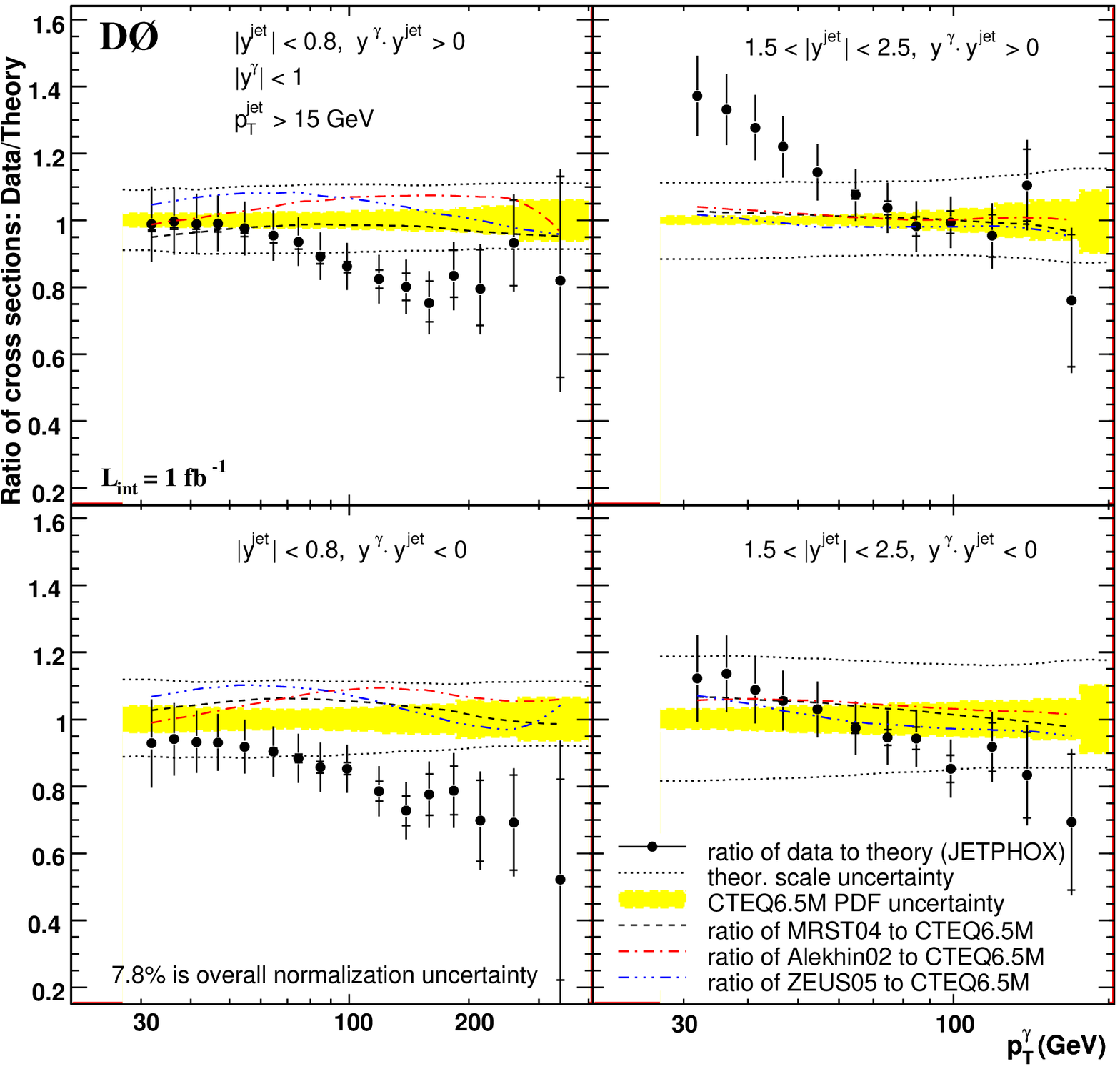,width=5.5cm}}
\put (5.0,2.0){\small (a)}
\put (10.5,2.5){\small (b)}
\put (15.0,3.8){\small (c)}
\end{picture}
\caption{\label{fig14}
{(a) Isolated-photon cross section as a function of the photon $p_T$;
  (b,c) isolated-photon plus jet cross sections as functions of photon
$p_T$ for different configurations of photon and
jet rapidities.}}
\end{figure}

Production of isolated photons in $\pp$ collisions at TeVatron are a
probe of QCD dynamics. Photons coming directly from the hard
interaction are largely independent of hadronisation corrections. The
understanding of these processes in QCD is crucial for searches of new
particles that decay into photons. The CDF Collaboration has
measured~\cite{cdf411} the inclusive cross section for isolated
photons as a function of the photon transverse momentum integrated
over the photon rapidity range $|\eta^{\gamma}|<1$. Figure~\ref{fig14}a 
shows the measurement together with the NLO predictions. The
calculations describe the data adequately within the experimental and
theoretical uncertainties.

The D\O\ Collaboration has studied~\cite{d0509} in more detail these
processes by measuring the cross section of isolated photons in
association with jets. The measurements were done as functions of the
photon transverse momentum for different configurations of photon and
jet rapidities. These measurements have the potential to constrain the
proton PDFs and different angular configurations give access to
different regions of $\q2$ and $x$. Figures~\ref{fig14}b and c show the
measurements together with the NLO calculations. The NLO predictions,
using different sets of proton PDFs can not describe the shape of the
cross sections simultaneously over the entire range
measured. Therefore, the theoretical understanding of these processes
needs to be improved before these data can be used to constrain the
proton PDFs.

\begin{figure}[h]
\setlength{\unitlength}{1.0cm}
\begin{picture} (18.0,7.0)
\put (3.0,3.3){\epsfig{figure=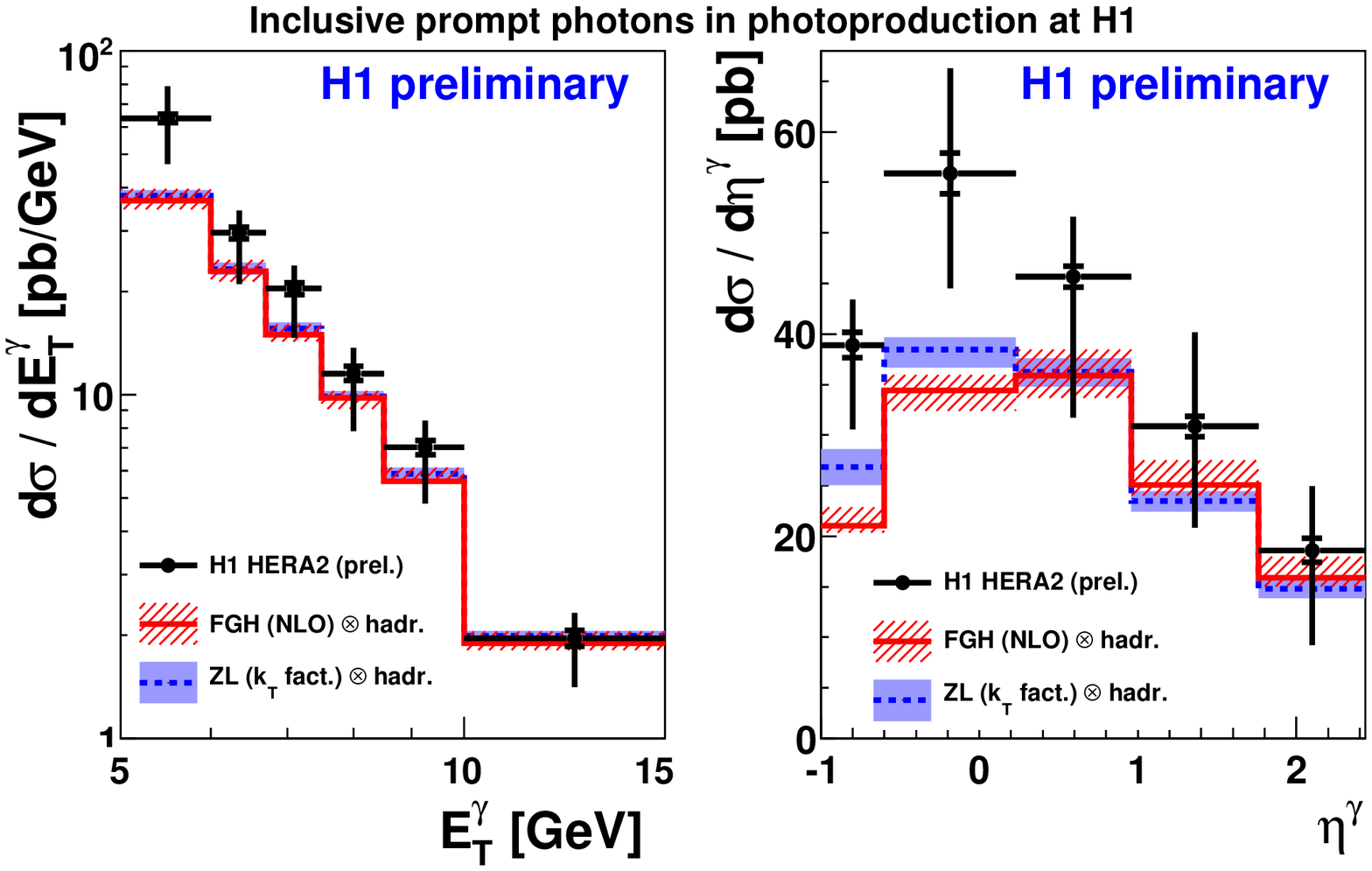,width=6.0cm}}
\put (3.0,-0.5){\epsfig{figure=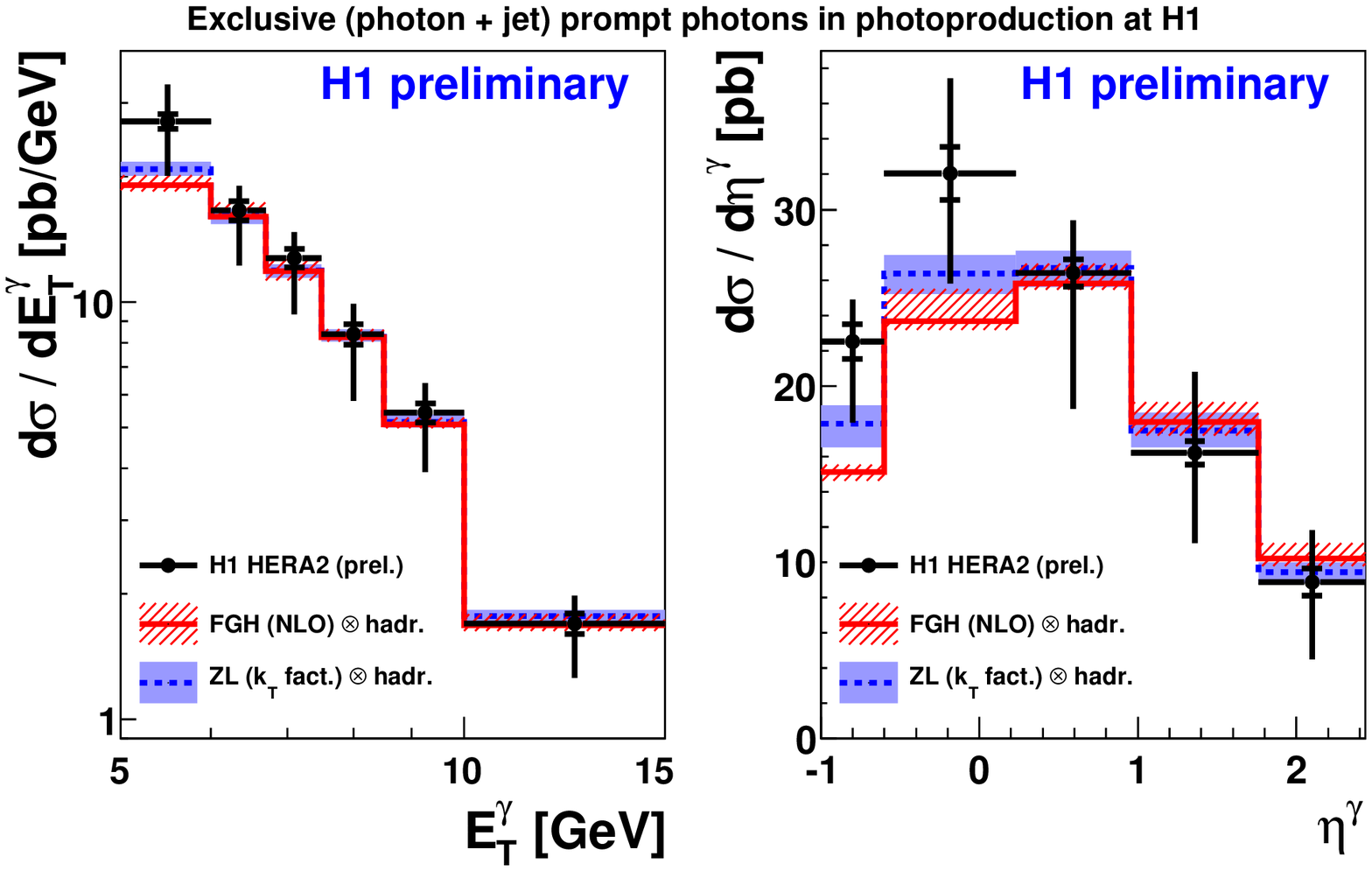,width=6.0cm}}
\put (0.0,-0.5){\epsfig{figure=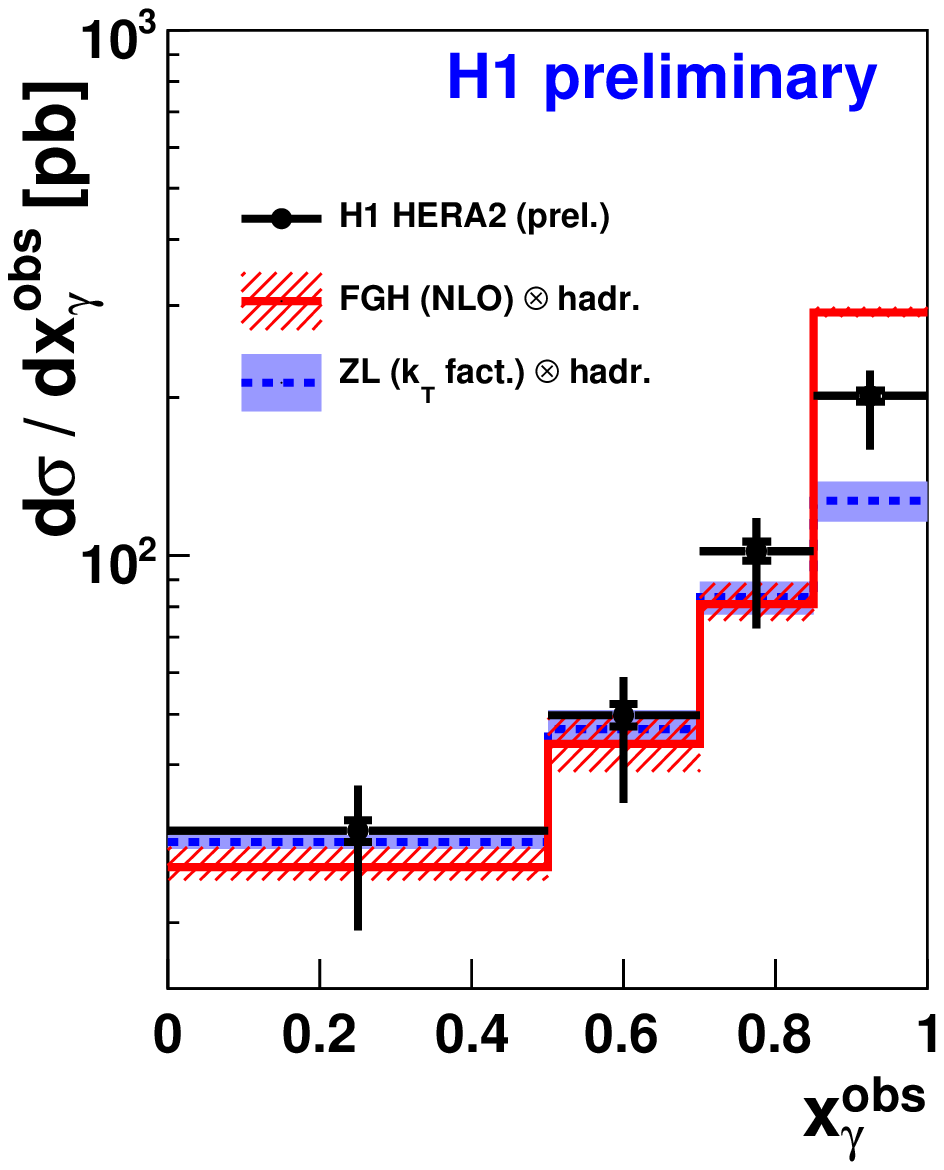,width=3.0cm}}
\put (12.5,3.3){\epsfig{figure=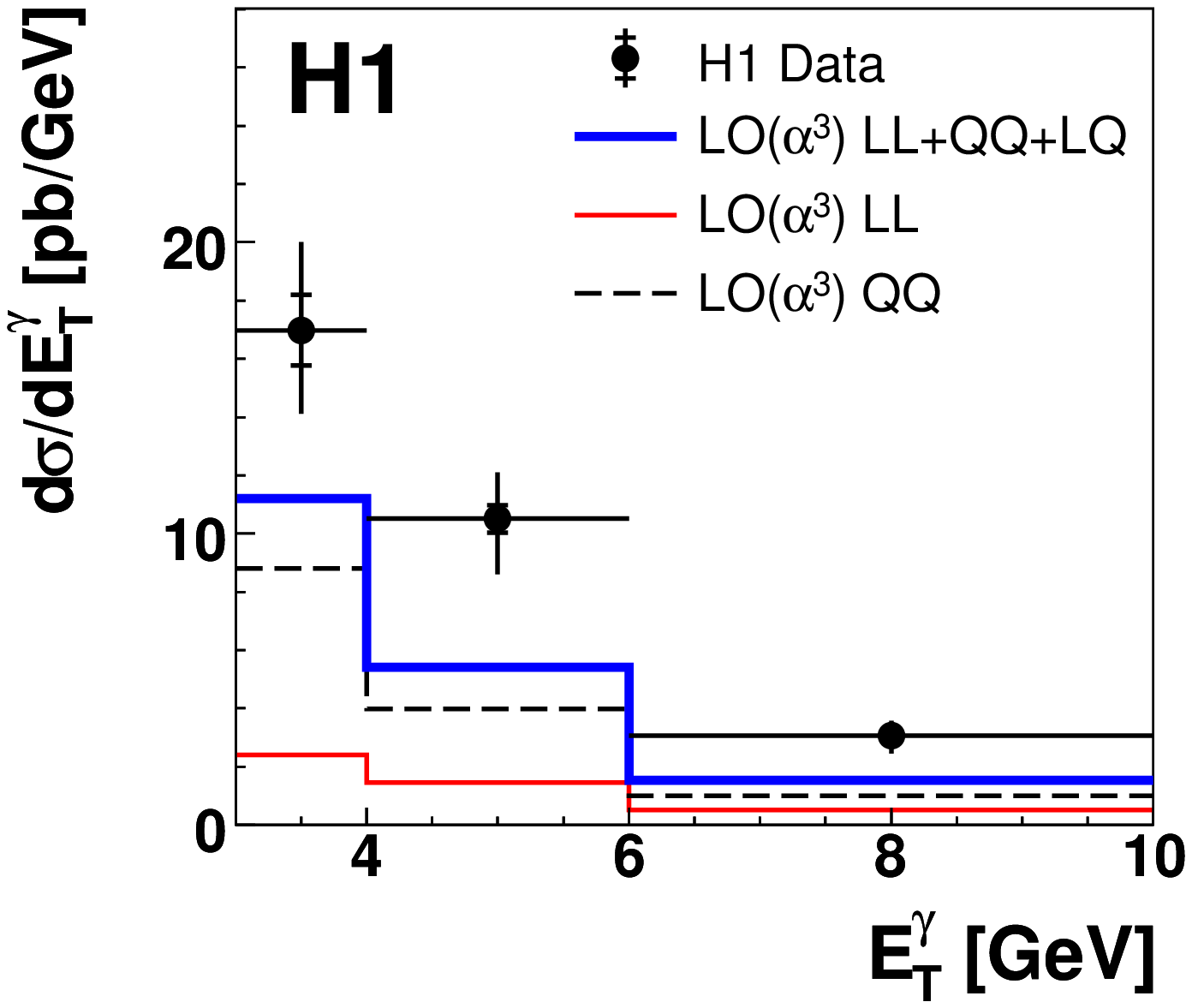,width=6.0cm}}
\put (12.5,-0.5){\epsfig{figure=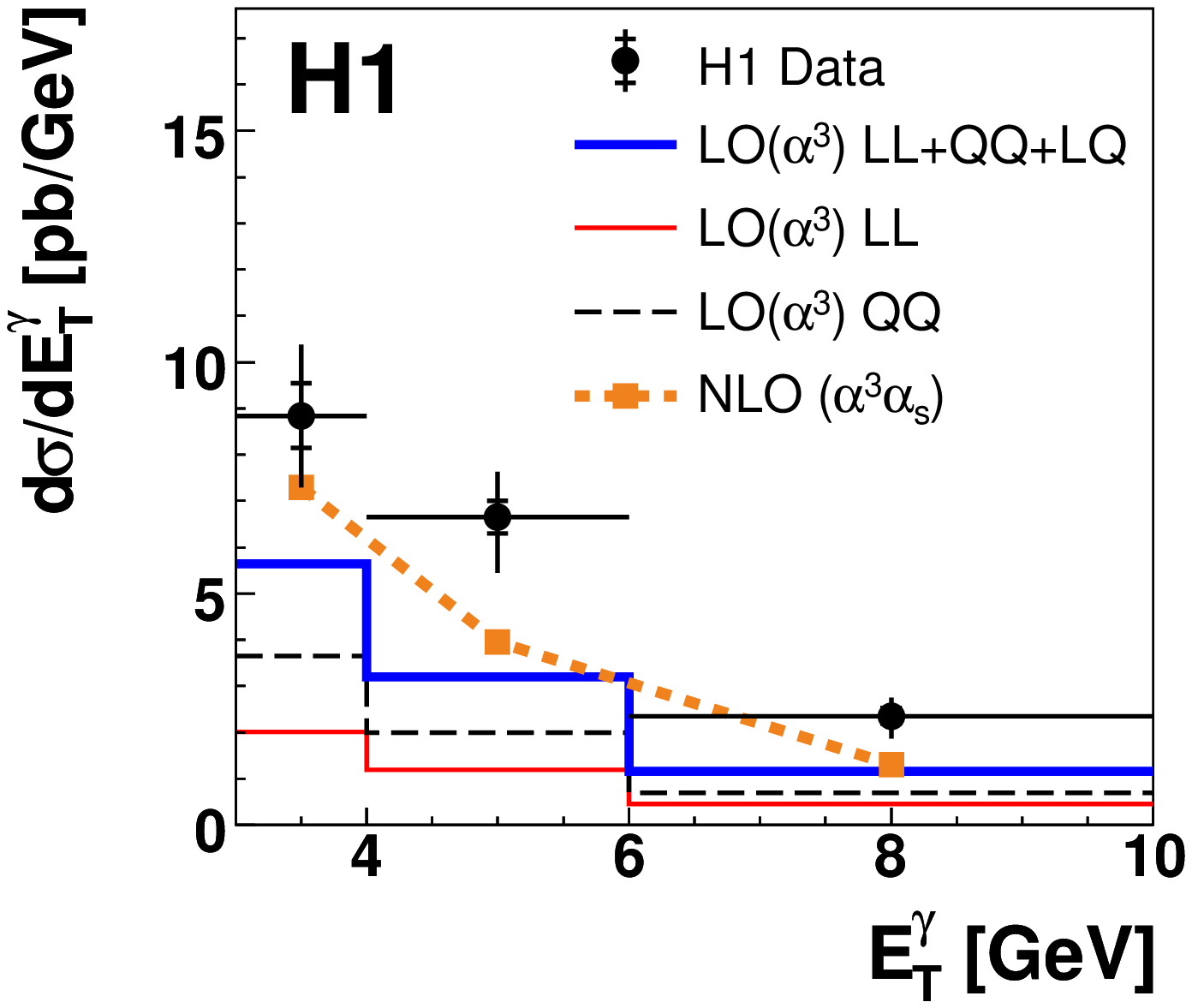,width=6.0cm}}
\put (8.3,-0.5){\epsfig{figure=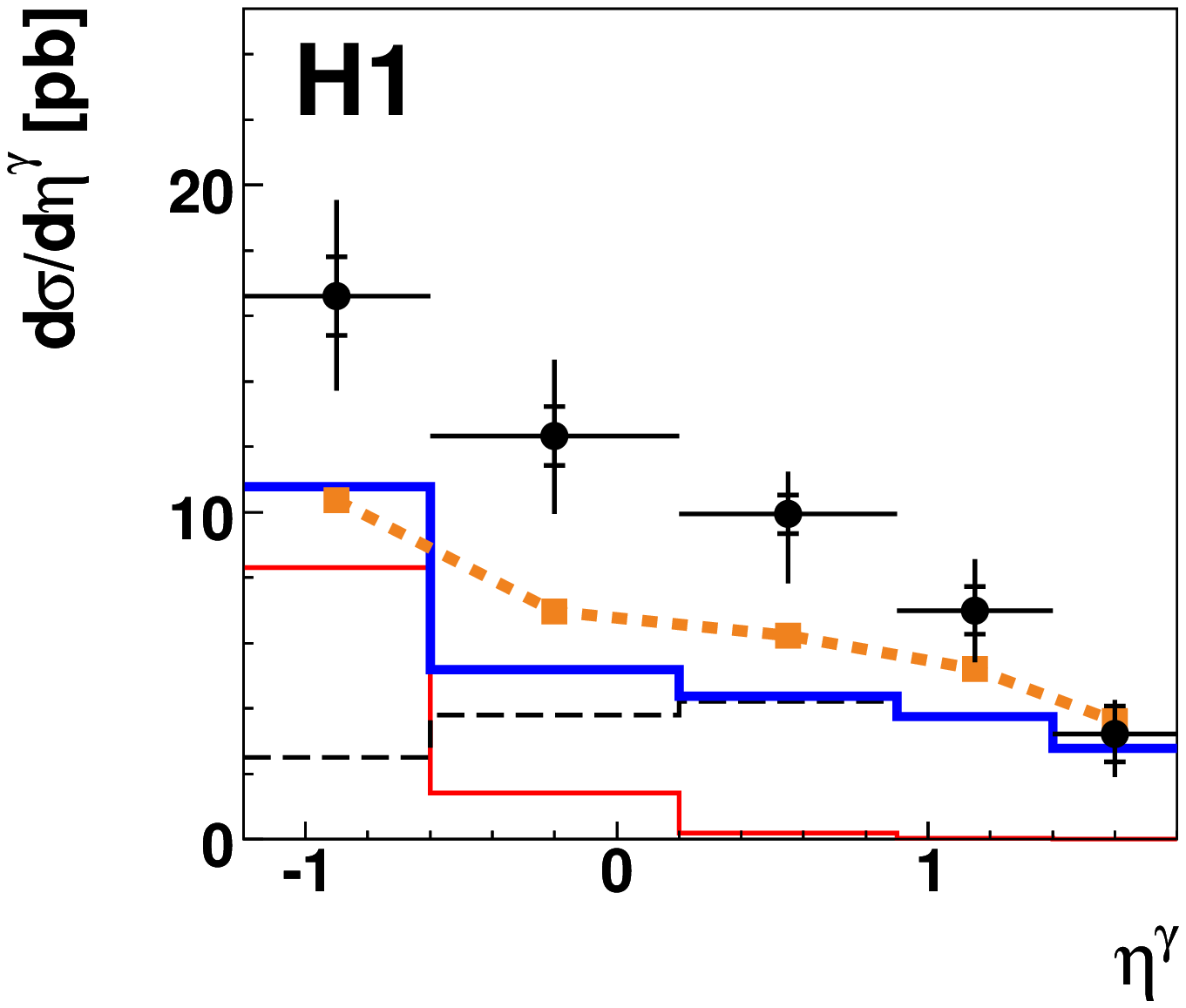,width=6.0cm}}
\put (5.0,6.5){\small (a)}
\put (8.0,6.5){\small (b)}
\put (2.0,2.5){\small (c)}
\put (5.0,2.5){\small (d)}
\put (8.0,2.5){\small (e)}
\put (16.0,5.5){\small (f)}
\put (12.0,2.5){\small (g)}
\put (16.5,2.3){\small (h)}
\end{picture}
\caption{\label{fig16}
{Inclusive-photon cross sections as functions of (a) $E_T^{\gamma}$
  and (b) $\eta^{\gamma}$ in photoproduction; exclusive-photon cross
  sections as functions of (c) $\xo$, (b) $E_T^{\gamma}$ and (c)
  $\eta^{\gamma}$ in photoproduction; (f) inclusive-photon cross
  section as a function of $E_T^{\gamma}$ in NC DIS; exclusive-photon
  cross sections as functions of (g) $\eta^{\gamma}$ and (h)
  $E_T^{\gamma}$ in NC DIS.}}
\end{figure}

Inclusive prompt photons and in association with jets have been
measured at HERA in photoproduction by the H1
Collaboration~\cite{h1846}. These processes are sensitive to the PDFs 
both in the proton and the photon. Inclusive prompt photon production
has been measured as a function of the transverse energy and 
pseudorapidity of the photon (see Figs.~\ref{fig16}a and
\ref{fig16}b). The NLO calculations are below the data, especially at
low $E_T^{\gamma}$ and low $\eta^{\gamma}$. The production of isolated
photons in association with jets has been measured as a function of
$E_T^{\gamma}$, $\eta^{\gamma}$ and $\xo$ (see Figs.~\ref{fig16}c to
\ref{fig16}e). The NLO calculations give a better description of these
data than for inclusive photons, except at high $\xo$.

Isolated photons have also been measured~\cite{h1797} in NC DIS by the H1
Collaboration. In this case, there are two major contributions to
photon emission, from the lepton and from the quark lines. The
measurements have been performed inclusively and in association with
jets as functions of photon transverse energy and pseudorapidity (see
Figs.~\ref{fig16}f, g and h). The LO calculations describe the shape
of the data but underestimate the normalisation by a factor of
approximately two, which can be attributed to an underestimation of the
quark-line contribution. The NLO calculations, which are only
available for photons in association with jets in this process, are
higher than the LO predictions, but still below the data. Therefore,
the theoretical understanding of these processes needs to be improved.

\begin{figure}[h]
\setlength{\unitlength}{1.0cm}
\begin{picture} (18.0,6.5)
\put (3.0,-0.3){\epsfig{figure=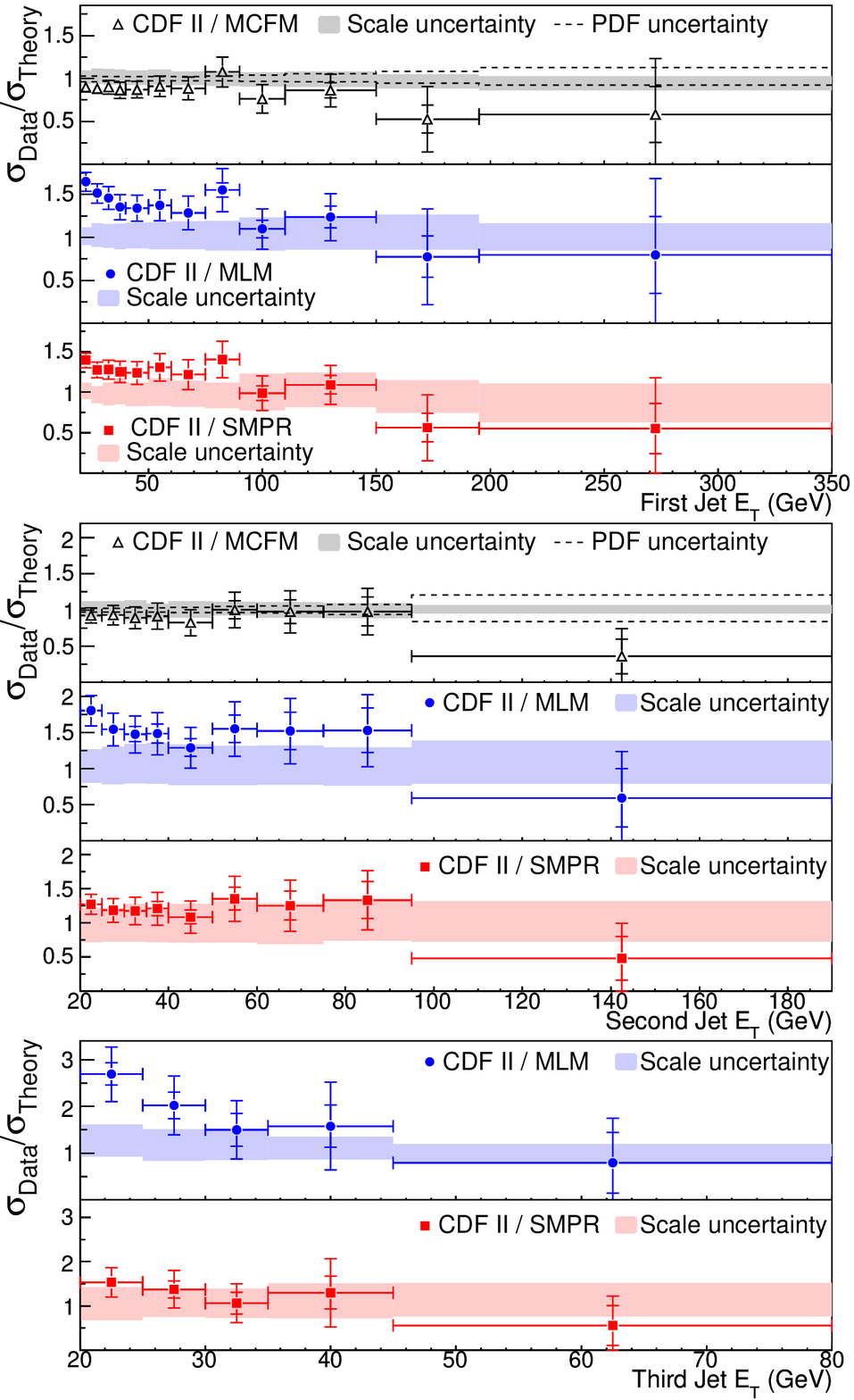,width=4cm}}
\put (9.0,-0.5){\epsfig{figure=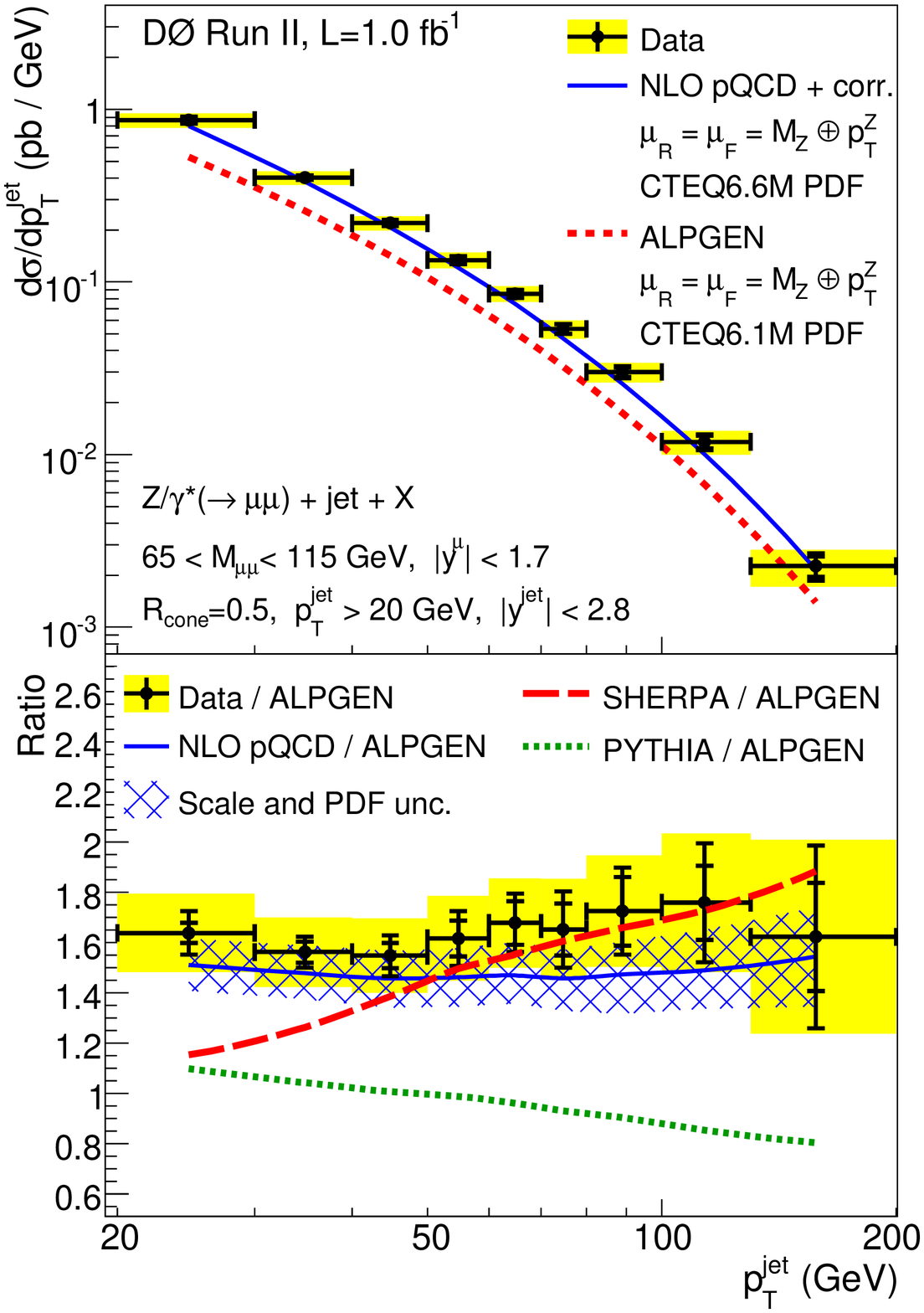,width=5cm}}
\put (7.0,4.5){\small (a)}
\put (14.0,4.5){\small (b)}
\end{picture}
\caption{\label{fig18}
{(a) $W$+jets cross sections as functions of $\etjet$; (b) $Z$+jets
  cross section as a function of $\etjet$.}}
\end{figure}

The production of vector bosons in association with jets in $\pp$
collisions is a key channel for studying top production within the SM
as well as for searches of Higgs and new physics. The study of the
production via QCD processes also provides a stringent test of the
theory. The CDF and D\O\ Collaborations have measured the production
of $W$~\cite{cdf416} and $Z$~\cite{d0604} bosons in association with
jets. Figure~\ref{fig18}a shows the measurements of $W+$jets from 
CDF, as a function of $\etjet$ of the first, second and third jets, and
Fig.~\ref{fig18}b shows the measurement of $Z+$jets from D\O, as a
function of $\etjet$. The pQCD calculations, which are NLO for up to
two jets, are compared with the measurements. The NLO calculations
give a good description of the total and differential cross sections.

\begin{figure}[h]
\setlength{\unitlength}{1.0cm}
\begin{picture} (18.0,5.0)
\put (0.5,-0.5){\epsfig{figure=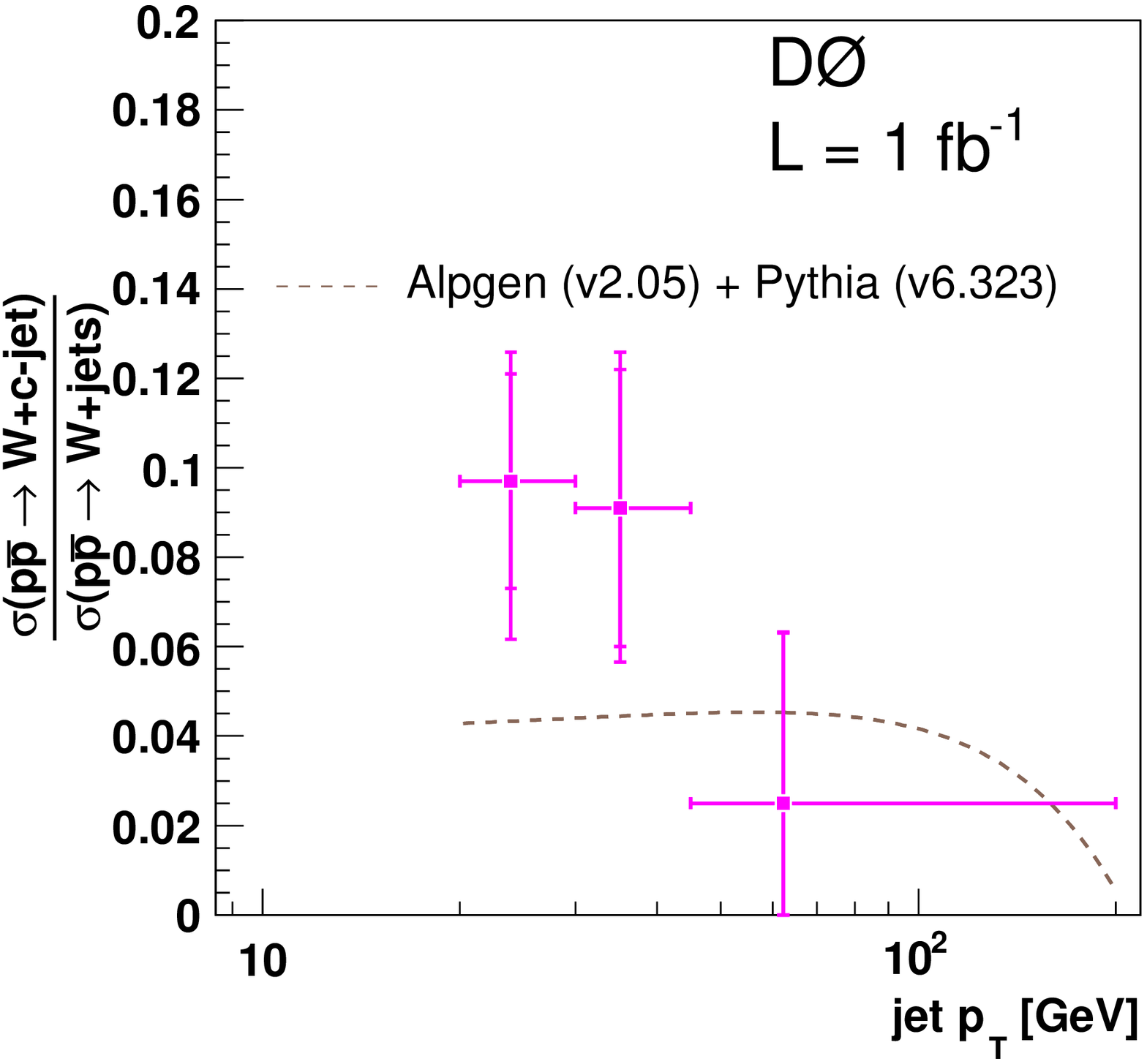,width=6cm}}
\put (6.5,-0.5){\epsfig{figure=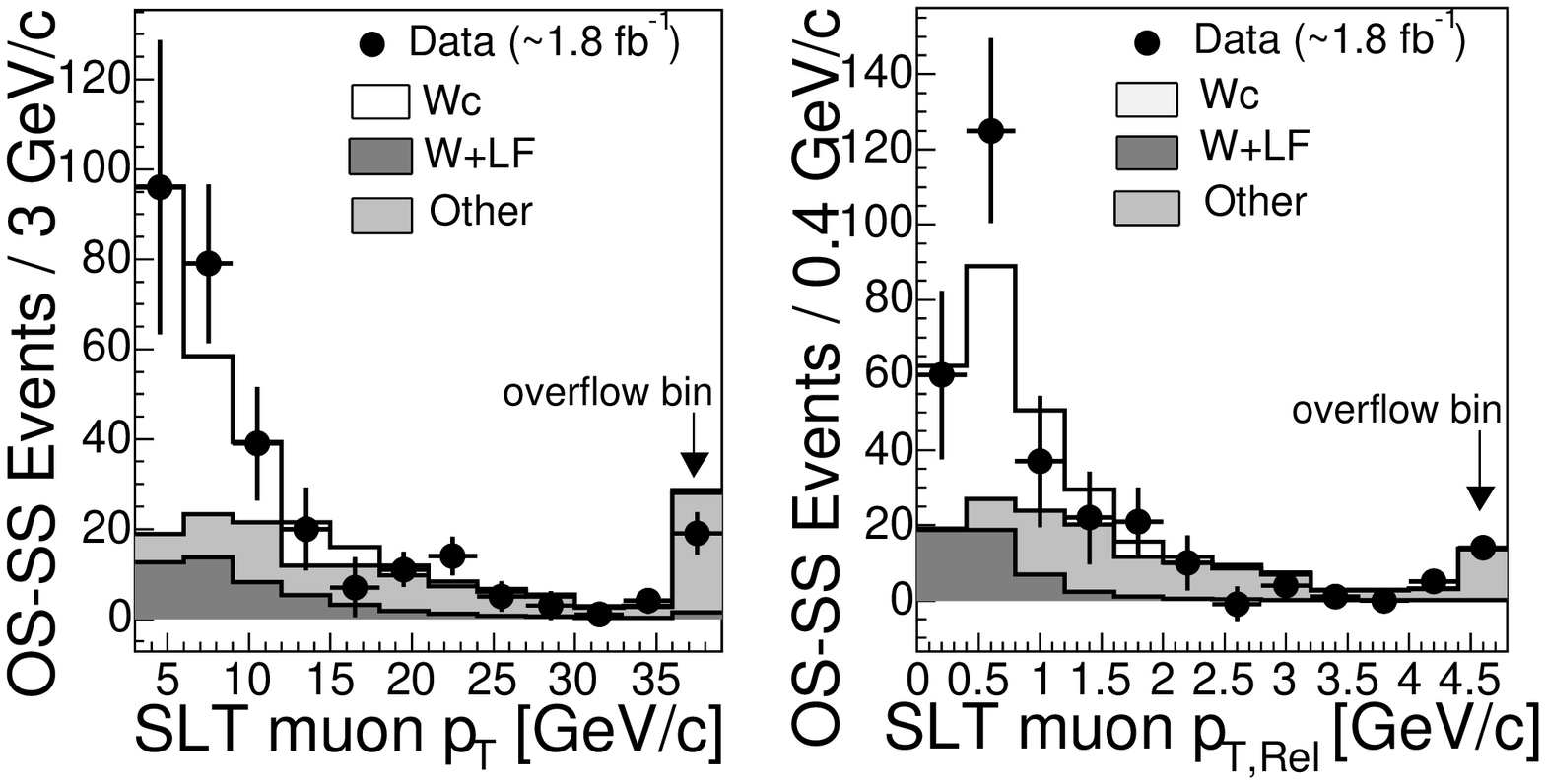,width=10cm}}
\put (5.5,4.0){\small (a)}
\put (10.5,4.0){\small (b)}
\put (15.5,4.0){\small (c)}
\end{picture}
\caption{\label{fig19}
{(a) Ratio of the $\sigma(W+c-{\rm jet})$ to $\sigma(W+{\rm jets})$;
  (b) muon $p_T$ and $p_{T,{\rm rel}}$ distributions.}}
\end{figure}

$W+c$-jet production at TevaTron is dominated by the $s$-gluon fusion
channel and so is sensitive to the $V_{\rm cs}$ matrix element. The
measurements are also sensitive to the $s$ and $g$ PDFs. This process
constitutes a background to top, Higgs and stop production as well as
to other searches for new physics. The D\O\ Collaboration has
measured~\cite{d0605} the total fraction of $W+c$-jet to $W+$jets,
$\frac{\sigma(W+c-{\rm jet})}{\sigma(W+{\rm jets})}$, to be 
$0.074\pm 0.019({\rm stat.})_{-0.014}^{+0.012}({\rm syst.})$,
as well as differentially as a function of $\etjet$ (see
Fig.~\ref{fig19}a). The CDF Collaboration has measured~\cite{cdf967}
the total cross section times the branching ratio of the 
$W\rightarrow l\nu$ channel, 
$\sigma_{W+c-{\rm jet}}\times Br(W\rightarrow l\nu)=9.8\pm 3.2$ pb.
Figures~\ref{fig19}b and \ref{fig19}c show the distributions of the
difference between same-sign and opposite-sign events as a function of
the lepton $p_T$, which shows clearly the signal. The predictions are
in reasonable agreement with the data. These measurements provide a
direct experimental evidence for the signal and constitute an
experimental validation of the $W+c$ theoretical prediction for use in
searches.

\begin{figure}[h]
\setlength{\unitlength}{1.0cm}
\begin{picture} (18.0,3.5)
\put (2.0,-0.5){\epsfig{figure=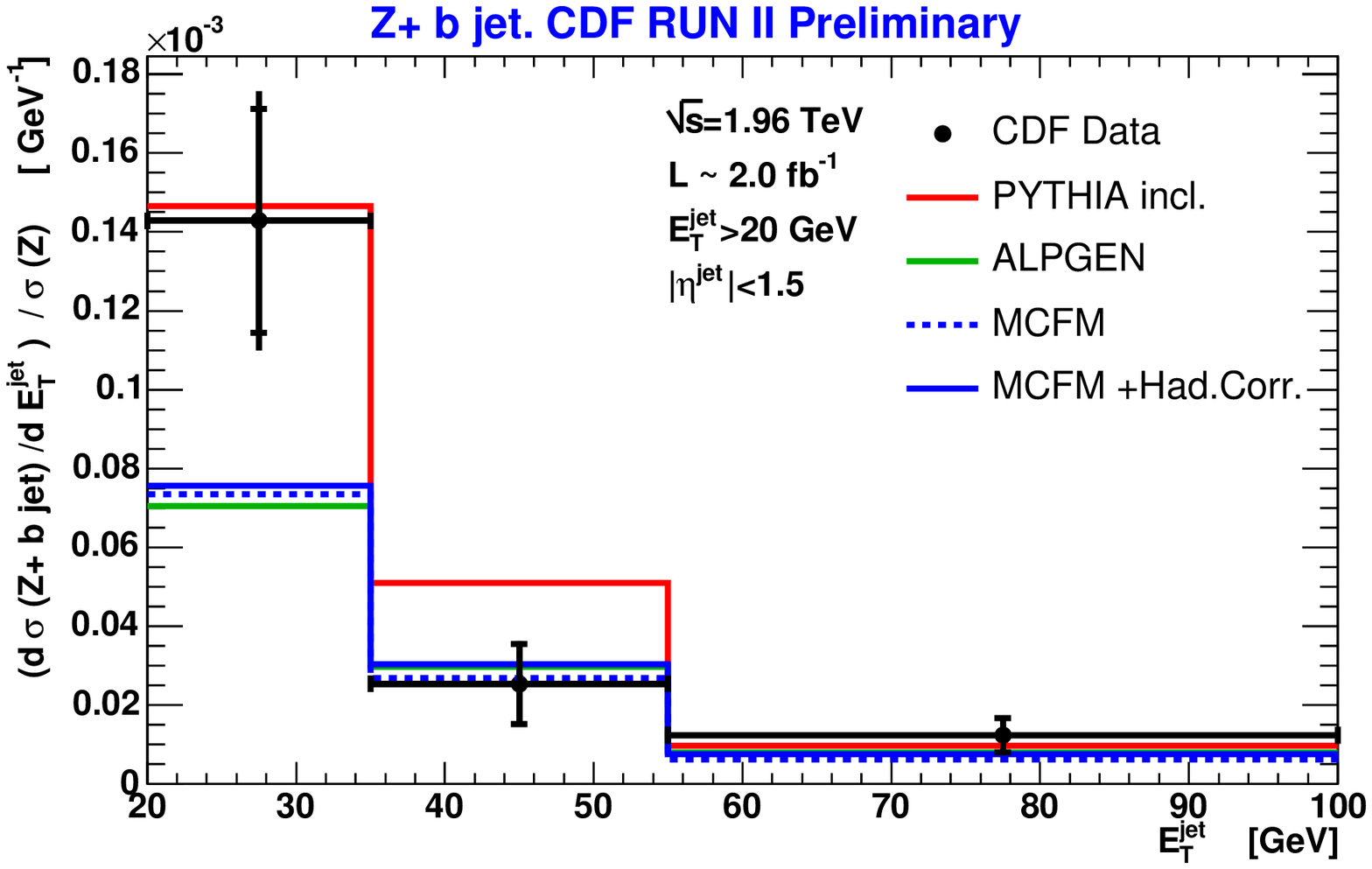,width=7cm}}
\put (9.0,-0.5){\epsfig{figure=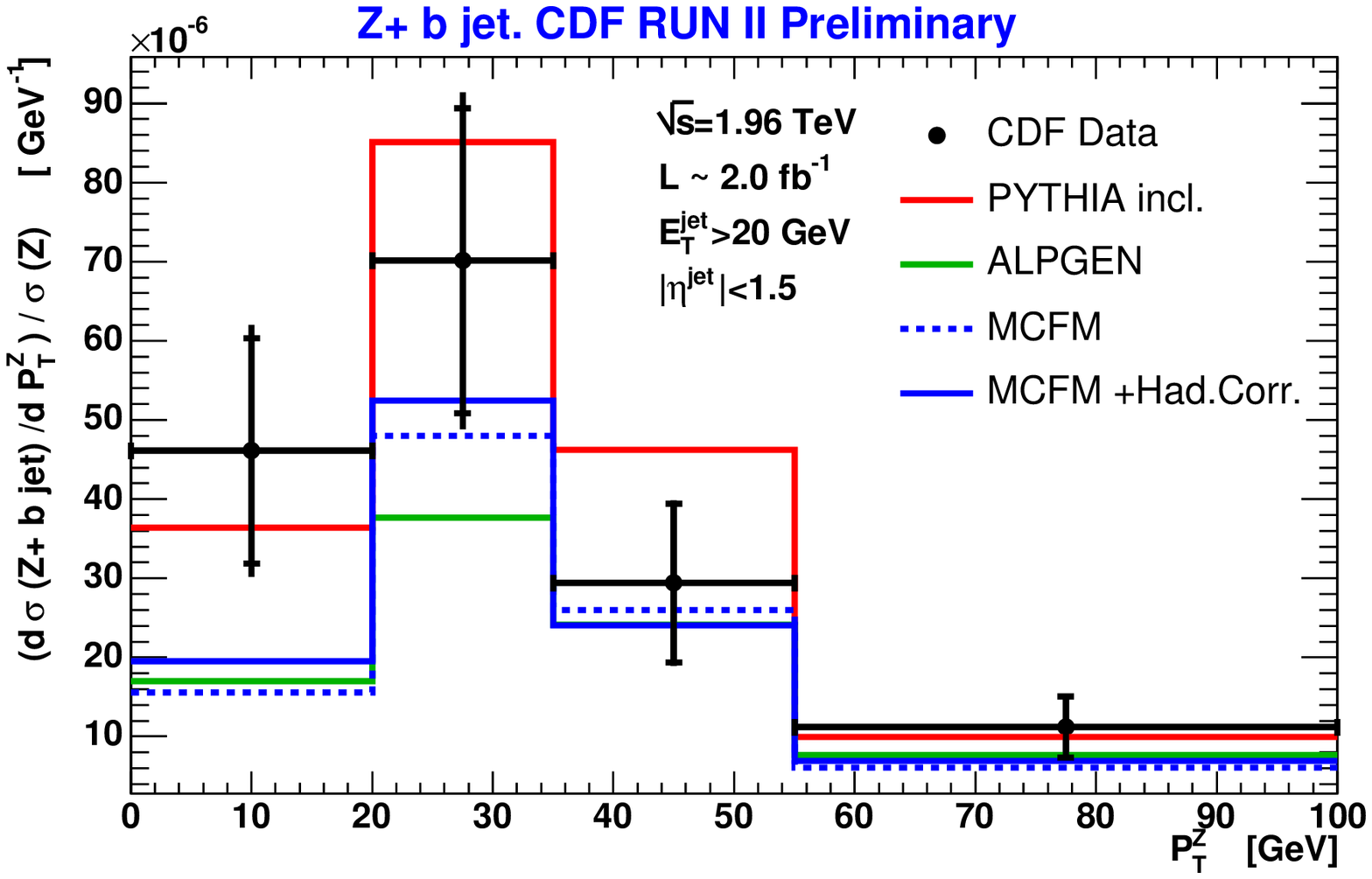,width=7cm}}
\end{picture}
\caption{\label{fig20}
{Normalised $Z+b$-jet cross sections as functions of $\etjet$ (left)
  and $p_T^Z$ (right).}}
\end{figure}

$W+b$-jet and $Z+b$-jet production are also important backgrounds to
Higgs searches and are sensitive to the $b$ parton density needed to
predict the production of Higgs, SUSY, top and other new
particles. The CDF Collaboration has measured~\cite{cdf415} the total
cross sections for $W+b$-jet and $Z+b$-jet production, 
$\sigma(Z\!+\!b- {\rm jet})=0.86\pm 0.14\pm 0.12$ pb and
$\sigma(W\!+\!b- {\rm jet})\!\times\! BR(W\rightarrow l\nu)=2.74\pm
0.27\pm 0.42$ pb.
The predictions are 0.53 pb (NLO) and 0.78 pb ({\sc Alpgen}),
respectively. The measured $\sigma(Z\!+\!b- {\rm jet})$ cross section
is somewhat higher than the NLO prediction. The normalised
differential cross sections as functions of $\etjet$ and $p_T^Z$ show
that the predictions fall below the data at low values but describe
the data well at high $\etjet$ and high $p_T^Z$ (see
Fig.~\ref{fig20}). For $W+b$-jet production, the LO prediction is
about 3.5 times lower than the data. The calculation has an
uncertainty of $30-40\%$, whereas the data has an uncertainty of only
$18\%$, so these measurements should be very helpful to constrain the
theory.

\begin{figure}[h]
\setlength{\unitlength}{1.0cm}
\begin{picture} (18.0,10.5)
\put (2.0,5.0){{\epsfig{figure=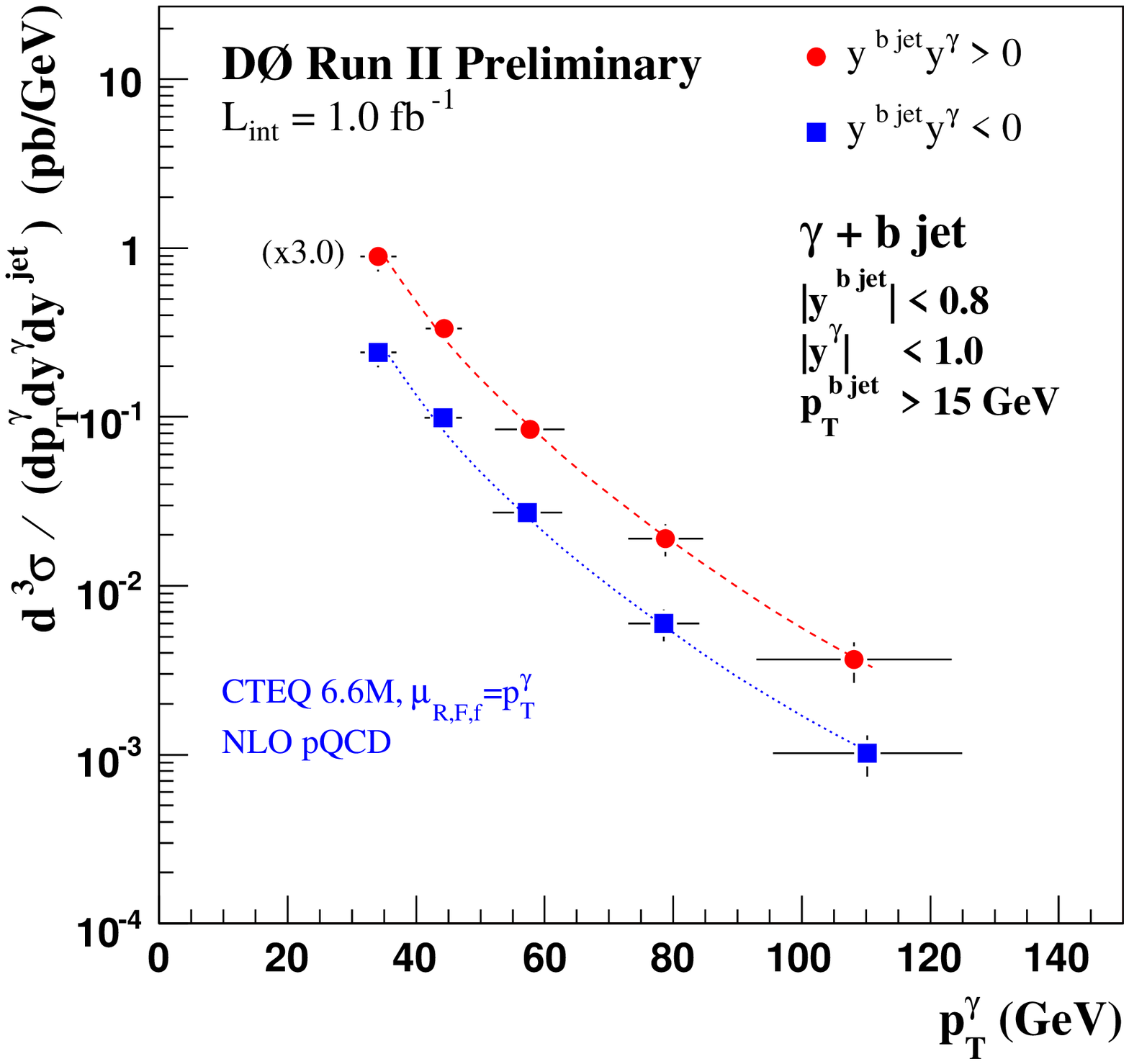,width=6.0cm}}}
\put (8.5,5.0){{\epsfig{figure=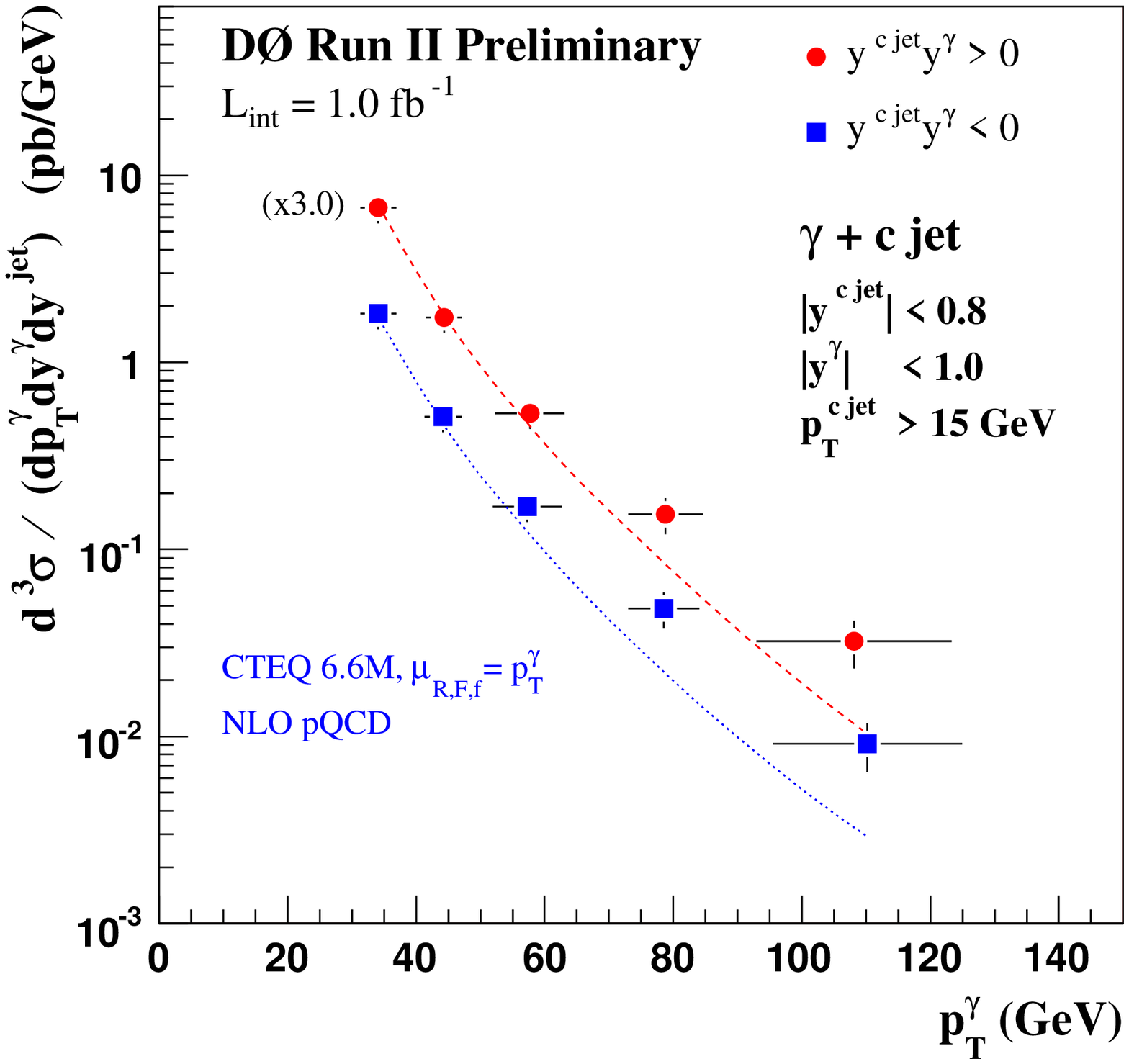,width=6.0cm}}}
\put (2.0,-0.5){{\epsfig{figure=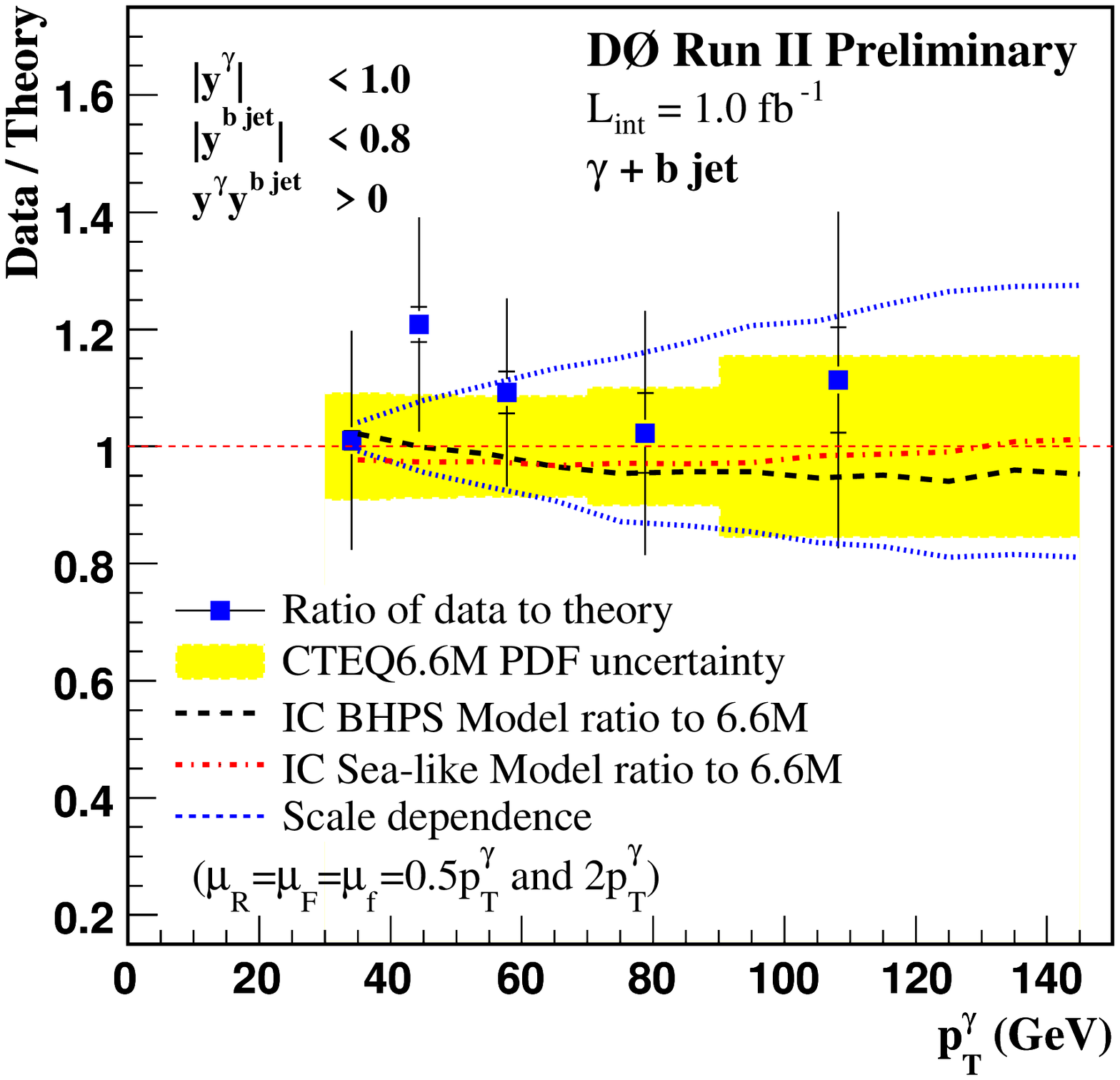,width=6.0cm}}}
\put (8.5,-0.5){{\epsfig{figure=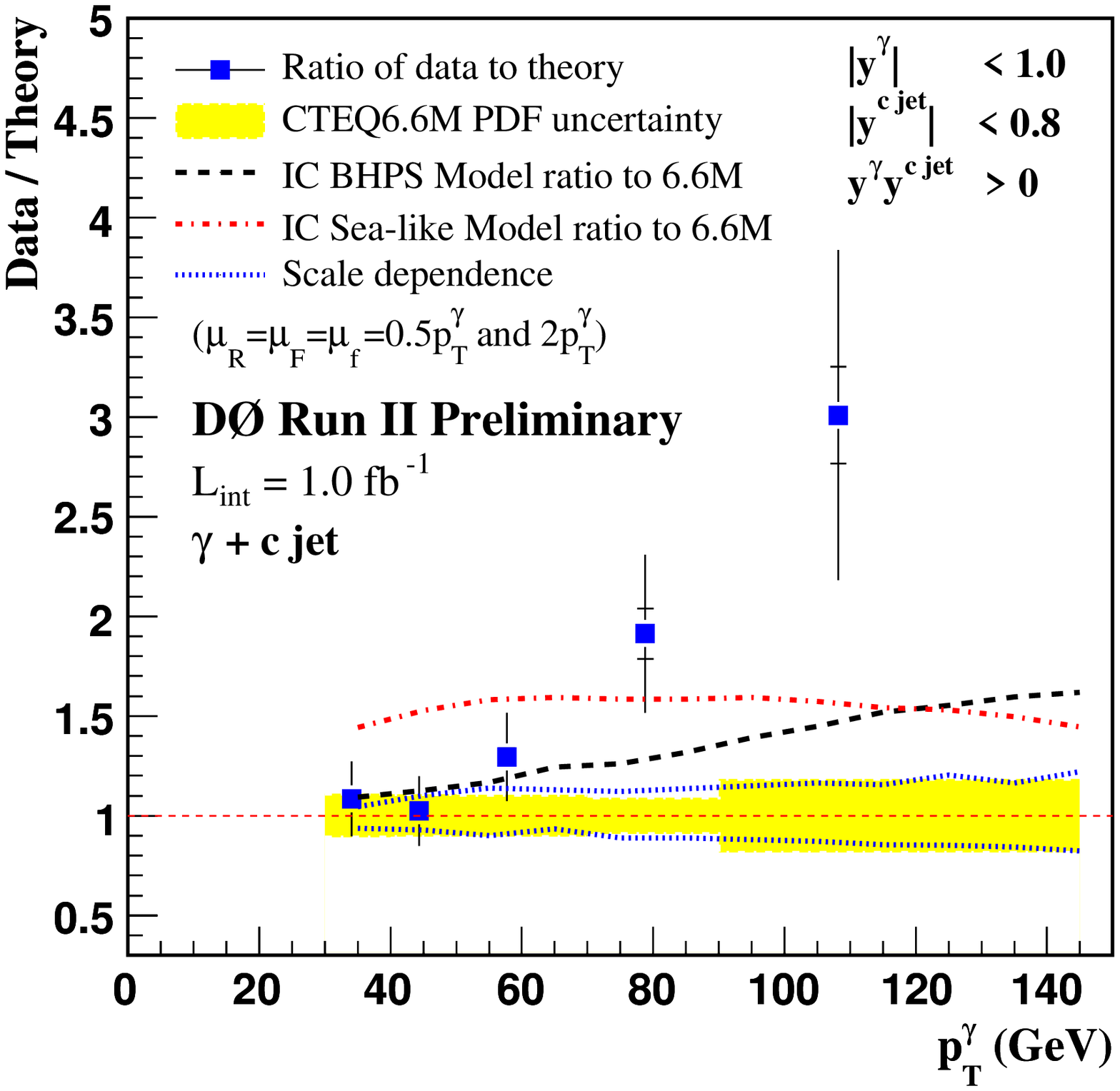,width=6.0cm}}}
\end{picture}
\caption{\label{fig21}
{Photon$+b$-jets (left) and photon$+c$-jets (right) cross sections as
  functions of $p_T^{\gamma}$.}}
\end{figure}

The production of photons in association with $b$- or $c$-jets also
provides a test of QCD. The D\O\ Collaboration has
measured~\cite{d0526} the cross sections for photons plus $b$- or
$c$-jets as functions of the photon transverse energy in different
rapidity configurations. Figure~\ref{fig21} shows the measurements
together with the NLO calculations. The predictions are in good
agreement with the data for photon$+b$-jets in all the $E_T^{\gamma}$
range and for photon$+c$-jets for $E_T^{\gamma}<50$ GeV. The
disagreement between the photon$+c$ measurements and the theory is
seen to grow with increasing $E_T^{\gamma}$. The origin of this
discrepancy is not yet clear; it could be attributed, for instance, to
intrinsic charm in the proton or to uncertainties in the splitting of
gluons into heavy-quark pairs.

\section{Parton dynamics at low $x$}

One of the main channels of Higgs production at LHC is 
$gg\rightarrow H$ via a top-quark loop. The predictions for this
process need information on the parton evolution at low $x$. This
information can be obtained from low-$x$ jet data at HERA. At high
scales, calculations at NLO using DGLAP evolution give a good
description of the data. However, DGLAP evolution is expected to break
down at low $x$. Other approaches to parton dynamics at low $x$
include the BFKL and CCFM evolution schemes. One way to study these
effects is the one proposed by M\"uller and Navelet, which consists of
analysing the production of jets close to the proton beam direction at
HERA or ``forward jets''. 

\begin{figure}[h]
\setlength{\unitlength}{1.0cm}
\begin{picture} (18.0,6.0)
\put (2.0,-0.5){\epsfig{figure=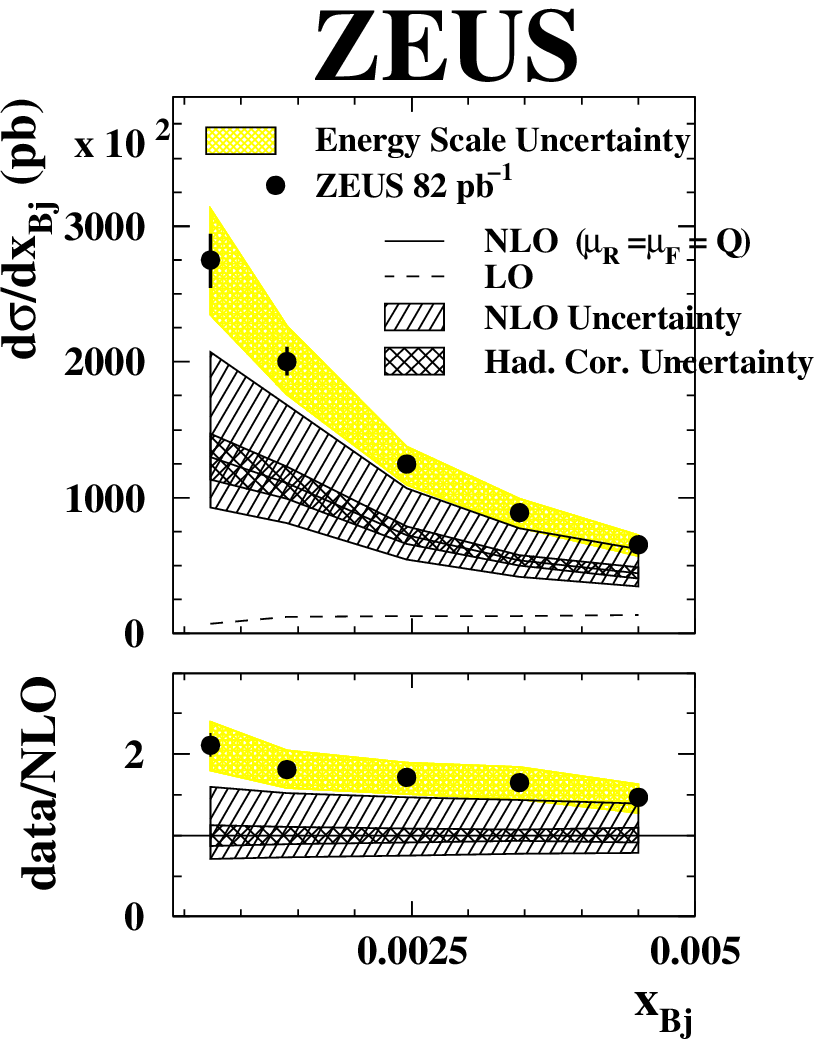,width=5cm}}
\put (8.0,-0.5){\epsfig{figure=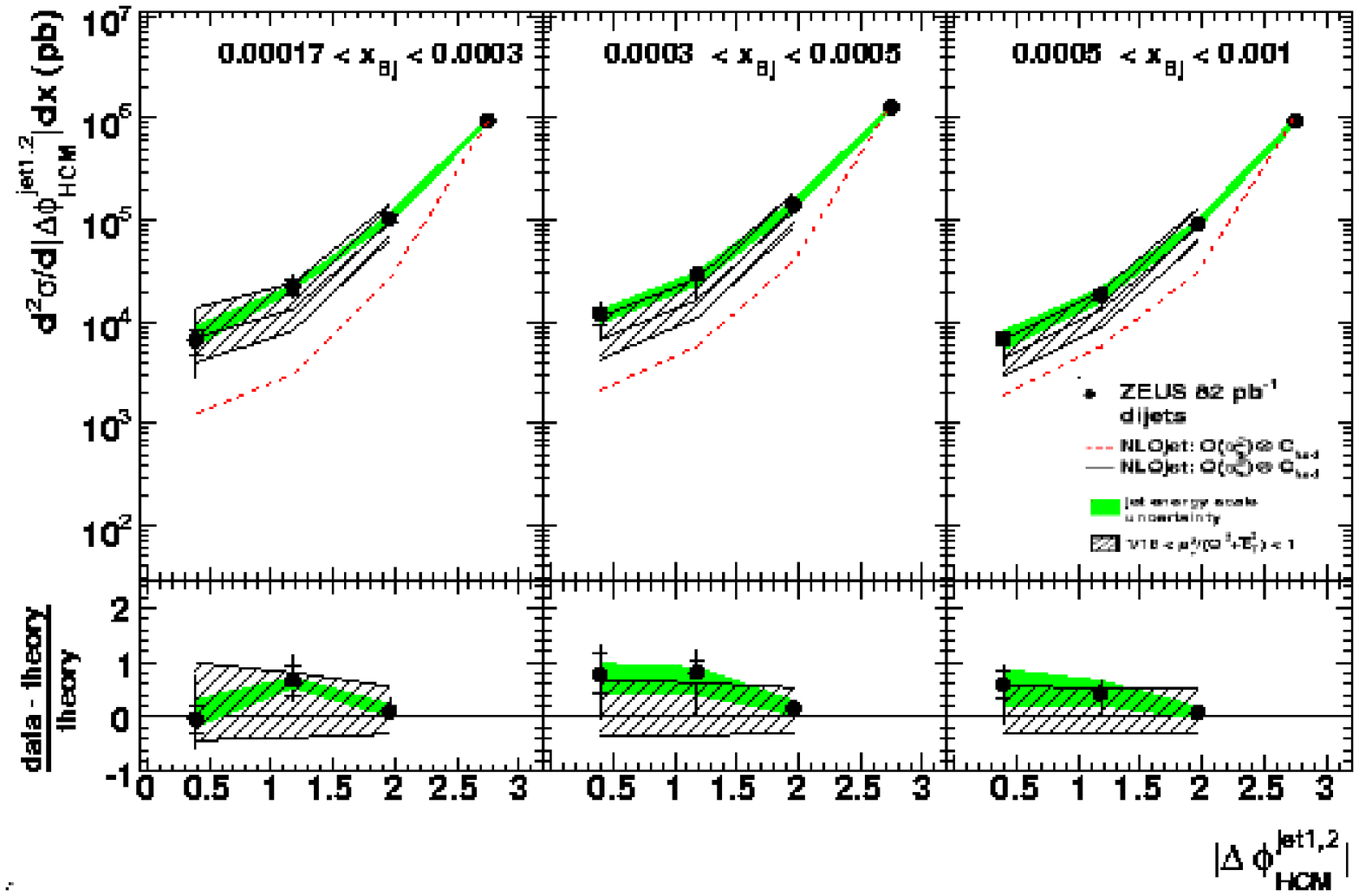,width=9cm}}
\put (5.0,3.0){\small (a)}
\put (9.5,4.5){\small (b)}
\end{picture}
\caption{\label{fig22}
{(a) Forward-jet cross section as a function of $x_{\rm Bj}$; (b)
  dijet cross sections as functions of $|\Delta\phi|$ in different
  $x_{\rm Bj}$ regions.}}
\end{figure}

To search for breakdown of DGLAP evolution, the ZEUS Collaboration has
measured~\cite{zeus126} forward-jet production at low
$x$. Figure~\ref{fig22}a shows the cross section as a function of $x$,
together with the pQCD predictions; the measured cross section
increases as $x$ decreases. The $\oas$ predictions are well below the
data, whereas the $\oass$ calculations are closer to the data, but
still fall short. The $\oass$ calculation is much larger than the
$\oas$ calculation and has large uncertainties; this indicates that
higher orders are important. This can be understood by the opening of
a new channel (gluon exchange in the $t$-channel) in these
calculations, so that the $\oass$ calculation becomes an effective LO
estimation.

\begin{figure}[h]
\setlength{\unitlength}{1.0cm}
\begin{picture} (18.0,4.5)
\put (2.0,-0.5){\epsfig{figure=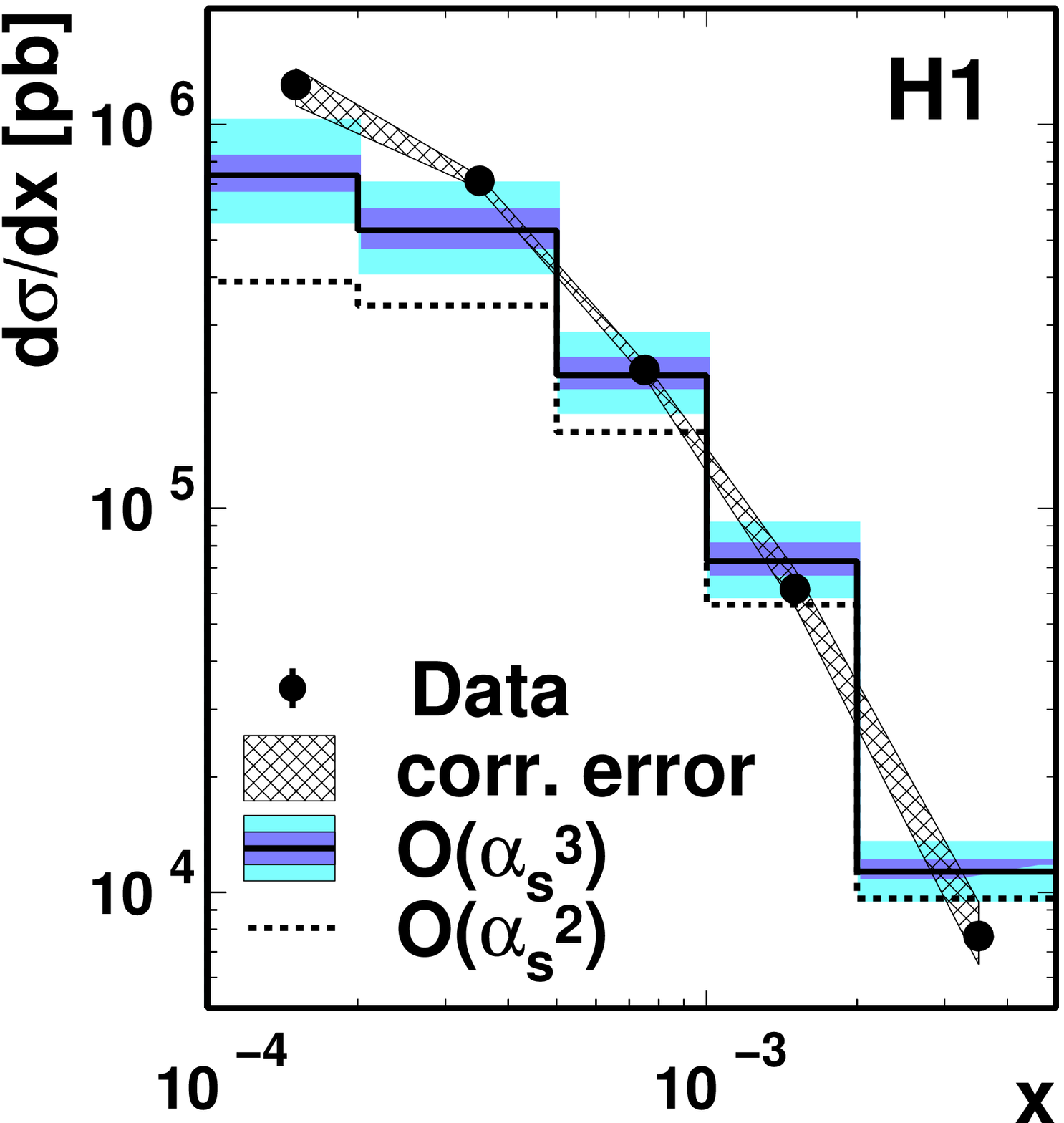,width=5cm}}
\put (8.0,-0.5){\epsfig{figure=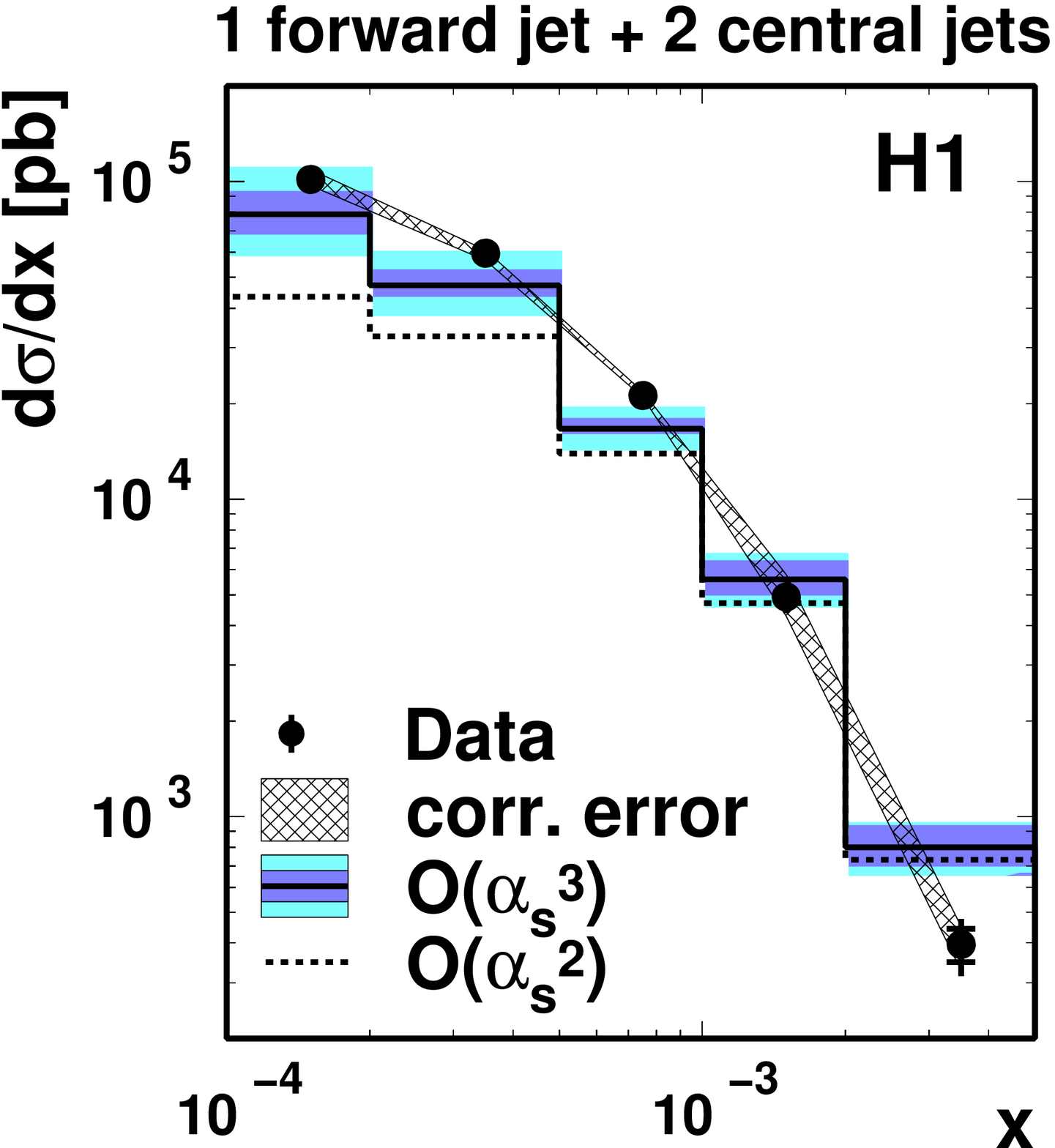,width=5cm}}
\end{picture}
\caption{\label{fig24}
{Three-jet cross section as a function of $x$ for all events (left) and
  events with two central jets and one forward jet (right).}}
\end{figure}

Multi-jet cross sections are better suited to test parton dynamics
at low $x$ since for dijet and three-jet cross sections a ``genuine''
NLO calculation can be performed. The ZEUS Collaboration
has measured~\cite{zeus122} the angular correlation as a function of
$\Delta\phi$ for dijets in different regions of $x$ (see
Fig.~\ref{fig22}b). The $\oass$ calculations are one order of magnitude
below the data for small jet angular separations, whereas the $\oasss$
calculations describe the data much better; this demonstrates the
importance of the higher orders at low $x$. The H1 Collaboration has 
measured~\cite{h1798} the $x$ distribution for three-jet events and
for the configuration of two central jets and one forward jet (see
Fig.~\ref{fig24}). The $\oasss$ calculation describes the data at low
$x$ reasonably well.

\section{Summary and conclusions}

A wealth of new measurements from HERA and TevaTron test QCD nowadays
with high precision. These data probe the theory up to the highest
available energies and down to the lowest possible $x$ values,
phase-space regions of special interest at LHC. The exploration of
these new regimes may well lead towards a new level of understanding
hard processes which will be crucial for interpreting any physics at
LHC.

\begin{figure}[h]
\setlength{\unitlength}{1.0cm}
\begin{picture} (18.0,7.0)
\put (1.0,-0.5){\centerline{\epsfig{figure=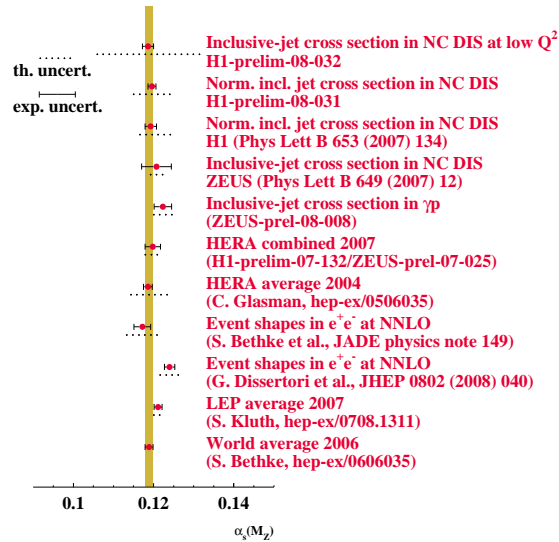,width=8cm}}}
\end{picture}
\caption{\label{fig25}
{Summary of recent $\asz$ determinations compared to the current world
average.}}
\end{figure}

In particular, the underlying event has been tested in all possible
environments, new high precision data will help to constrain further
the proton PDFs, more and more precise determinations of the strong
coupling are being obtained (see Fig.~\ref{fig25}) and successful
tests of colour dynamics at the highest available energies and down to
the lowest possible $x$ values have been performed.

For further progress in understanding QCD and take full advantage of
the available data, higher-orders corrections will be extremely useful.

\vspace{0.5cm}
\noindent {\bf Acknowledgments}.
I would like to thank the organisers of the ICHEP08 conference for
providing me with the opportunity of giving this talk and for a well
organised conference. I would like to thank C. Diaconu, 
G. Dissertori, K. Hatakeyama, G. Hesketh, A. Juste, S. Kluth, C. Pahl,
S. Soldner-Rembold, A. Specka, J. Terr\oo n and M. Wobisch for their
help in preparing the talk. Special thanks to \mbox{Prof. J. Terr\'on}
for a critical reading of this manuscript.


\end{document}